\pdfoutput=1

\documentclass[]{INCLUDES/llncs}
\usepackage{TOOLS/dcpic,pictexwd}
\usepackage{graphicx}
\usepackage{subfig}
\usepackage{url}
\usepackage{verbatim} 
\usepackage{listings}
\lstnewenvironment{codex}
    {\lstset{}%
      \csname lst@SetFirstLabel\endcsname}
    {\csname lst@SaveFirstLabel\endcsname}
\lstnewenvironment{code}
    {\lstset{}%
      \csname lst@SetFirstLabel\endcsname}
    {\csname lst@SaveFirstLabel\endcsname}
\newtheorem{prop}{Proposition}
\newtheorem{conj}{Conjecture}

\newtheorem{df}{Definition}
\newtheorem{lem}{Lemma}

\newcommand{\BI}[0]{\begin{itemize}}
\newcommand{\EI}[0]{\end{itemize}}

\newcommand{\BE}[0]{\begin{enumerate}}
\newcommand{\EE}[0]{\end{enumerate}}

\newcommand{\BX}[0]{\begin{codex}}
\newcommand{\EX}[0]{\end{codex}}
\newcommand{\BV}[0]{\begin{verbatim}}
\newcommand{\EV}[0]{\end{verbatim}}
    
\def \bscale1 {0.25}
\def \bscale {0.25}

\newcommand{\FIG}[4]{
\begin{figure}[htbp]
\centering
{\includegraphics[scale=#3]{./figs/#4}}
\caption{#2}
\label{#1}
\end{figure}
}

\newcommand{\VFIGS}[6]{
\begin{figure}[htbp]
  \begin{center}
    {\includegraphics[scale=0.40]{./figs/#5}}
    {\includegraphics[scale=0.40]{./figs/#6}}
  \caption{#2: {\em #3} and {\em #4}}
  \label{#1}
  \end{center}
\end{figure}
}

\title{
    Isomorphic Data Encodings in Haskell and their
    Generalization to Hylomorphisms on Hereditarily Finite Data Types
}
\author{Paul Tarau}
\institute{
   Department of Computer Science and Engineering\\
   University of North Texas\\
   {\em E-mail: tarau@cs.unt.edu}
}

\begin{document}
\maketitle
\date{}

\begin{abstract}
This paper is an exploration in a functional programming framework
of {\em isomorphisms} between elementary
data types (natural numbers, sets, multisets, finite functions, permutations
binary decision diagrams, graphs, hypergraphs,
parenthesis languages, dyadic
rationals, primes, DNA sequences etc.) and their extension to hereditarily
finite universes through {\em hylomorphisms} derived
from {\em ranking/unranking} and
{\em pairing/unpairing} operations.

An embedded higher order {\em combinator language} provides
any-to-any encodings automatically.

Besides applications to experimental mathematics,
a few examples of
``free algorithms'' obtained by transferring 
operations between data types 
are shown. Other applications range from
stream iterators on
combinatorial objects to
self-delimiting codes,
succinct data representations and generation
of random instances.

The paper covers 59 data types and, through the use of
the embedded combinator language, provides 3540 distinct
bijective transformations between them.

The self-contained source code of the paper, as generated from a
literate Haskell program, is available at
\url{http://logic.csci.unt.edu/tarau/research/2008/fISO.zip}.

A short, 5 page version of the paper,
published as \cite{sac09fISO} describes
the idea of organizing various data transformations
as encodings to sequences of natural numbers and
gives a few examples of hylomorphisms that lift the
encodings to related hereditarily finite universes.

\end{abstract}

\section{Introduction}

Analogical/metaphorical thinking routinely shifts entities and
operations from a field to another hoping to uncover similarities in representation
or use \cite{lakoff}.

Compilers convert programs from human centered to machine centered
representations - sometime reversibly.

Complexity classes are defined through compilation with limited resources
(time or space) to similar problems
\cite{Cook04theoriesfor,Cook93functionalinterpretations}.

Mathematical theories often borrow proof patterns and reasoning
techniques across close and sometime not so close fields.

A relatively small number of universal data types are used as basic building
blocks in programming languages and their runtime interpreters,
corresponding to a few well tested mathematical abstractions like sets,
functions, graphs, groups, categories etc.

A less obvious leap is that if heterogeneous objects can be seen
in some way as isomorphic, then we can share them and
compress the underlying informational universe by collapsing
isomorphic encodings of data or programs whenever possible.

Sharing heterogeneous data objects faces two problems:
\begin{itemize}
\item some form of equivalence needs to be proven between two objects A and B before A can 
replace B in a data structure, a possibly tedious and error prone task

\item the fast growing diversity of data types makes harder and harder to
recognize sharing opportunities.
\end{itemize}

Besides, this rises the question: what
guaranties do we have that sharing across heterogeneous
data types is useful and safe?

The techniques introduced in this paper provide a generic solution to these
problems, through isomorphic mappings between heterogeneous data types, 
such that unified internal representations 
make equivalence checking and sharing possible. 
The added benefit of these ``shapeshifting" 
data types is that the functors transporting their data content will also transport their 
operations, resulting in  shortcuts that provide, for free, implementations of 
interesting algorithms. The simplest instance is the case of
isomorphisms -- reversible mappings that also transport operations.
In their simplest form such isomorphisms show up as {\em encodings}
to some simpler and easier to manipulate representation, for
instance natural numbers. 

Such encodings can be traced back to G\"{o}del
numberings \cite{Goedel:31,conf/icalp/HartmanisB74} associated to formulae,
but a wide diversity of common computer operations, ranging
from data compression and serialization to
wireless data transmissions and cryptographic codes qualify.
 
Encodings between data types provide a variety of services ranging from
free iterators and random objects to data compression and succinct
representations. Tasks like serialization and persistence are facilitated
by simplification of reading or writing operations without the need of
special purpose parsers. Sensitivity to internal data representation
format or size limitations can be circumvented without extra programming
effort.

\section{An Embedded Data Transformation Language}

We will start by designing an embedded transformation language
as a set of operations on a groupoid of isomorphisms. We will then
extended it with a set of higher order combinators mediating the
composition of the encodings and the transfer of operations between 
data types.

\subsection{The Groupoid of Isomorphisms}

We implement an isomorphism between two objects X and Y as a 
Haskell data type encapsulating a bijection $f$ and its inverse $g$. 
We will call the {\em from} function the first component (a {\em
section} in category theory parlance) and
the {\em to} function the second component (a {\em retraction}) defining
the isomorphism.
We can organize isomorphisms as a {\em groupoid} as follows:

\begindc{\commdiag}[5]
\obj(14,11){$X$}
\obj(39,11){$Y$}
\mor(14,12)(39,12){$f=g^{-1}$}
\mor(39,10)(14,10){$g=f^{-1}$}
\enddc

\begin{code}
data Iso a b = Iso (a->b) (b->a)

from (Iso f _) = f
to (Iso _ g) = g

compose :: Iso a b -> Iso b c -> Iso a c
compose (Iso f g) (Iso f' g') = Iso (f' . f) (g . g')
itself = Iso id id
invert (Iso f g) = Iso g f
\end{code}
Assuming that for any pair of type {\tt Iso a b},  $f \circ g = id_a$ and $g
\circ f=id_b$, we can now formulate {\em laws} about isomorphisms
that can be used to test correctness of implementations with tools like
QuickCheck \cite{DBLP:journals/sigplan/ClaessenH02}.
\begin{prop} 
The data type Iso has a groupoid structure, i.e. the {\em compose} operation,
when defined, is associative, {\em itself} acts as an identity element
and {\em invert} computes the inverse of an isomorphism.
\end{prop}

We can transport operations from an object to another
with {\em borrow} and {\em lend} combinators defined as follows:

\begin{code}
borrow :: Iso t s -> (t -> t) -> s -> s
borrow (Iso f g) h x = f (h (g x))
borrow2 (Iso f g) h x y = f (h (g x) (g y))
borrowN (Iso f g) h xs = f (h (map g xs))

lend :: Iso s t -> (t -> t) -> s -> s
lend = borrow . invert
lend2 = borrow2 . invert
lendN = borrowN . invert
\end{code}

The combinators {\tt fit} and {\tt retrofit} 
just transport an object {\tt x}
through an isomorphism and and apply 
to it an operation {\tt op} available on
the other side:
\begin{code}
fit :: (b -> c) -> Iso a b -> a -> c
fit op iso x = op ((from iso) x)

retrofit :: (a -> c) -> Iso a b -> b -> c
retrofit op iso x = op ((to iso) x)
\end{code}

We can see the combinators {\tt from, to, compose, itself, invert, borrow,
lend, fit etc.} as part of an {\em embedded data transformation language}.
Note that in this design we borrow from our
strongly typed host programming language
its abstraction layers and safety mechanisms
that continue to check the semantic validity of
the embedded language constructs.

\subsection{Choosing a Root}
To avoid defining $n(n-1)/2$ isomorphisms between $n$ objects,
we choose a {\em Root} object to/from which we will actually
implement isomorphisms. We will extend our embedded
combinator language using the groupoid structure of the isomorphisms
to connect any two objects through isomorphisms to/from
the {\em Root}.

Choosing a {\em Root} object is somewhat arbitrary, but it makes sense to
pick a representation that is relatively easy convertible to various
others, efficiently implementable and, last but not least, scalable to
accommodate large objects up to the runtime system's 
actual memory limits.

We will choose as our {\em Root} object {\em finite sequences of natural
numbers}. They can be seen as {\tt finite functions} from an initial 
segment of $Nat$, say $[0..n]$, to $Nat$.
 This implies that a finite
function can be seen as an array or a list of natural 
numbers except that we do not limit the size of 
the representation of its values.
We will represent them as lists i.e. their Haskell type is $[Nat]$.
\begin{code}
type Nat = Integer
type Root = [Nat]
\end{code}
We can now define an {\em Encoder} as an isomorphism
connecting an object to {\em Root} 
\begin{code}
type Encoder a = Iso a Root
\end{code}
together with the combinators {\em with} and {\em as}
providing an {\em embedded transformation language} for routing
isomorphisms through two {\em Encoders}.
\begin{code}  
with :: Encoder a->Encoder b->Iso a b
with this that = compose this (invert that)

as :: Encoder a -> Encoder b -> b -> a
as that this thing = to (with that this) thing
\end{code}
The combinator {\tt with} turns two Encoders
into an arbitrary isomorphism, i.e. acts as a connection hub between
their domains. The combinator {\tt as} adds a more convenient syntax
such that converters between A and B can be designed as:
\begin{codex}
a2b x = as A B x
b2a x = as B A x
\end{codex}
\vskip 0.30cm
\begindc{\commdiag}[5]
\obj(26,0){$Root$}
\obj(14,11){$A$}
\obj(39,11){$B$}

\mor(26,0)(39,10){$b$}
\mor(26,0)(14,10){$a^{-1}$}
\mor(39,10)(26,0){$b^{-1}$}
\mor(14,10)(26,0){$a$}
\mor(14,12)(39,12){$a2b=as~B~A$}
\mor(39,10)(14,10){$b2a=as~A~B$}
\enddc
\vskip 0.30cm
A particularly useful combinator that
transports binary operations from an Encoder to another, {\tt
borrow\_from}, can be defined as follows:
\begin{code}
borrow_from :: Encoder a -> (a -> a -> a) -> Encoder b -> b -> b -> b
borrow_from other op this x y = borrow2 (with other this) op x y
\end{code}
Note that one can also use the more intuitive equivalent definition
\begin{code}
borrow_from' other op this x y = z where
  x' = as other this x
  y' = as other this y
  z' = op x' y'
  z =  as this other z' 
\end{code}
given that the following equivalence always holds:
\begin{equation}
borrow\_from \equiv borrow\_from'
\end{equation}
We will provide extensive use cases for these combinators
as we populate our groupoid of isomorphisms.
\noindent Given that $[Nat]$ has been chosen as the root, we will define our
finite function data type {\em fun} simply as the identity isomorphism 
on sequences in $[Nat]$.
\begin{code}  
fun :: Encoder [Nat]
fun = itself
\end{code}

\section{Extending the Groupoid of Isomorphisms}
We will now populate our groupoid of isomorphisms with
combinators based on a few primitive converters.

\subsection{An Isomorphism between Finite Multisets and Finite Functions}
\label{msetiso}

Multisets \cite{multisetOver} are unordered collections with repeated
elements. Non-decreasing sequences provide a canonical representation for
multisets of natural numbers. 
The isomorphism between finite multisets and finite functions is specified with
two bijections {\tt mset2fun} and {\tt fun2mset}.
\begin{code}
mset :: Encoder [Nat]
mset = Iso mset2fun fun2mset
\end{code}
While finite multisets and sequences representing finite functions share a
common representation $[Nat]$, multisets are subject to the implicit constraint that their
order is immaterial \footnote{Such
constraints can be regarded as {\em laws} that we assume about 
a given data type, when needed, restricting it to the
appropriate domain of the underlying mathematical concept.}.
This suggest that a multiset like $[4,4,1,3,3,3]$ could be
represented by first ordering it as $[1,3,3,3,4,4]$ and then compute the 
differences between consecutive elements i.e.
$[x_0 \ldots x_i, x_{i+1} \ldots] \rightarrow [x_0 \ldots x_{i+1}-x_i
\ldots]$.
This gives $[1,2,0,0,1,0]$, with
the first element $1$ followed by the increments $[2,0,0,1,0]$,
as implemented by {\tt mset2fun}:
\begin{code}
mset2fun = to_diffs . sort . (map must_be_nat)

to_diffs xs = zipWith (-) (xs) (0:xs)

must_be_nat n | n>=0 = n
\end{code}
It can now be verified easily that incremental sums of the
numbers in such a sequence
return the original set
in sorted form, as implemented by {\tt fun2mset}:
\begin{code}
fun2mset ns = tail (scanl (+) 0 (map must_be_nat ns)) 
\end{code}
The resulting isomorphism {\tt mset} can be applied directly using its two
components {\tt mset2fun} and {\tt fun2mset}. Equivalently, it can be
expressed more ``generically'' by using the {\tt as} combinator, as
follows:
\begin{codex}
*ISO> mset2fun [1,3,3,3,4,4]
[1,2,0,0,1,0]
*ISO> fun2mset [1,2,0,0,1,0]
[1,3,3,3,4,4]
*ISO> as fun mset [1,3,3,3,4,4]
[1,2,0,0,1,0]
*ISO> as mset fun [1,2,0,0,1,0]
[1,3,3,3,4,4]
\end{codex}

\subsection{An Isomorphism to Finite Sets of Natural Numbers}
While finite sets and sequences share a common representation $[Nat]$,
sets are subject to the implicit constraints that all their elements
are distinct and order is immaterial.
Like in the case of multisets, this suggest that a set like $\{7,1,4,3\}$ could
be represented by first ordering it as $\{1,3,4,7\}$ and then compute the 
differences between consecutive elements. This gives $[1,2,1,3]$, with
the first element $1$ followed by the increments $[2,1,3]$. To turn
it into a bijection, including $0$ as a possible member of a sequence,
another adjustment is needed: elements in the sequence of increments should
be replaced by their predecessors. This gives $[1,1,0,2]$ as implemented
by {\tt set2fun}:
\begin{code}
set2fun xs | is_set xs = shift_tail pred (mset2fun xs)

shift_tail _ [] = []
shift_tail f (x:xs) = x:(map f xs)

is_set ns = ns==nub ns
\end{code}
It can now be verified easily that predecessors of the incremental sums of the
successors of numbers in such a sequence, return the original set
in sorted form, as implemented by {\tt fun2set}:
\begin{code}
fun2set = (map pred) . fun2mset . (map succ)
\end{code}
The {\em Encoder} (an isomorphism with {\tt fun}) can be specified with the two
bijections {\tt set2fun} and {\tt fun2set}.
\begin{code}
set :: Encoder [Nat]
set = Iso set2fun fun2set
\end{code}
The Encoder ({\tt set}) is now ready to interoperate 
with another Encoder:
\begin{codex}
*ISO> as fun set [0,2,3,4,9]
[0,1,0,0,4]
*ISO> as set fun [0,1,0,0,4]
[0,2,3,4,9]
*ISO> as mset set [0,2,3,4,9]
[0,1,1,1,5]
*ISO> as set mset [0,1,1,1,5]
[0,2,3,4,9]
\end{codex}
As the example shows,the Encoder {\tt set} connects arbitrary lists of
natural numbers representing finite functions
to strictly increasing sequences
of (distinct) natural numbers representing sets.
Then, through the use of the combinator {\tt as}, sets represented by {\tt set}
are connected to multisets represented by {\tt mset}. This connection is
(implicitly) routed through a connection to {\tt fun}, as if
\begin{codex}
*ISO> as mset fun [0,1,0,0,4]
[0,1,1,1,5]
\end{codex}
were executed.

\subsection{Folding Sets into Natural Numbers} \label{natset}
We can fold a set, represented as a list of distinct
natural numbers into a single natural number,
reversibly, by observing that it can be seen
as the list of exponents of {\tt 2} in the number's
base {\tt 2} representation.

\begin{code}
nat_set = Iso nat2set set2nat 

nat2set n | n>=0 = nat2exps n 0 where
  nat2exps 0 _ = []
  nat2exps n x = 
    if (even n) then xs else (x:xs) where
      xs=nat2exps (n `div` 2) (succ x)

set2nat ns | is_set ns = sum (map (2^) ns)
\end{code}

We will standardize this pair of operations as an {\em Encoder} 
for a natural number using our Root as a mediator: 
\begin{code}
nat :: Encoder Nat
nat = compose nat_set set
\end{code}
Given that {\tt nat} is an isomorphism with the Root {\tt fun}, one can use
directly its {\tt from} and {\tt to} components:
\begin{codex}
*ISO> from nat 2008
[3,0,1,0,0,0,0]
*ISO> to nat it
2008
\end{codex}
Moreover, the resulting Encoder ({\tt nat}) is now ready to interoperate 
with any Encoder, in a generic way:
\begin{codex}
*ISO> as fun nat 2008
[3,0,1,0,0,0,0]
*ISO> as set nat 2008
[3,4,6,7,8,9,10]
*ISO> as nat set [3,4,6,7,8,9,10]
2008
*ISO> lend nat reverse 2008
1135
*ISO> lend nat_set reverse 2008
2008
*ISO> borrow nat_set succ [1,2,3]
[0,1,2,3]
*ISO> as set nat 42
[1,3,5]
*ISO> fit length nat 42
3
*ISO> retrofit succ nat_set [1,3,5]
43
\end{codex}

The reader might notice at this point that we have
already made full circle - as finite sets can be seen as
instances of finite sequences. 
Injective functions that are not surjections with wider and wider gaps can be
generated using the fact that one of the representations is information theoretically
``denser'' than the other, for a given range:
\begin{codex}
*ISO> as set fun [0,1,2,3]
[0,2,5,9]
*ISO> as set fun $ as set fun [0,1,2,3]
[0,3,9,19]
*ISO> as set fun $ as set fun $ as set fun [0,1,2,3]
[0,4,14,34]
\end{codex}
One can now define, for instance, a mapping from natural numbers to multi-sets
simply as:
\begin{code}
nat2mset = as mset nat
mset2nat = as nat mset 
\end{code}
but we will not explicitly need such definitions as the the equivalent
function is clearly provided by the combinator {\tt as}.
One can now borrow operations between {\tt set} and {nat} as follows:
\begin{codex}
*ISO> borrow_from set union nat 42 2008
2042
*ISO> 42 .|. 2008 :: Nat
2042
*ISO> borrow_from set intersect nat 42 2008
8
*ISO> 42 .&. 2008 :: Nat
8
*ISO> borrow_from nat (*) set [1,2,3] [4,5]
[5,7,9]
*ISO> borrow_from nat (+) set [1,2,3] [3,4,5]
[1,2,6]
\end{codex}
and notice that operations like union and intersection of sets map to boolean
operations on numbers as expected, while other operations are not necessarily
meaningful at first sight. We will show next a few cases where such
``shapshiftings'' of operations reveal more interesting analogies.

\subsection{Encoding Finite Multisets with Primes}

A factorization of a natural number is uniquely
described as multi-set or primes. We will use the fact that each prime number 
is uniquely associated to its position in the infinite stream of primes
to obtain a bijection from multisets of natural numbers to natural numbers.
We assume defined a prime generator {\tt primes} and a factoring function
{\tt to\_factors} (see Appendix).

The function {\tt nat2pmset} maps a natural number to the multiset of prime
positions in its factoring. Note that we treat {\tt 0} as {\tt []} and shift
{\tt n} to {\tt n+1} to accomodate {\tt 0} and {\tt 1}, to which prime factoring
operations do not apply.
\begin{code}
nat2pmset 0 = []
nat2pmset n = map (to_pos_in (h:ts)) (to_factors (n+1) h ts) where
  (h:ts)=genericTake (n+1) primes
  
to_pos_in xs x = fromIntegral i where
   Just i=elemIndex x xs
\end{code}

The function {\tt pmset2nat} maps back a multiset of positions of primes to
the result of the product of the corresponding primes. Again, we map {\tt []} to
{\tt 0} and shift back by {\tt 1} the result.
\begin{code}
pmset2nat [] = 0
pmset2nat ns = (product ks)-1 where
  ks=map (from_pos_in ps) ns
  ps=primes

from_pos_in xs n = xs !! (fromIntegral n)
\end{code}
We obtain the Encoder:
\begin{code}
pmset :: Encoder [Nat]
pmset = compose (Iso pmset2nat nat2pmset) nat
\end{code}
working as follows:
\begin{codex}
*ISO> as pmset nat 2008
[3,3,12]
*ISO> as nat pmset it
2008
*ISO> map (as pmset nat) [0..7]
[[],[0],[1],[0,0],[2],[0,1],[3],[0,0,0]]
\end{codex}
Note that the mappings from a set or sequence to a number work in time and
space linear in the bitsize of the number. On the other hand, as prime number
enumeration and factoring are involved in the mapping from numbers to multisets
this encoding is intractable for all but small values.

We are now ready to ``shapeshift'' between data types while watching for
interesting landscapes to show up.

\subsection{Exploring the analogy between multiset decompositions and factoring}

As natural numbers can be uniquely represented as a multiset
of prime factors and, independently, they can also be represented as a multiset
with the Encoder {\tt mset} described in subsection \ref{msetiso}, the following
question arises naturally:

{\em Can in any way the ``easy to reverse'' encoding {\tt mset} emulate or
predict properties of the the difficult to reverse factoring operation?}

The first step is to define an analog of the multiplication operation in terms
of the computationally easy multiset encoding {\tt mset}. Clearly, it makes
sense to take inspiration from the fact that factoring of an ordinary product of 
two numbers can be computed by concatenating the multisets of 
prime factors of its operands.

\begin{code}
mprod = borrow_from mset (++) nat
\end{code}
\begin{prop}
$<N,mprod,0>$ is a commutative monoid i.e. {\tt mprod} is defined for all pairs of natural
numbers and it is associative, commutative
and has 0 as an identity element.
\end{prop}
After rewriting the definition of {\tt mprod} as the equivalent:
\begin{code}
mprod_alt n m = as nat mset ((as mset nat n) ++ (as mset nat m))
\end{code}
the proposition follows immediately from the associativity of the
concatenation operation and the order independence of the multiset
encoding provided by {\tt mset}.

We can derive an exponentiation operation as a repeated application of
{\tt mprod}:
\begin{code}
mexp n 0 = 0
mexp n k = mprod n (mexp n (k-1))
\end{code}

Here are a few examples comparing {\tt mprod} to ordinary multiplication and
exponentiation:
\begin{codex}
*ISO> mprod 41 (mprod 33 88)
3539
*ISO> mprod (mprod 41 33)  88
3539
*ISO> mprod 33 46
605
*ISO> mprod 46 33
605
*ISO> mprod 0 712
712
*ISO> mprod 5513 0
5513

*ISO> (41*33)*88
119064
*ISO> 41*(33*88)
119064
*ISO> 33*46
1518
*ISO> 46*33
1518
*ISO> 1*712
712
*ISO> 5513*1
5513
*ISO> map (\x->mexp x 2) [0..15]
[0,3,6,15,12,27,30,63,24,51,54,111,60,123,126,255]
*ISO> map (\x->x^2) [0..15]
[0,1,4,9,16,25,36,49,64,81,100,121,144,169,196,225]
\end{codex}
Note also that any multiset encoding of natural numbers can be
used to define a similar commutative monoid structure. In the case of {\tt
pmset} we obtain:
\begin{code}
pmprod n m = as nat pmset ((as pmset nat n) ++ (as pmset nat m))
\end{code}
If one defines:
\begin{code}
pmprod' n m = (n+1)*(m+1)-1
\end{code}
it follows immediately from the definition of {\tt mprod} that:
\begin{equation}
pmprod \equiv pmprod'
\end{equation}
This is useful as computing {\tt pmprod'} is easy while computing {\tt mprod} is
intractable for large values. This brings us back to observe that:
\begin{prop}
$<N,pmprod,0>$ is a commutative monoid i.e. {\tt pmprod} is defined for all pairs of
natural numbers and it is associative, commutative
and has 0 as an identity element.
\end{prop}

Fig. \ref{mprodfig} compares the shapes of {\tt pmprod'} (virtually the same as
ordinary multiplication) and mprod for operands in $[0..2^7-1]$. 
One can see the contrast between the regular shape of ordinary multiplication
and the recursively ``self-similar'' landscape induced by {\tt mprod}.
\VFIGS{mprodfig}
{multiplication vs mprod}
{pmprod'}{mprod}{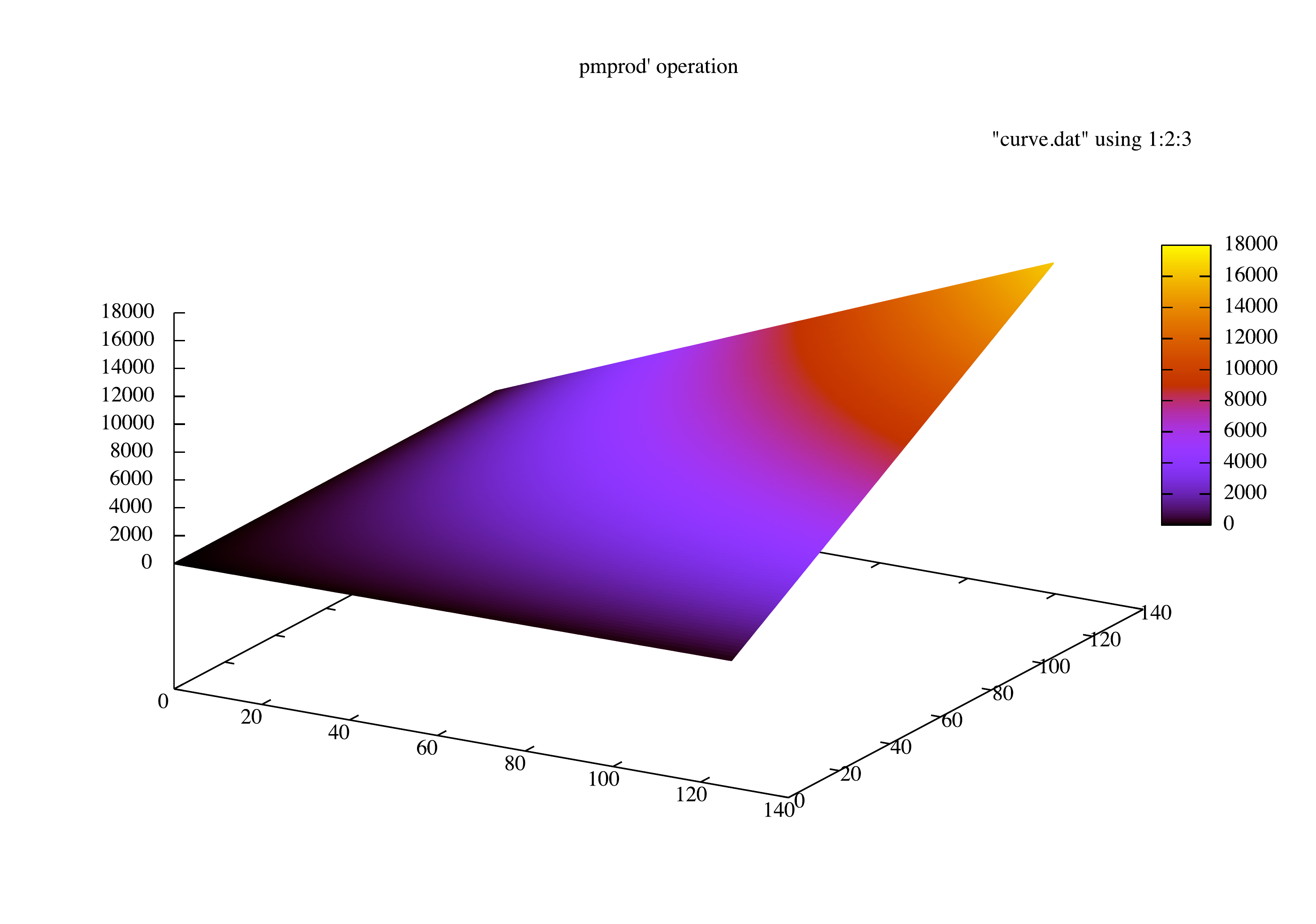}{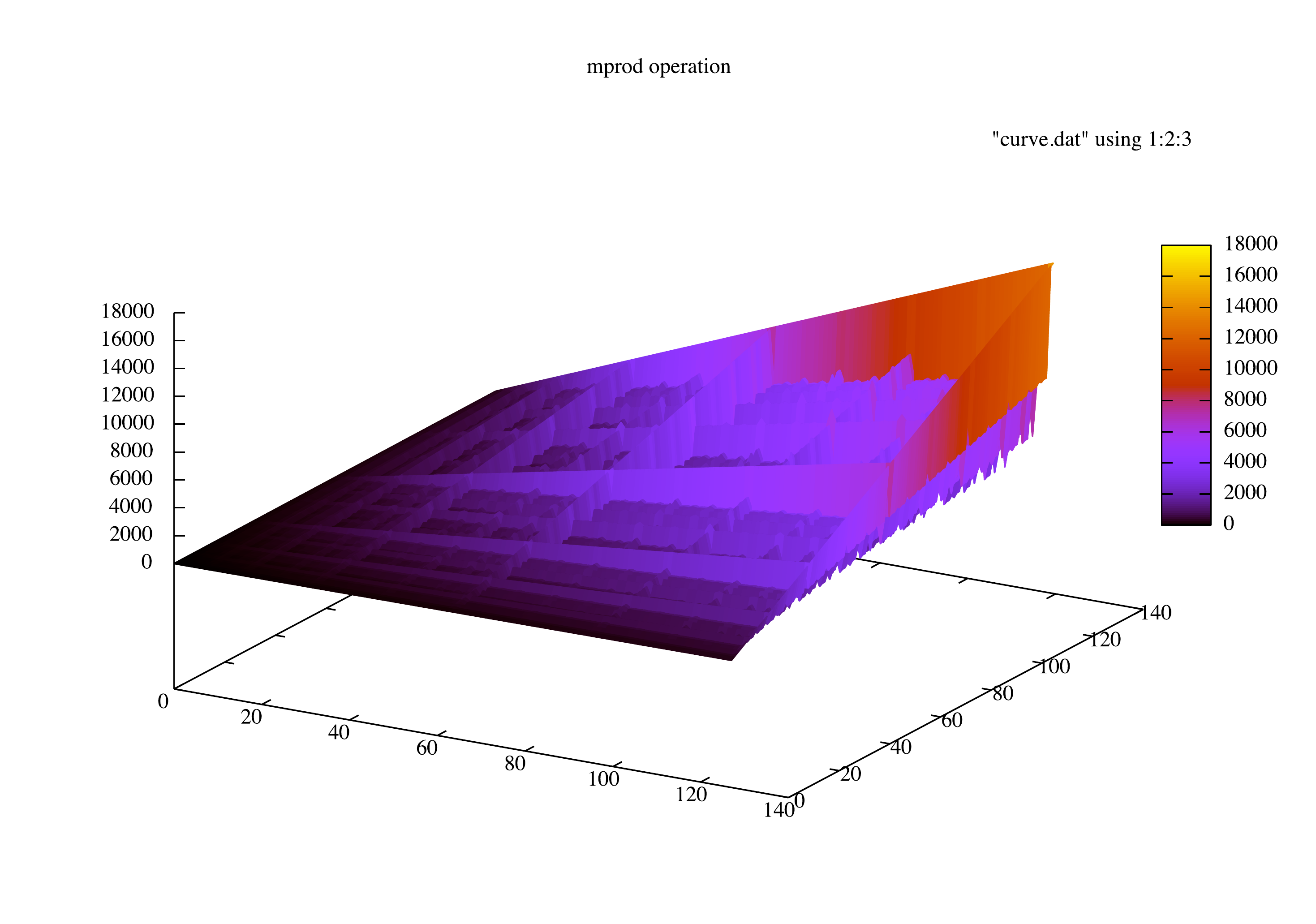}

One can also bring {\tt mprod} closer to ordinary multiplication by defining
\begin{code}
mprod' 0 _ = 0
mprod' _ 0 = 0
mprod' m n = (mprod (n-1) (m-1)) + 1

mexp' n 0 = 1
mexp' n k = mprod' n (mexp' n (k-1))
\end{code}
and by observing that they correlate as follows:
\begin{codex}
*ISO> map (\x->mexp' x 2) [0..16]
[0,1,4,7,16,13,28,31,64,25,52,55,112,61,124,127,256]
*ISO> map (\x->x^2) [0..16]
[0,1,4,9,16,25,36,49,64,81,100,121,144,169,196,225,256]
[0,1,8,15,64,29,120,127,512,57,232,239,960,253,1016,1023,4096]
*ISO> map (\x->x^3) [0..16]
[0,1,8,27,64,125,216,343,512,729,1000,1331,1728,2197,2744,3375,4096]
\end{codex}
Fig. \ref{isoExpMexp} shows that values for {\tt mexp'} follow from below those
of the $x^2$ function and that equality only holds when x is a power of 2.
\FIG{isoExpMexp}{Square vs. mexp' n 2 }{0.40}{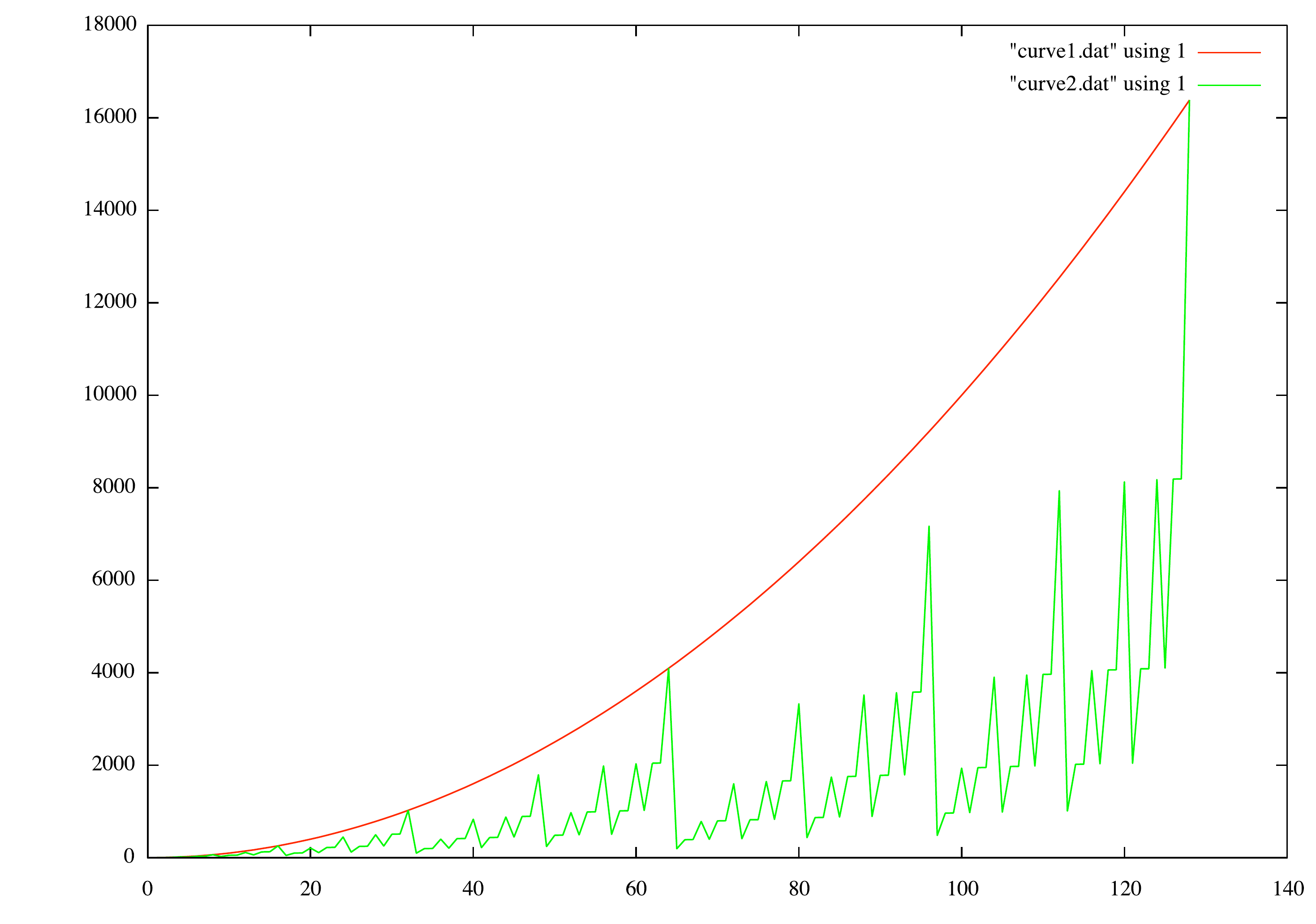}

Note that the structure induced by {\tt mprod'} is similar to ordinary
multiplication:
\begin{prop}
$<N,mprod',1>$ is a commutative monoid i.e. {\tt mprod'} is defined for all pairs of
natural numbers and it is associative, commutative
and has {\tt 1} as an identity element.
\end{prop}
Interestingly, {\tt mprod'} coincides with ordinary multiplication if one of the
operands is a power of 2. More precisely, the following holds:
\begin{prop}
$mprod'~x~y = x * y$ if and only if 
$\exists n \geq 0$ such that $x=2^n$ or $y=2^n$.
Otherwise, $mprod'~x~y <  x * y$.
\end{prop}
Fig. \ref{isoMdivP} shows the (scaled up by 1000) self-similar landscape
generated by the $[0..1]$-valued function {\tt (mprod' x y) / (x*y)}
\FIG{isoMdivP}{Ratio between mprod' and product}{0.40}{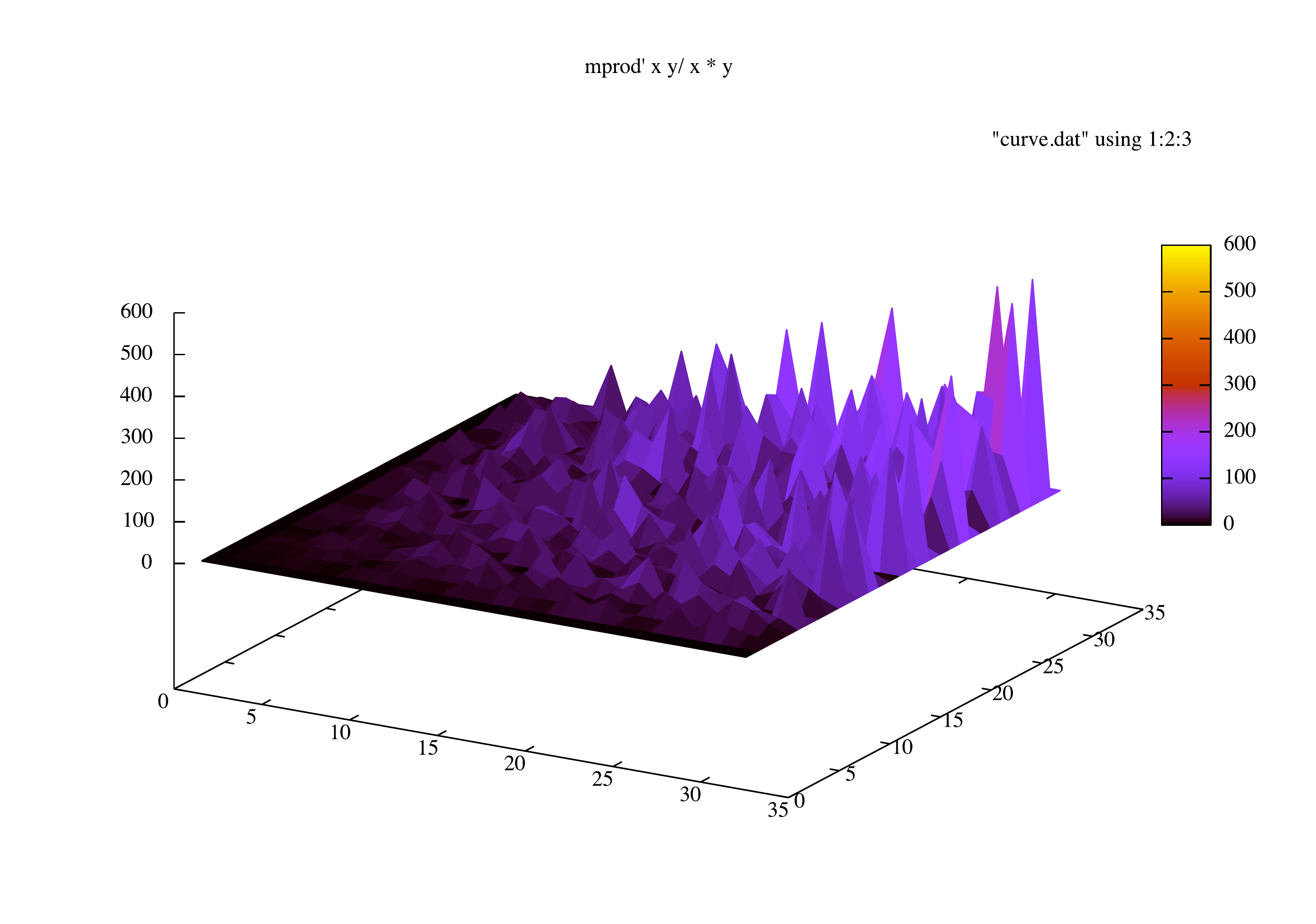}

Besides the connection with products, natural mappings worth investigating are
the analogies between {\em multiset intersection} and {\tt gcd} of the
corresponding numbers or between {\em multiset
union} and the {\tt lcm} of the corresponding numbers. Assuming the
definitions of multiset operations provided in the Appendix, one can define:

\begin{code}
mgcd :: Nat -> Nat -> Nat
mgcd = borrow_from mset msetInter nat

mlcm :: Nat -> Nat -> Nat
mlcm = borrow_from mset msetUnion nat

mdiv :: Nat -> Nat -> Nat
mdiv = borrow_from mset msetDif nat
\end{code}
and note that properties similar to usual arithmetic operations hold:
\begin{equation}
mprod (mgcd~x~y) (mlcm~x~y)  \equiv mprod~x~y
\end{equation}
\begin{equation}
mdiv (mprod~x~y)~y \equiv x
\end{equation}
\begin{equation}
mdiv (mprod~x~y)~x \equiv y
\end{equation}
While {\tt mprod,mprod',pmprod'} and {\tt pmprod} are not distributive with
ordinary addition, it looks like an interesting problem to find for each of
them compatible additive operations.

\subsection{Unfolding Natural Numbers into Bitstrings} \label{bits}
The isomorphism between natural numbers and bitstring is well known, except
that it is usually ignored that conventional bit representations
of integers need a twist to be mapped one-to-one to
{\em arbitrary} sequences of {\tt 0}s and {\tt 1}s.
As the usual binary representation always has
{\tt 1} as its highest digit, {\tt nat2bits}
will drop this bit, given
that the length of the list of digits is 
(implicitly) known. This transformation (a variant of the so called {\em
bijective base n} representation), brings us an isomorphism between $Nat$ and
the regular language $\{0,1\}^*$.
\begin{code}
bits :: Encoder [Nat]
bits = compose (Iso bits2nat nat2bits) nat

nat2bits = drop_last . (to_base 2) . succ

drop_last bs=
    genericTake ((genericLength bs)-1) bs
    
to_base base n = d : 
  (if q==0 then [] else (to_base base q)) where
     (q,d) = quotRem n base
    
bits2nat bs = pred (from_base 2 (bs ++ [1]))

from_base base [] = 0
from_base base (x:xs) | x>=0 && x<base = 
   x+base*(from_base base xs)
\end{code}
Note also that, strictly speaking, this is only an isomorphism
when the digits in the bitlist are in $\{0,1\}$,
therefore we shall assume this constraint
as a {\em law} governing this Encoder.
The following examples show two conversion operations and
$bits$ borrowing a multiplication operation from $nat$.
\begin{codex}
*ISO> as bits nat 42
[1,1,0,1,0]
*ISO> as nat bits [1,1,0,1,0]
42
*ISO> borrow2 (with nat bits) (*) [1,1,0] [1,0,1,1]
[1,0,0,1,1,0,0,0]
\end{codex}

The reader might notice at this point that we have
made full circle again - as bitstrings can be seen as
instances of finite sequences. 
Injective functions that are not surjections 
with wider and wider gaps can be generated
by composing the {\tt as} combinators:
\begin{codex}
*ISO> as bits fun [1,1]
[1,1,0]
*ISO> as bits fun (as bits fun [1,1])
[1,1,0,1]
*ISO> as bits fun $ as bits fun $ as bits fun [1,1]
[1,1,0,1,1,0]
\end{codex}

\subsection{Encoding Signed Integers}
To encode signed integers one can map positive numbers to even numbers and
strictly negative numbers to odd numbers. This gives the Encoder:

\begin{code}
type Z = Integer
z:: Encoder Z
z = compose (Iso z2nat nat2z) nat

nat2z n = if even n then n `div` 2 else (-n-1) `div` 2
z2nat n = if n<0 then -2*n-1 else 2*n
\end{code}
working as follows:
\begin{codex}
*ISO> as set z (-42)
[0,1,4,6]
*ISO> as z set [0,1,4,6]
-42
\end{codex}

\subsection{Functional Binary Numbers}

Church numerals are well known as a functional
representation for Peano arithmetic. While
benefiting from lazy evaluation, they
implement a form of unary arithmetic
that uses $O(n)$ space to represent $n$.
This suggest devising a functional representation
that mimics binary numbers. We will do this 
following the model described in subsection \ref{bits}
to provide an isomorphism between $Nat$ 
and the functional equivalent of
the regular language $\{0,1\}^*$. We will
view each bit as a $Nat \rightarrow Nat$ 
transformer:
\begin{code}
b x = pred x -- begin
o x = 2*x+0  -- bit 0
i x = 2*x+1  -- bit 1
e = 1        -- end
\end{code}
As the following example shows, composition
of functions $o$ and $i$
closely parallels the corresponding bitlists: 
\begin{codex}
*ISO> b$i$o$o$i$i$o$i$i$i$i$e
2008
*ISO> as bits nat 2008
[1,0,0,1,1,0,1,1,1,1]
\end{codex}

We can follow the same model with an abstract data type:
\begin{code}
data D = E | O D | I D deriving (Eq,Ord,Show,Read)
data B = B D deriving (Eq,Ord,Show,Read)
\end{code}
from which we can generate functional
bitstrings as an instance of a {\em fold} operation:
\begin{code}
funbits2nat :: B -> Nat
funbits2nat = bfold b o i e

bfold fb fo fi fe (B d) = fb (dfold d) where
  dfold E = fe
  dfold (O x) = fo (dfold x)
  dfold (I x) = fi (dfold x)
\end{code}
Dually, we can reverse the effect of the functions $b,o,i,e$ as:
\begin{code}
b' x = succ x
o' x | even x = x `div` 2
i' x | odd x = (x-1) `div` 2
e' = 1
\end{code}
and define a generator for our data type as an {\em unfold} operation:
\begin{code}
nat2funbits :: Nat -> B
nat2funbits = bunfold b' o' i' e'

bunfold fb fo fi fe x = B (dunfold (fb x)) where
  dunfold n | n==fe = E
  dunfold n | even n = O (dunfold (fo n))
  dunfold n | odd n = I (dunfold (fi n))
\end{code}
The two operations form an isomorphism:
\begin{codex}
*ISO> funbits2nat (B $ I $ O $ O $ I $ I $ O $ I $ I $ I $ I $ E)
2008
*ISO> nat2funbits it
B (I (O (O (I (I (O (I (I (I (I E))))))))))
\end{codex}
We can define our Encoder as follows:
\begin{code}
funbits :: Encoder B
funbits = compose (Iso funbits2nat nat2funbits) nat
\end{code}

Arithmetic operations can now be performed directly on this representation.
For instance, one can define a successor function as:
\begin{code}
bsucc (B d) = B (dsucc d) where
  dsucc E = O E
  dsucc (O x) = I x
  dsucc (I x) = O (dsucc x) 
\end{code}
Equivalently arithmetics can be borrowed from $Nat$:
\begin{codex}
*ISO> bsucc (B $ I $ O $ O $ I $ I $ O $ I $ I $ I $ I $ E)
B (O (I (O (I (I (O (I (I (I (I E))))))))))
*ISO> as nat funbits it
2009
*ISO> borrow (with nat funbits) 
             succ (B $ I $ O $ O $ I $ I $ O $ I $ I $ I $ I $ E)
B (O (I (O (I (I (O (I (I (I (I E))))))))))
*ISO> as nat funbits it
2009
\end{codex}
While Haskell's C-based arbitrary length integers are likely
to be more efficient for most operations, this representation, like Church
numerals, has the benefit of supporting partial or delayed computations
through lazy evaluation.

\section{Generic Unranking and Ranking Hylomorphisms}
\label{unrank}

The {\em ranking problem} for a family of
combinatorial objects is finding a unique 
natural number associated to it,
called its {\em rank}.
The inverse {\em unranking problem} consists of generating a unique
combinatorial object associated to each natural number. 

\subsection{Pure Hereditarily Finite Data Types}

The unranking operation is seen here as an instance of a generic
{\em anamorphism} mechanism (an {\em unfold} operation), while the ranking
operation is seen as an instance of the corresponding catamorphism (a {\em
fold} operation) \cite{DBLP:journals/jfp/Hutton99,DBLP:conf/fpca/MeijerH95}.
Together they form a mixed transformation called {\em hylomorphism}. 
We will use such hylomorphisms to lift isomorphisms between lists
and natural numbers to isomorphisms between a derived ``self-similar'' tree
data type and natural numbers.
In particular we will derive Ackermann's encoding
from Hereditarily Finite Sets to Natural Numbers.

The data type representing hereditarily finite structures will be
a generic multi-way tree with a single leaf type {\tt []}.

\begin{code}
data T = H [T] deriving (Eq,Ord,Read,Show)
\end{code}
The two sides of our hylomorphism 
are parameterized by two transformations
{\tt f} and {\tt g} forming
an isomorphism {\tt Iso f g}:
\begin{code}
unrank f n = H (unranks f (f n))
unranks f ns = map (unrank f) ns

rank g (H ts) = g (ranks g ts)
ranks g ts = map (rank g) ts
\end{code}
Both combinators can be seen as a form of ``structured recursion''
that propagate a simpler operation guided by the structure of
the data type. For instance, the size of a tree of type $T$ is
obtained as:
\begin{code}
tsize = rank (\xs->1 + (sum xs))
\end{code}
Note also that {\tt unrank} and {\tt rank}
work on $T$ in cooperation
with {\tt unranks} and {\tt ranks} 
working on $[T]$. 

We can now combine an 
anamorphism+catamorphism pair into
an isomorphism {\tt hylo} defined
with {\tt rank} and {\tt unrank} 
on the corresponding
hereditarily finite data types:
\begin{code}
hylo :: Iso b [b] -> Iso T b
hylo (Iso f g) = Iso (rank g) (unrank f)

hylos :: Iso b [b] -> Iso [T] [b]
hylos (Iso f g) = Iso (ranks g) (unranks f)
\end{code}

\subsubsection{Hereditarily Finite Sets}
Hereditarily Finite Sets will be represented
as an Encoder for the tree type {\tt T}:
\begin{code}
hfs :: Encoder T
hfs = compose (hylo nat_set) nat
\end{code}
The {\tt hfs} Encoder can now borrow operations from
sets or natural numbers as follows:
\begin{code}
hfs_union = borrow2 (with set hfs) union
hfs_succ = borrow (with nat hfs) succ
hfs_pred = borrow (with nat hfs) pred
\end{code}
\begin{codex}
*ISO> hfs_succ (H [])
H [H []]
*ISO> hfs_union (H [H []] ) (H [])
H [H []]
\end{codex}
Otherwise, hylomorphism induced isomorphisms
work as usual with our embedded
transformation language:
\begin{codex}
*ISO> as hfs nat 42
H [H [H []],H [H [],H [H []]],H [H [],H [H [H []]]]]
*ISO> as hfs nat 2008
H [H [H [],H [H []]],H [H [H [H []]]],H [H [H []],
   H [H [H []]]],H [H [],H [H []],H [H [H []]]],
   H [H [H [],H [H []]]],H [H [],H [H [],H [H []]]],
   H [H [H []],H [H [],H [H []]]]]
\end{codex}
One can notice that we have just derived as a
``free algorithm'' Ackermann's encoding
\cite{ackencoding,DBLP:journals/tplp/PiazzaP04}
from Hereditarily Finite Sets to Natural Numbers:
\vskip 0.30cm
$f(x)$ = {\tt if} $x=\{\}$ {\tt then} $0$ {\tt else} $\sum_{a \in x}2^{f(a)}$
\vskip 0.30cm
\noindent together with its inverse:
\begin{code}
ackermann = as nat hfs
inverse_ackermann = as hfs nat
\end{code}
One can represent the action of a hylomorphism unfolding a natural number into
a hereditarily finite set as a directed graph with outgoing edges
induced by by applying the {\tt inverse\_ackermann} function as shown
in Fig. \ref{f1}.

\FIG{f1}{2008 as a HFS}{0.60}{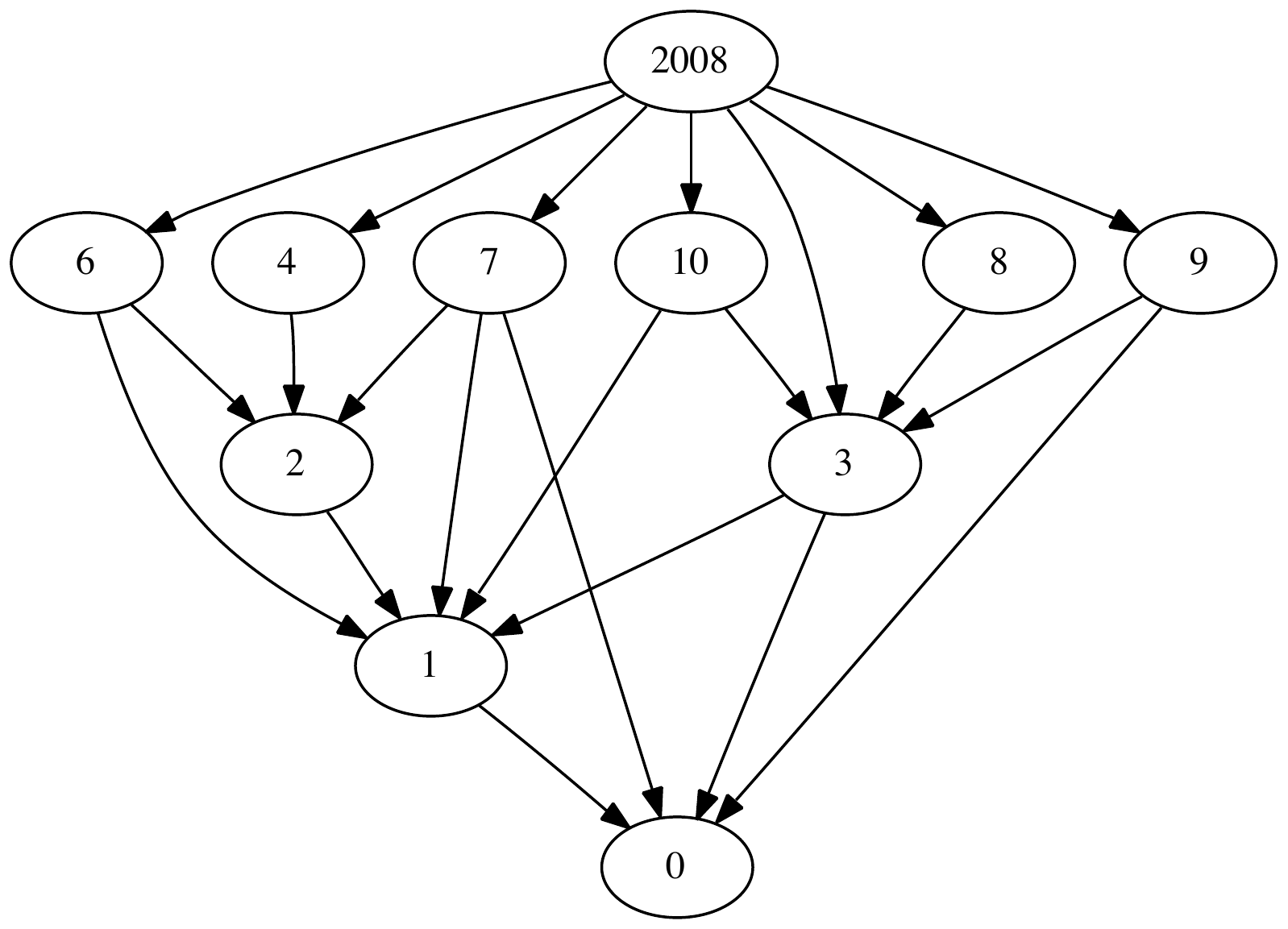}

\subsubsection{Hereditarily Finite Functions}
The same tree data type can host a hylomorphism
derived from finite functions instead of
finite sets:
\begin{code}
hff :: Encoder T
hff = compose (hylo nat) nat
\end{code}
The {\tt hff} Encoder can be seen as another ``free algorithm'', providing
data compression/succinct representation for Hereditarily 
Finite Sets. Note, for instance,
the significantly smaller tree size in:
\begin{codex}
*ISO> as hff nat 42
H [H [H []],H [H []],H [H []]]
ISO> as hff nat 2008
H [H [H [],H []],H [],H [H []],H [],H [],H [],H []]
\end{codex}
As the cognoscenti might observe 
this is explained by the fact that
{\tt hff} provides higher information density
than {\tt hfs}, by incorporating order information
that matters in the case of a sequence and
is ignored in the case of a set.
One can represent the action of a hylomorphism unfolding a natural number into
a hereditarily finite function as a directed ordered multi-graph as shown
in Fig. \ref{f2}. Note that as the mapping {\tt as fun nat} generates
a sequence where the order of the edges matters, this order is
indicated integers starting from {\tt 0} labeling the edges.
\FIG{f2}{2008 as a HFF}{0.60}{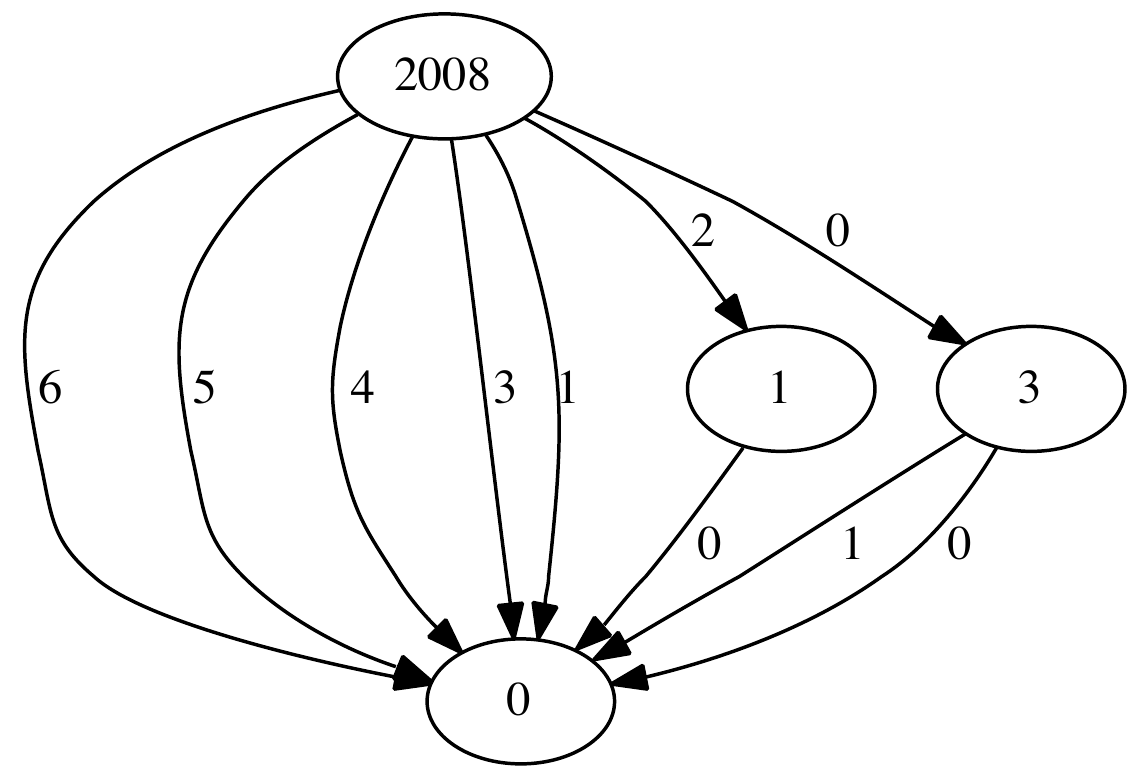}

It is also interesting to connect sequences and HFF directly - in case one
wants to represent giant ``sparse numbers'' that correspond to sequences
that would overflow memory if represented as natural numbers but have a
relatively simple structure as formulae used to compute them. We obtain the
Encoder:
\begin{code}
hffs :: Encoder T
hffs = Iso hff2fun fun2hff

fun2hff ns = H (map (as hff nat) ns)
hff2fun (H hs) = map (as nat hff) hs
\end{code}
which can be used to generate HFFs associated to very large numbers:
\begin{codex}
*ISO> as hffs fun [2^65,2^131]
H [H [H [H [],H [H [],H [H []]]]],H [H [H [],H [],H [H [],H [H []]]]]]
\end{codex}

\subsection{Hereditarily Finite Multisets}
In a similar way, one can derive an Encoder for Hereditarily Finite Multisets
based on either the {\tt mset} or the {\tt pmset} isomorphisms:
\begin{code}
nat_mset = Iso nat2mset mset2nat

hfm :: Encoder T
hfm = compose (hylo nat_mset) nat

nat_pmset = Iso nat2pmset pmset2nat

hfpm :: Encoder T
hfpm = compose (hylo nat_pmset) nat
\end{code}
working as follows:
\begin{codex}
*ISO> as hfm nat 2008
H [H [H [],H []],H [H [],H []],H [H [H [H []]]],H [H [H [H []]]],
   H [H [H [H []]]],H [H [H [H []]]],H [H [H [H []]]]]
*ISO> as nat hfm it
2008

*ISO> as hfpm nat 2008
H [H [H [],H []],H [H [],H []],H [H [H [],H [H []]]]]
*ISO> as nat hfpm it
2008
\end{codex}
After implementing this encoding some Google search revealed that it is
essentially the same as \cite{DBLP:journals/jct/Gobel80} where
it appears as an encoding of {\em rooted trees}.

\subsection{A Hylomorphism with Atoms/Urelements}
A similar construction can be carried out for the
more practical case when Atoms ({\em Urelements}
in Set Theory parlance) are present.
Hereditarily Finite Sets with Urelements are
represented as generic multi-way trees with a
leaf type holding urelements/atoms:
\begin{code}
data UT a = A a | F [UT a] deriving (Eq,Ord,Read,Show)
\end{code}
Atoms will be mapped to natural numbers in {\tt [0..ulimit-1]}.
Assuming for simplicity that {\tt ulimit} is fixed, we 
denote this set $A$ and 
denote $UT$ the set of trees of type $UT$ with atoms in $A$.

\paragraph*{Unranking} As an adaptation of the {\em unfold} 
operation, natural numbers will be mapped to elements of $UT$ with a generic 
higher order function {\tt unrankU f}, defined from $Nat$ to $UT$, 
parameterized by the natural number {\tt ulimit}
and the transformer function {\tt f}:
\begin{code}
ulimit = 4
\end{code}

\begin{code}
unrankU = unrankUL ulimit
unranksU = unranksUL ulimit

unrankUL l _ n | n>=0 && n<l = A n
unrankUL l f n = F (unranksUL l f (f (n-l)))

unranksUL l f ns =  map (unrankUL l f) ns
\end{code}

\paragraph*{Ranking} Similarly, as an adaptation of {\em fold}, a generic
inverse mapping {\tt rankU} is defined as:
\begin{code}
rankU = rankUL ulimit
ranksU = ranksUL ulimit

rankUL l _ (A n) | n>=0 && n<l = n
rankUL l g (F ts) = l+(g (ranksUL l g ts))

ranksUL l g ts = map (rankUL l g) ts
\end{code}
where {\tt rankU g} maps trees to numbers and
{\tt ranksU g} maps lists of trees to lists of numbers.

The following proposition describes conditions under which
{\tt rankU} and {\tt unrankU} can be used to lift isomorphisms
between $[Nat]$ and $Nat$ to isomorphisms involving
hereditarily finite structures:

\begin{prop}
If the transformer function $f:Nat \rightarrow [Nat]$ is a bijection 
with inverse $g$, such that 
$n \geq ulimit \wedge f(n)=[n_0,...n_i,...n_k] \Rightarrow n_i<n$, 
then $(unrankU~f) : Nat \rightarrow UT$ 
is a bijection  
with inverse $(rankU~ g) : UT \rightarrow Nat$ 
and the recursive computations defining both functions
terminate in a finite number of steps.
\end{prop}

\begin{proof} Note that {\tt unrankU} terminates as its arguments strictly
decrease at each step and {\tt rankU} terminates as leaf nodes are eventually
reached. That both are bijections, follows by induction on the structure of $Nat$ and $UT$,
given that {\tt map} preserves bijections and that adding/subtracting 
{$ulimit$} ensures that encodings of atoms and sets never overlap.
\end{proof}

The resulting hylomorphisms are defined as previously:
\begin{code}
hyloU (Iso f g) = Iso (rankU g) (unrankU f)
hylosU (Iso f g) = Iso (ranksU g) (unranksU f)
\end{code}
An Encoder for Hereditarily Finite Sets with Urelements is defined as:
\begin{code}
uhfs :: Encoder (UT Nat)
uhfs = compose (hyloU nat_set) nat
\end{code}
Note that this encoder provides a generalization of Ackermann's mapping,
to Hereditarily Finite Sets with Urelements in $[0..u-1]$ defined as:

\vskip 0.30cm
$f_{u}(x)$ = 
{\tt if} $x<u$ 
{\tt then} $x$
{\tt else} $u+\sum_{a\in x}2^{f_{u}(a)}$ 
\vskip 0.30cm

A similar Encoder for Hereditarily Finite Functions with Urelements is defined
as:
\begin{code}
uhff :: Encoder (UT Nat)
uhff = compose (hyloU nat) nat
\end{code}

\subsection{Extending the encoding for the case of an infinite 
set of Atoms/Urelements} 

An adaptation of the previous construction for the case when an infinite supply
of atoms/urelements is needed (i.e. when their number is not known in  advance)
follows.

\paragraph*{Unranking} As an adaptation of the {\em unfold} 
operation, natural numbers will be mapped to elements of $UT$ with a generic 
higher order function {\tt unrankIU f}, defined from $Nat$ to $UT$, 
parameterized by the transformer function {\tt f}:

\begin{code}
unrankIU _ n | even n = A (n `div` 2) 
unrankIU f n = F (unranksIU f (f  ((n-1) `div` 2)))

unranksIU f ns =  map (unrankIU f) ns
\end{code}
Note that (an infinite supply of) even numbers provides codes for atoms, while
odd numbers are used to encode the non-leaf structure of the trees in {\tt UT}.

\paragraph*{Ranking} Similarly, as an adaptation of {\em fold}, a generic
inverse mapping {\tt rankIU g} is defined as:
\begin{code}
rankIU _ (A n) = 2*n
rankIU g (F ts) = 1+2*(g (ranksIU g ts))

ranksIU g ts = map (rankIU g) ts
\end{code}
where {\tt rankIU g} maps trees to numbers and
{\tt ranksIU g} maps lists of trees to lists of numbers.

The resulting hylomorphisms are defined as previously:
\begin{code}
hyloIU (Iso f g) = Iso (rankIU g) (unrankIU f)
hylosIU (Iso f g) = Iso (ranksIU g) (unranksIU f)
\end{code}
An Encoder for Hereditarily Finite Sets with an infinite supply of Urelements is
defined as:
\begin{code}
iuhfs :: Encoder (UT Nat)
iuhfs = compose (hyloIU nat_set) nat
\end{code}

A similar Encoder for Hereditarily Finite Functions with and infinite
supply of Urelements is defined as:
\begin{code}
iuhff :: Encoder (UT Nat)
iuhff = compose (hyloIU nat) nat
\end{code}

\section{Permutations and Hereditarily Finite Permutations} \label{perm}
We have seen that finite sets and their derivatives represent
information in an {\em order} independent way, focusing exclusively
on information {\em content}. 
We will now look at data representations that focus exclusively on {\em order}
in a {\em content} independent way - finite permutations and their hereditarily
finite derivatives.

To obtain an encoding for finite permutations
we will first review a ranking/unranking mechanism for permutations that
involves an unconventional numeric representation, {\em factoradics}.

\subsection{The Factoradic Numeral System}
The factoradic numeral system \cite{knuth_art_1997-1} replaces digits
multiplied by a power of a base $n$ with digits that multiply successive values
of the factorial of $n$. In the increasing order variant {\tt fr} the first
digit $d_0$ is 0, the second is $d_1 \in \{0,1\}$ and the $n$-th is $d_n \in
[0..n]$. For instance, $42=0*0!+0*1!+0*2!+3*3!+1*4!$.
The left-to-right, decreasing order variant {\tt fl} 
is obtained by reversing the digits of {\tt fr}.
\begin{codex}
fr 42
  [0,0,0,3,1]
rf [0,0,0,3,1]
  42
fl 42
  [1,3,0,0,0]
lf [1,3,0,0,0]
  42
\end{codex}
\noindent The function {\tt fr} 
generating the factoradics of n, right to left,
handles the special case of $0$ and
calls a local function {\tt f} which recurses and divides with increasing values
of $n$ while collecting digits with {\tt mod}:
\begin{code}
fr 0 = [0]
fr n = f 1 n where
   f _ 0 = []
   f j k = (k `mod` j) : 
           (f (j+1) (k `div` j))
\end{code}
The function {\tt fl}, with digits left to right is obtained as follows:
\begin{code}
fl = reverse . fr
\end{code}
The function {\tt lf} (inverse of {\tt fl}) converts back to decimals by
summing up results while computing the factorial progressively:
\begin{code}
rf ns = sum (zipWith (*) ns factorials) where
  factorials=scanl (*) 1 [1..]
\end{code}
Finally, {\tt lf}, the inverse of {\tt fl} is obtained as:
\begin{code}
lf = rf . reverse
\end{code}

\subsection{Ranking and unranking permutations of given size with Lehmer codes
and factoradics} 
The Lehmer code of a permutation $f$ of size $n$ is defined as the sequence
$l(f)=(l_1(f) \ldots l_i(f) \ldots l_n(f))$ 
where $l_i(f)$ is the number
of elements of the set $\{j>i|f(j)<f(i)\}$
\cite{DBLP:journals/dmtcs/MantaciR01}.
 \begin{prop}
 The Lehmer code of a permutation determines the permutation uniquely.
 \end{prop} 
The function {\tt perm2nth} computes a {\tt rank} 
for a permutation {\tt ps} of {\tt size>0}. 
It starts by first computing its Lehmer code {\tt ls} with 
{\tt perm2lehmer}. Then  it associates a unique natural 
number {\tt n} to {\tt ls}, 
by converting it with the function {\tt lf} 
from factoradics to decimals. 
Note that the Lehmer code {\tt Ls} is used as the list of digits
in the factoradic representation.
\begin{code}
perm2nth ps = (l,lf ls) where 
  ls=perm2lehmer ps
  l=genericLength ls

perm2lehmer [] = []
perm2lehmer (i:is) = l:(perm2lehmer is) where
  l=genericLength [j|j<-is,j<i]  
\end{code}

The function {\tt nat2perm} provides the matching {\em unranking}
operation associating a permutation {\tt ps} to a given {\tt size>0} 
and a natural number {\tt n}. It generates the $n$-th permutation of a given
size.
\begin{code}
nth2perm (size,n) = 
  apply_lehmer2perm (zs++xs) [0..size-1] where 
    xs=fl n
    l=genericLength xs
    k=size-l
    zs=genericReplicate k 0
\end{code}
The following function extracts 
a permutation from a ``digit'' list
in factoradic representation.
\begin{code}
apply_lehmer2perm [] [] = []
apply_lehmer2perm (n:ns) ps@(x:xs) = 
   y : (apply_lehmer2perm ns ys) where
   (y,ys) = pick n ps

pick i xs = (x,ys++zs) where 
  (ys,(x:zs)) = genericSplitAt i xs
\end{code}
Note also that {\tt apply\_lehmer2perm} is used this time
to reconstruct the permutation {\tt ps} from its Lehmer code,
which in turn is computed from the permutation's 
factoradic representation.

One can try out this bijective mapping as follows:
\begin{codex}
nth2perm (5,42)
  [1,4,0,2,3]
perm2nth [1,4,0,2,3]
  (5,42)
nth2perm (8,2008)
  [0,3,6,5,4,7,1,2]
perm2nth [0,3,6,5,4,7,1,2]
  (8,2008)
\end{codex}

\subsection{A bijective mapping from permutations to natural numbers}
Like in the case of BDDs, one more step is needed to to extend the mapping
between permutations of a given length to a bijective 
mapping from/to $Nat$: we will have to ``shift
towards infinity'' the starting point of each new bloc of permutations in $Nat$
as permutations of larger and larger sizes are enumerated.

First, we need to know by how much - so we compute the sum of
all factorials up to $n!$.
\begin{code}
sf n = rf (genericReplicate n 1)
\end{code}
This is done by noticing that the factoradic representation of
[0,1,1,..] does just that.

What we are really interested into, is decomposing {\tt n} into
the distance to the
last sum of factorials smaller than {\tt n}, {\tt n\_m}
and the its index in the sum, {\tt k}.
\begin{code}
to_sf n = (k,n-m) where 
  k=pred (head [x|x<-[0..],sf x>n])
  m=sf k
\end{code}
{\em Unranking} of an arbitrary permutation is now easy - the index {\tt k}
determines the size of the permutation and {\tt n-m} determines
the rank. Together they select the right permutation with {\tt nth2perm}.
\begin{code}
nat2perm 0 = []
nat2perm n = nth2perm (to_sf n)
\end{code}
{\em Ranking} of a permutation is even easier: we first compute
its size and its rank, then we shift the rank by 
the sum of all factorials up to its size, enumerating the
ranks previously assigned.
\begin{code}
perm2nat ps = (sf l)+k where 
  (l,k) = perm2nth ps
\end{code}
It works as follows:
\begin{codex}
nat2perm 2008
  [0,2,3,1,4]
perm2nat [0,2,3,1,4]
  42
nat2perm 2008
  [1,4,3,2,0,5,6]
perm2nat [1,4,3,2,0,5,6]
  2008
\end{codex}

We can now define the Encoder as:
\begin{code}
perm :: Encoder [Nat]
perm = compose (Iso perm2nat nat2perm) nat
\end{code}
The Encoder works as follows:
\begin{codex}
*ISO> as perm nat 2008
[1,4,3,2,0,5,6]
*ISO> as nat perm it
2008
*ISO> as perm nat 1234567890
[1,6,11,2,0,3,10,7,8,5,9,4,12]
*ISO> as nat perm it
1234567890
\end{codex}

\subsection{Hereditarily Finite Permutations}

By using the generic {\tt unrank} and {\tt rank} functions defined 
in section \ref{unrank} we can extend the isomorphism defined by {\tt nat2perm}
and {\tt perm2nat} to encodings of Hereditarily Finite Permutations ($HFP$).
\begin{code}
nat2hfp = unrank nat2perm
hfp2nat = rank perm2nat
\end{code}
The encoding works  as follows:
\begin{codex}
*ISO> nat2hfp 42
H [H [],H [H [],H [H []]],H [H [H []],H []],
   H [H []],H [H [],H [H []],H [H [],H [H []]]]]
*ISO> hfp2nat it
  42
\end{codex}
We can now define the Encoder as:
\begin{code}
hfp :: Encoder T
hfp = compose (Iso hfp2nat nat2hfp) nat
\end{code}
The Encoder works as follows:
\begin{codex}
*ISO> as hfp nat 42
H [H [],H [H [],H [H []]],H [H [H []],H []],
   H [H []],H [H [],H [H []],H [H [],H [H []]]]]
*ISO> as nat hfp it
42
*JFISO> as hfp nat 2008
H [H [H []],H [H [],H [H []],H [H [],H [H []]]],H [H [H []],H []],
  H [H [],H [H []]],H [],H [H [],H [H [],H [H []]],H [H []]],
  H [H [H []],H [],H [H [],H [H []]]]]
*ISO> as nat hfp it
2008
\end{codex}
As shown in Fig \ref{f3} an ordered digraph (with labels starting from 0
representing the order of outgoing edges) can be used to represent the unfolding
of a natural number to the associated hereditarily finite permutation.
\FIG{f3}{2008 as a HFP}{0.60}{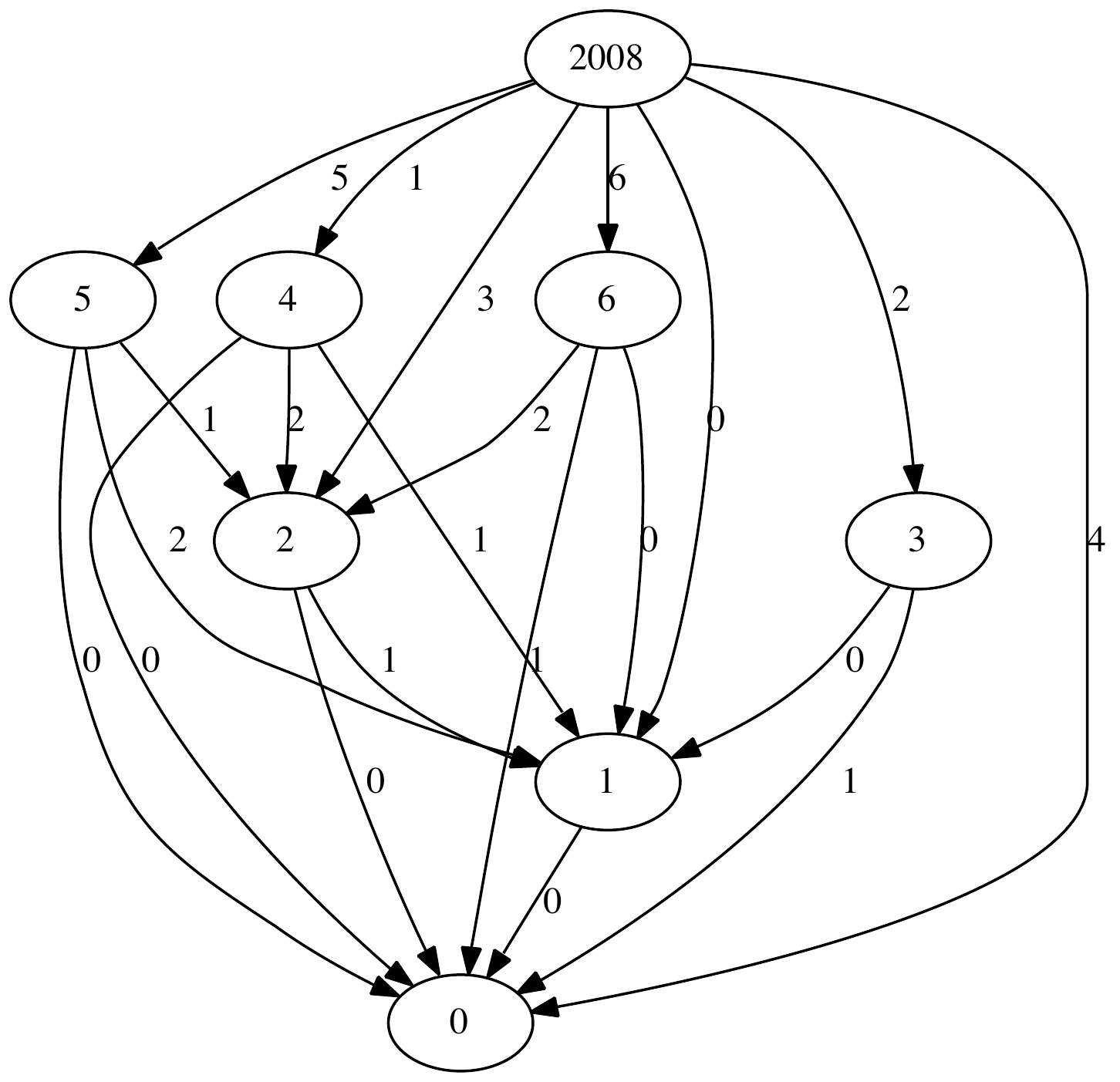}
An interesting property of graphs associated to hereditarily finite permutations
is that moving from a number n to its successor typically only induces a
reordering of the labeled edges, as shown in Fig. \ref{f4}.
\FIG{f4}{2009 as a HFP}{0.60}{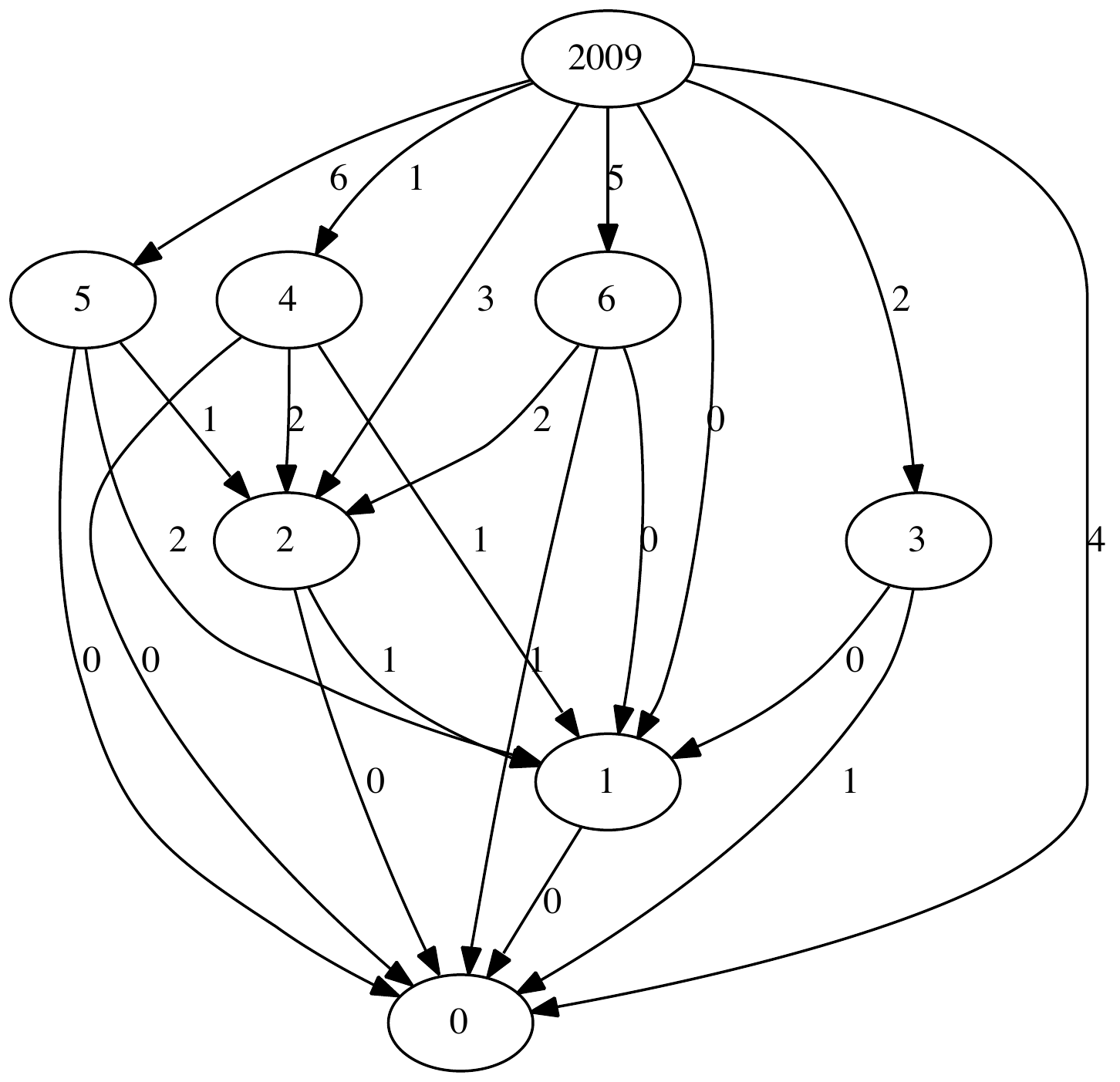}

\section{Hereditary base-k represenations and Goodstein sequences}

\begin{df}
Hereditary base-k representation of a number x is obtained by representing x as
a sum of powers of k followed by expression of each of the exponents with
nonzero coeficients as a sum of powers of k, recursively.
\end{df}
First we express a single step of this transformation to/from a polynomial in
base {\tt k} as a pair of bijections:
\begin{code}
nat2kpoly k n = filter (\p->0/=fst p) ps where 
  ns=to_base k n
  l=genericLength ns
  is=[0..l-1]
  ps=zip ns is 

kpoly2nat k ps = sum (map (\(d,e)->d*k^e) ps)
\end{code}
The transformation works as follows:
\begin{codex}
*ISO> nat2kpoly 3 2009
[(2,0),(1,2),(2,3),(2,5),(2,6)]
*ISO> kpoly2nat 3 it
2009
\end{codex}
The recursive process generates a tree, with coeficients of each
expansion labeling nodes. We can host this expansion in the data type {\tt HB}:
\begin{code}
data HB a = HB a [HB a]  deriving (Eq,Ord,Show,Read)
\end{code}
We will define, for each base k, two isomorphisms {\tt nat2hb k} and {\tt
hb2nat k} between natural numbers and polynomials:
\begin{code}
nat2hb :: Nat->Nat->[HB Nat]

nat2hb _k 0 = [] 
nat2hb k n | n<k = [HB n []]
nat2hb k n = gs where 
  ps'=nat2kpoly k n
  gs=map (nat2hb1 k) ps'
  nat2hb1 k (d,e) = HB d (nat2hb k e)
 
hb2nat :: Nat -> [HB Nat] -> Nat
 
hb2nat k [] = 0
hb2nat k ts = kpoly2nat k ps where
  ps=map (hb2nat1 k) ts
  hb2nat1 k (HB d ts) = (d,hb2nat k ts)
\end{code}
We can now define a family of {\tt Encoders}, one for each base {\tt k}, as
follows:
\begin{code}
hb :: Nat->Encoder [HB Nat]
hb k = compose (Iso (hb2nat k) (nat2hb k)) nat
\end{code}
The new concept here is working with a parametric family of Encoders. With
a small adaptation, the syntax of the {\tt as} combinator scales up naturally:
\begin{codex}
*ISO> as  (hb 3) nat 42
[HB 2 [HB 1 []],HB 1 [HB 2 []],HB 1 [HB 1 [HB 1 []]]]
*ISO> as nat (hb 3) it
42
\end{codex}
Note that the base does not occur as such in the hereditary base-k
expression obtained with the Encoder {\tt hb}. This property can be used to
obtain {\tt Goodstein sequences} by {\em bumping the base} from {\tt k} to
{\tt k+1} i.e. interpreting a {\tt (hb k)} expression as a {\tt (hb (k+1))}
expression and then subtracting 1 from the result, i.e:
\begin{code}
goodsteinStep k n = (hb2nat (k+1) (nat2hb k n)) - 1

goodsteinSeq _ 0 = []
goodsteinSeq k n = n:(goodsteinSeq (k+1) m) where 
  m=goodsteinStep k n
  
goodstein m = goodsteinSeq 2 m
\end{code}
\begin{codex}
*ISO> goodstein 3
[3,3,3,2,1]
*ISO> take 12 (goodstein 4)
[4,26,41,60,83,109,139,173,211,253,299,348]
\end{codex}
Goodstein's Theorem (provable in second order arithmetics) states that this
sequence always terminates at 0. The remarkable thing about it is that it is 
an undecidable statement in first order Peano arithmetics, that in contrast to
G\"{o}del's therorem, involves only ``conventional'' numerical relations.

\section{Pairing/Unpairing}

A {\em pairing} function is an isomorphism $f:Nat \times Nat
\rightarrow Nat$. Its inverse is called {\em unpairing}.

\subsection{The Pepis-Kalmar-Robinson Pairing Function}
An classic pairing function is {\bf pepis\_J}, together with its left and right
unpairing companions {\bf pepis\_K} and {\bf pepis\_L} that have been used, by Pepis, Kalmar and Robinson 
together with Cantor's functions, in some fundamental work on recursion theory, 
decidability and Hilbert's Tenth Problem in
\cite{pepis,kalmar1,kalmar2,kalmar3,robinson50,robinson55,robinson68a,robinsons68b,robinson67}.
The function {\bf pepis\_J}
combines two numbers reversibly by multiplying
a power of 2 derived from the first and
an odd number derived from the second:
\begin{equation}
f(x,y)=2^x*(2*y+1)-1
\end{equation}
Its Haskell implementation, together with its inverse is:
\begin{code}
pepis_J x y  = pred ((2^x)*(succ (2*y)))

pepis_K n = two_s (succ n)

pepis_L n = (pred (no_two_s (succ n))) `div` 2
 
two_s n | even n = succ (two_s (n `div` 2))
two_s _ = 0

no_two_s n = n `div` (2^(two_s n))
\end{code}
This pairing function (slower in the second argument) works as follows:
\begin{codex}
pepis_J 1 10
  41
pepis_J 10 1
  3071
[pepis_J i j|i<-[0..3],j<-[0..3]]
  [0,2,4,6,1,5,9,13,3,11,19,27,7,23,39,55]
\end{codex}
As Haskell provides a built-in ordered pair, it is convenient to regroup the
functions {\tt J, K, L} (given in Julia Robinson's original notation) as
mappings to/from built-in ordered pairs:
\begin{code}
pepis_pair (x,y) = pepis_J x y
pepis_unpair n = (pepis_K n,pepis_L n)
\end{code}
Observing that the number of {\tt 0}s in front of the
representation of a natural number {\tt n} as a sequence equals {\tt pepis\_K n}, an alternative
implementation could be:
\begin{code}
pepis_pair' (x,y) = (fun2nat (x:(nat2fun y)))-1

pepis_unpair' n = (x,fun2nat ns) where 
  (x:ns)=nat2fun (n+1)  

fun2nat = set2nat . fun2set
nat2fun = set2fun . nat2set  
\end{code} 
Note also that {\tt pepis\_unpair} is ``asymmetrical'' in the sense that its
first component grows much slower than the second, when applied to {\tt [0..]}.
Sometimes it is more useful to have the opposite behavior
\begin{code}
rpepis_pair (x,y) = pepis_pair (y,x)
rpepis_unpair n = (y,x) where (x,y)=pepis_unpair n
\end{code}
After defining
\begin{code}
type Nat2 = (Nat,Nat)
\end{code}
we obtain the encoder
\begin{code}
pnat2 :: Encoder Nat2
pnat2 = compose (Iso pepis_pair pepis_unpair) nat

rpnat2 :: Encoder Nat2
rpnat2 = compose (Iso rpepis_pair rpepis_unpair) nat
\end{code}

\subsection{A Bitwise Pairing/Unpairing Function}
We will now introduce an unusually simple pairing function 
(also mentioned in \cite{pigeon}, p.142).

The function {\tt bitpair} works by splitting a 
number's big endian bitstring
representation into odd and even bits, 
while its inverse {\tt bitunpair}
blends the odd and even bits back together.

\begin{code}
bitpair ::  Nat2 -> Nat
bitpair (i,j) = 
  set2nat ((evens i) ++ (odds j)) where
    evens x = map (2*) (nat2set x)
    odds y = map succ (evens y)

bitunpair :: Nat->Nat2  
bitunpair n = (f xs,f ys) where 
  (xs,ys) = partition even (nat2set n)
  f = set2nat . (map (`div` 2))
\end{code}

The transformation of the bitlists
is shown in the following example 
with bitstrings aligned:
\begin{codex}
*ISO> bitunpair 2008
  (60,26)

-- 2008:[0, 0, 0, 1, 1, 0, 1, 1, 1, 1, 1]
--   60:[0,    0,    1,    1,    1,    1]
--   26:[   0,    1,    0,    1,    1   ]
\end{codex}

We can derive the following Encoder:
\begin{code}
nat2 :: Encoder Nat2
nat2 = compose (Iso bitpair bitunpair) nat
\end{code}
working as follows:
\begin{codex}
*ISO> as nat2 nat 2008
(60,26)
*ISO> as nat nat2 (60,26)
2008
\end{codex}

In a way similar to hereditarily finite trees generated by unfoldings one can
apply strictly decreasing\footnote{except for 0 and 1, typically} unpairing
functions recursively. Figures \ref{iso2008p} and \ref{iso2008b} show the
directed graphs describing recursive application of {\tt bitunpair} and {\tt
pepis\_unpair}.

\FIG{iso2008p}{Graph obtained by recursive application of {\tt pepis\_unpair}
for 2008}{0.40}{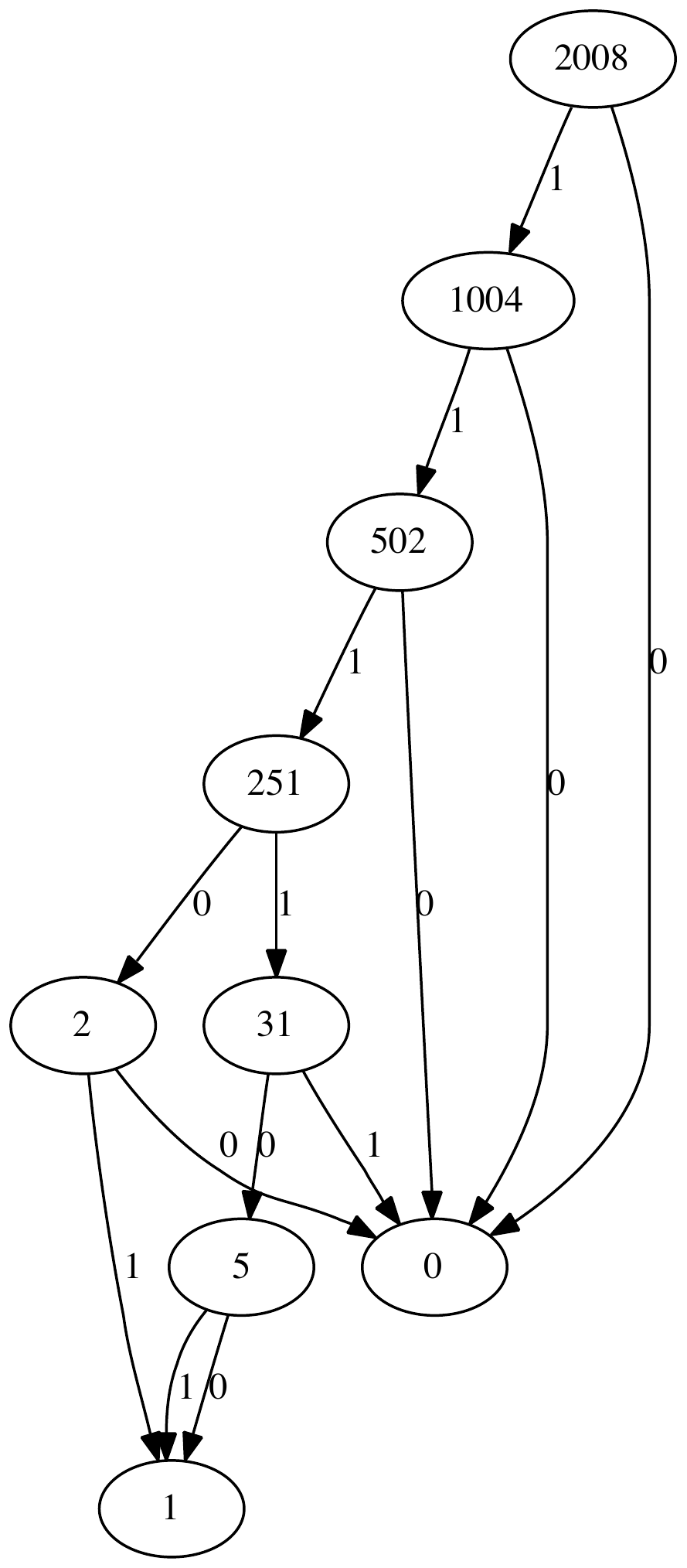}

\FIG{iso2008b}{Graph obtained by recursive application of {\tt bitunpair}
for 2008}{0.40}{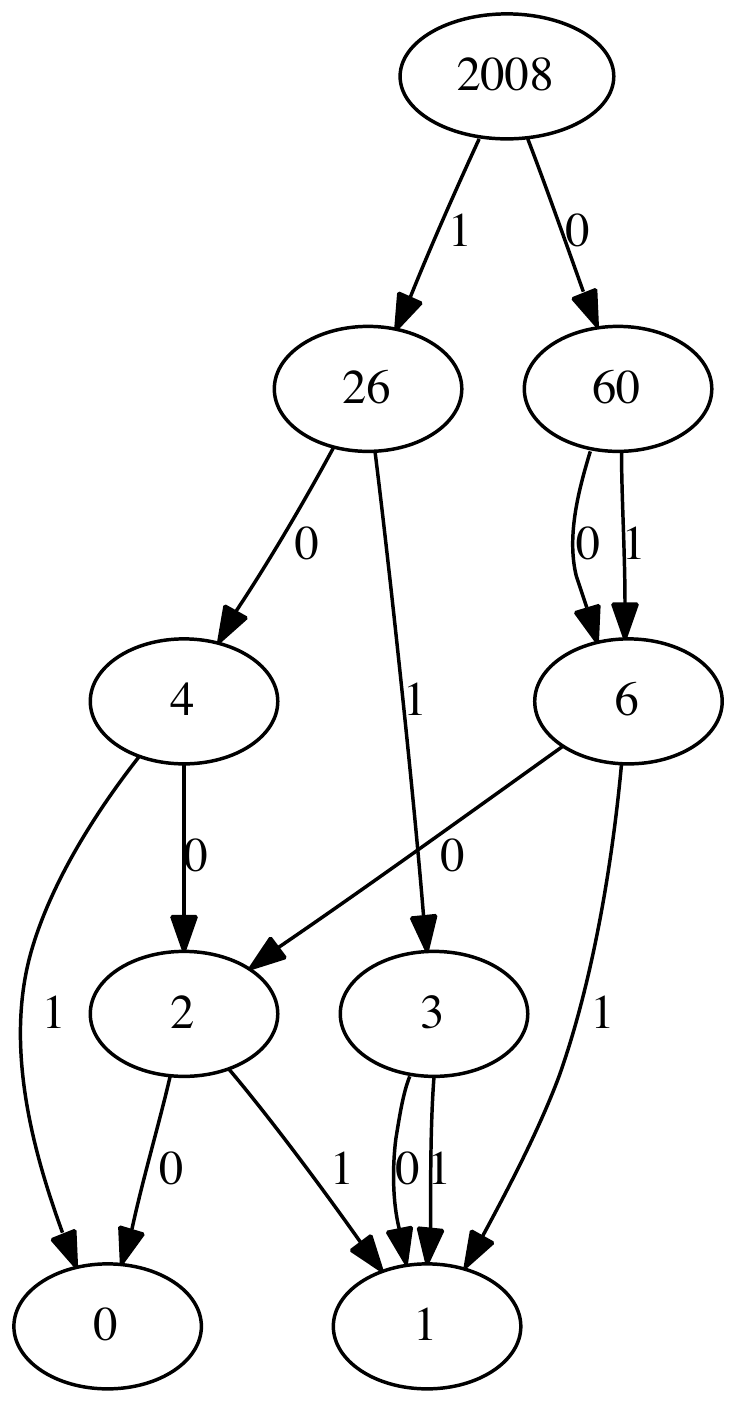}

Given that unpairing functions are bijections from $Nat$ to $Nat \times Nat$
they will progressively cover all points having natural number coordinates in
their range in the plane. Figures \ref{isounpair1}, \ref{isounpair2} 
show the curves generated by {\tt bitunpair} and {\tt pepis\_unpair}.

\FIG{isounpair1}
{2D curve connecting values of {\tt bitunpair n} for $n \in [0..2^{10}-1]$}
{0.40}{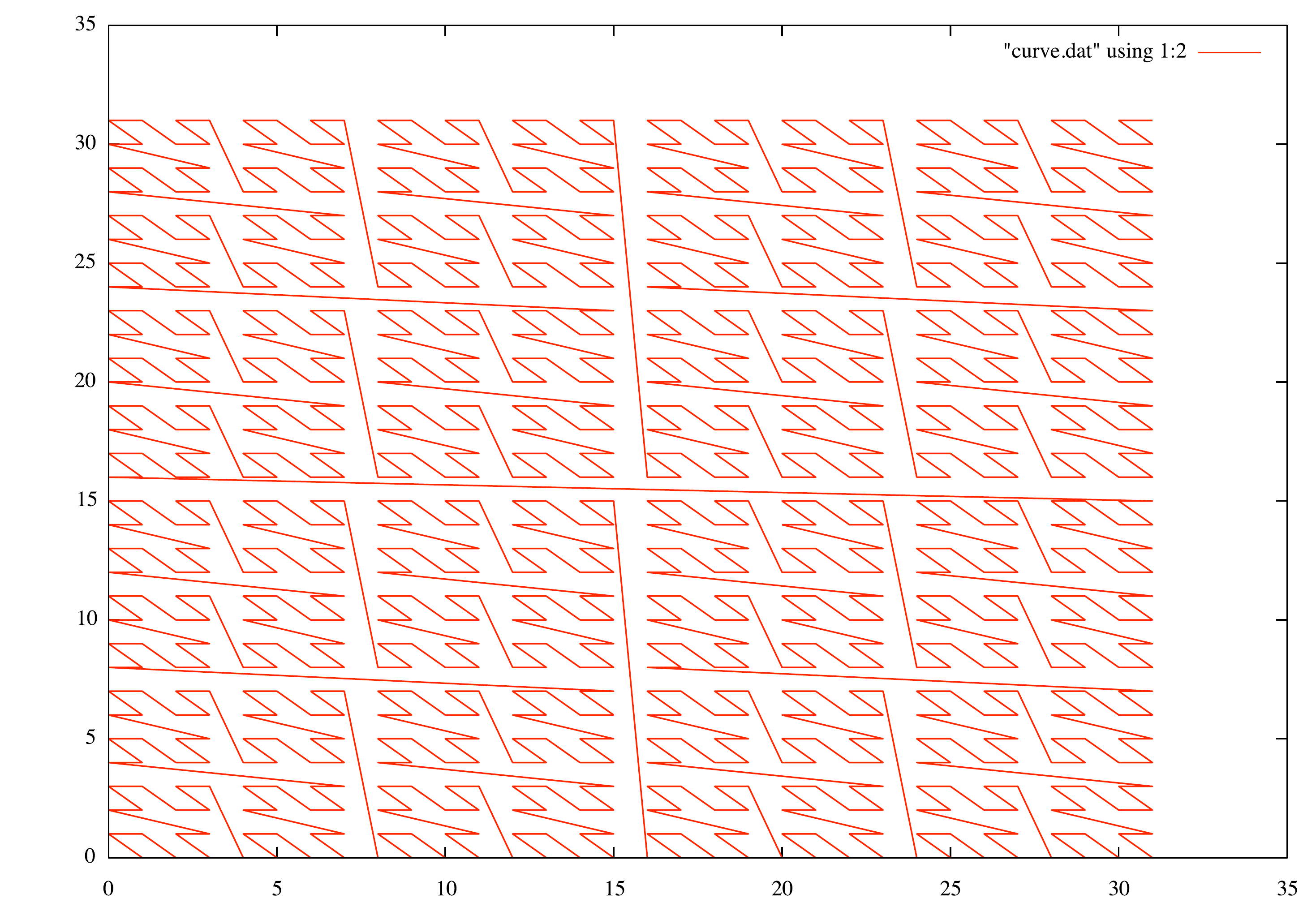}

\FIG{isounpair2}
{2D curve connecting values of {\tt pepis\_unpair n} for $n \in [0..2^{10}-1]$}
{0.40}{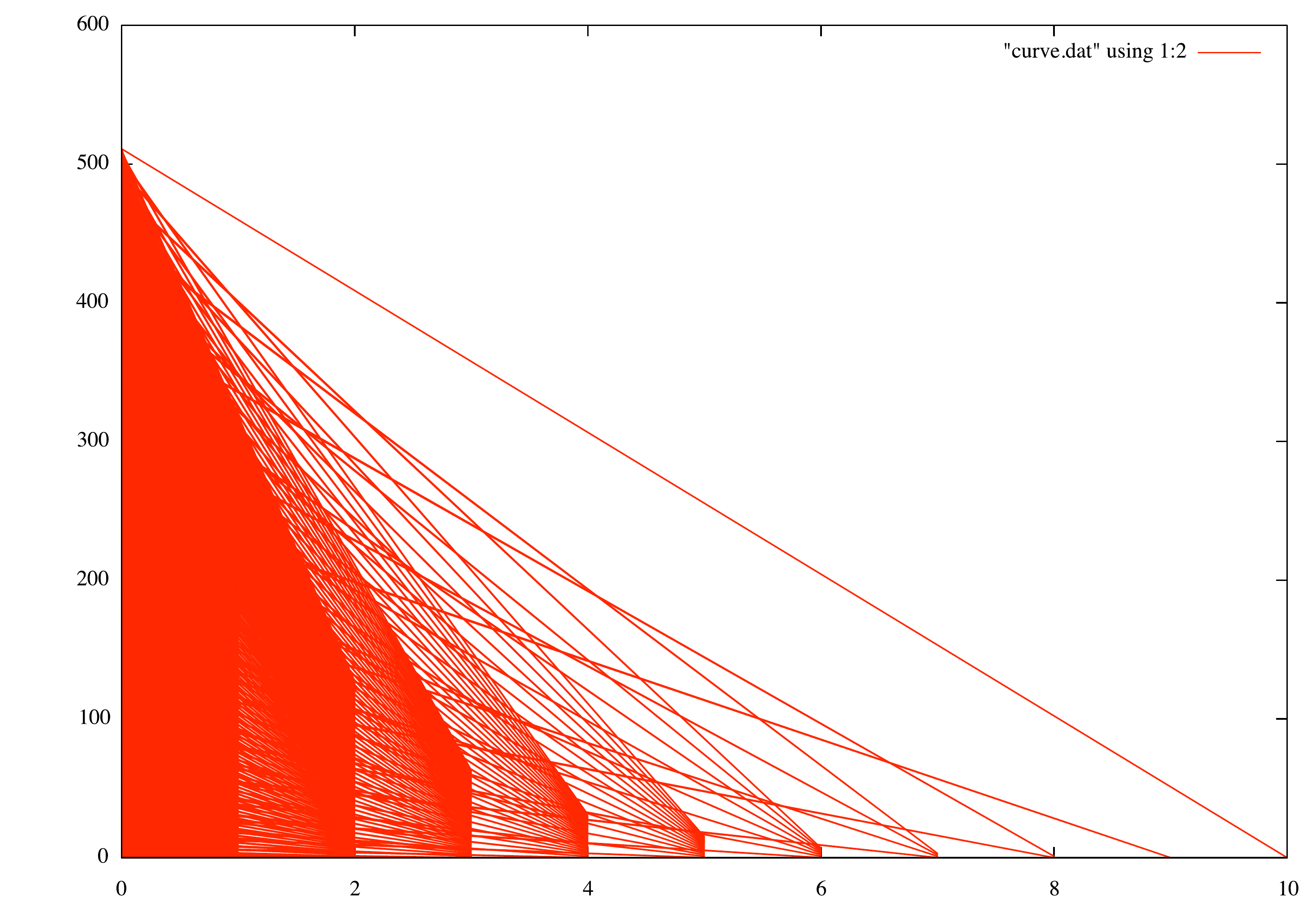}

Fig. \ref{f6} shows the action of the pairing function {\tt bitpair}
on its two arguments arguments in [0..63].

\FIG{f6}{Values of bitpair x y with x,y in [0..63]}{0.40}{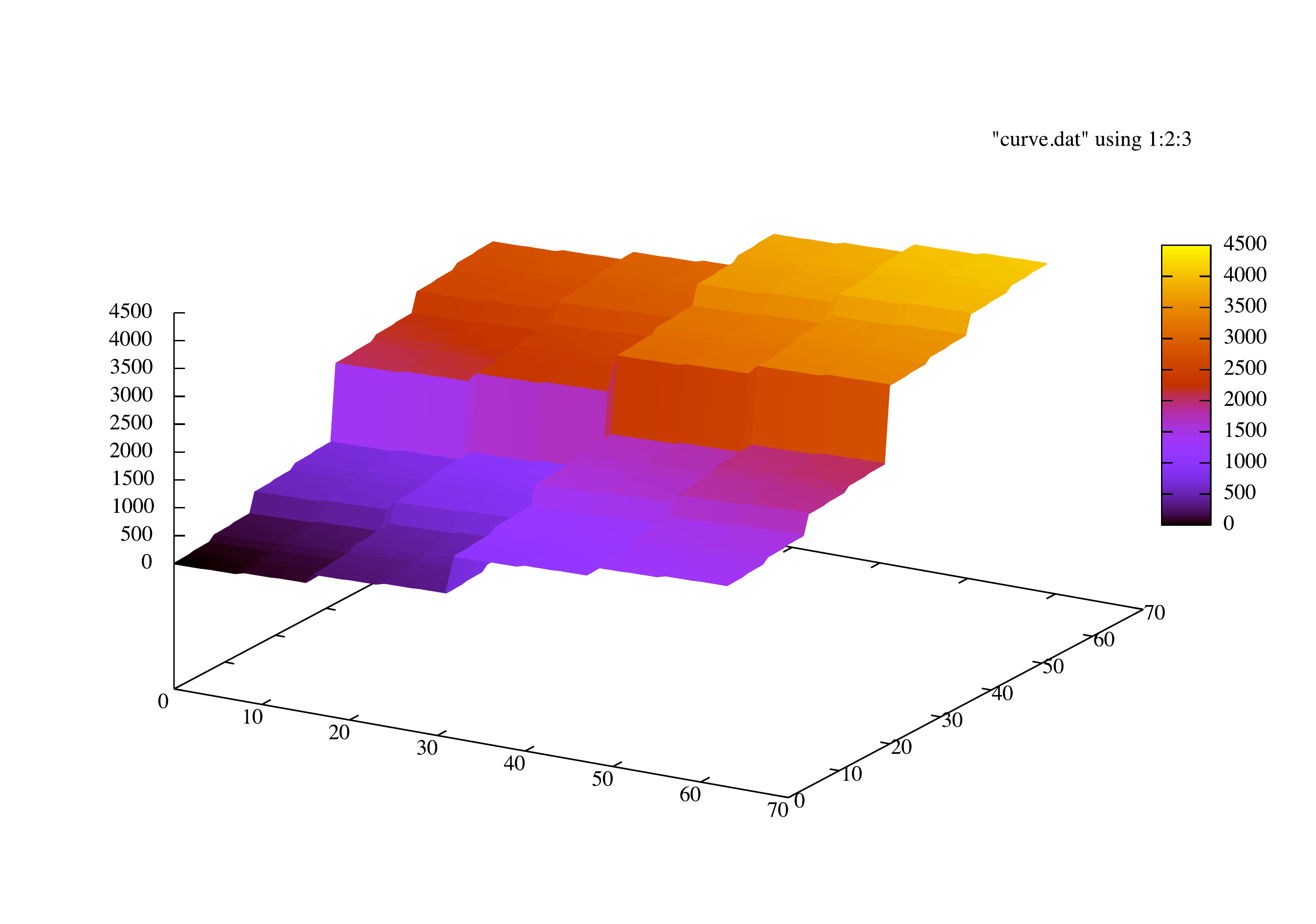}

\subsection{Encoding Unordered Pairs}
To derive an encoding of unordered pairs, i.e. 2 element sets, one
can combine pairing/unpairing with conversion between sequences and
sets:
\begin{code}
pair2unord_pair (x,y) = fun2set [x,y]
unord_pair2pair [a,b] = (x,y) where 
  [x,y]=set2fun [a,b]   

unord_unpair = pair2unord_pair . bitunpair
unord_pair = bitpair . unord_pair2pair
\end{code}
We can derive the following equivalent Encoders:
\begin{code}
set2 :: Encoder [Nat]
set2 = compose (Iso unord_pair2pair pair2unord_pair) nat2
\end{code}
that goes through {\tt nat2}, working as follows:
\begin{codex}
*ISO> as set2 nat 2008
[60,87]
*ISO> as nat set2 it
2008
\end{codex}
and
\begin{code}
set2' :: Encoder [Nat]
set2' = compose (Iso unord_pair unord_unpair) nat
\end{code}
that goes through {\tt nat}, working as follows:
\begin{codex}
*ISO> as set2' nat 2008
[60,87]
*ISO> as nat set2' [60,87]
2008
*ISO> as nat set2' [87,60]
2008
\end{codex}

\subsection{Encodings Multiset Pairs}
To derive an encoding of 2 element multisets, one
can combine pairing/unpairing with conversion between sequences and
multisets:
\begin{code}
pair2mset_pair (x,y) = (a,b) where [a,b]=fun2mset [x,y]
mset_unpair2pair (a,b) = (x,y) where [x,y]=mset2fun [a,b]

mset_unpair = pair2mset_pair . bitunpair
mset_pair = bitpair . mset_unpair2pair
\end{code}
We can derive the following Encoder:
\begin{code}
mset2 :: Encoder Nat2
mset2 = compose (Iso mset_unpair2pair pair2mset_pair) nat2
\end{code}
working as follows:
\begin{codex}
*ISO> as mset2 nat 2008
(60,86)
*ISO> as nat mset2 it
2008
\end{codex}

Figure \ref{isounpair3} shows the curve generated by {\tt mset\_unpair}
covering the lattice of points in its range.

\FIG{isounpair3}
{2D curve connecting values of {\tt mset\_unpair n} for $n \in [0..2^{10}-1]$}
{0.40}{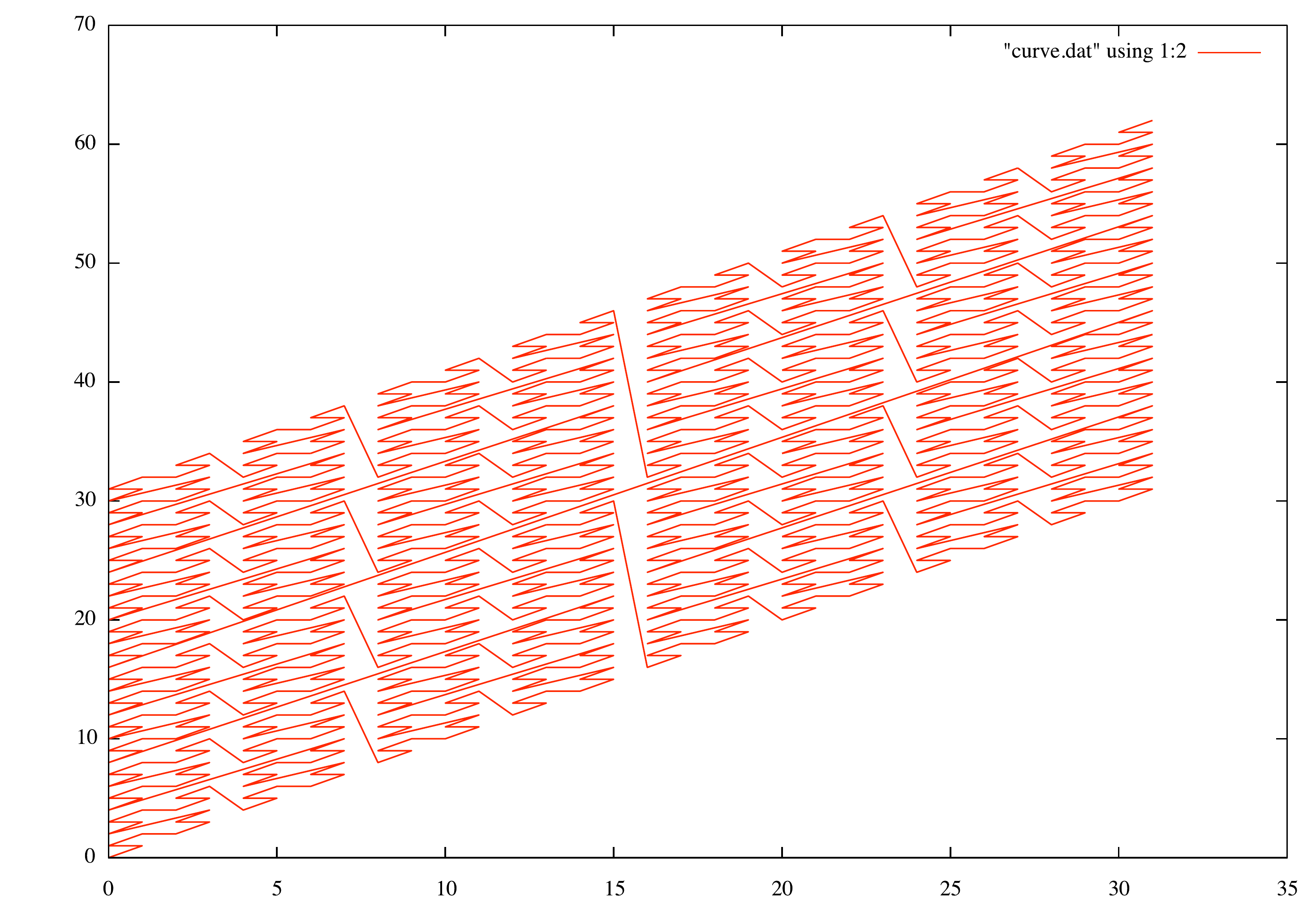}

\subsection{Extending Pairing/Unpairing to Signed Integers}
Given the bijection from $nat$ to $z$ one can easily extend pairing/unpairing
operations to signed integers. We obtain the Encoder:
\begin{code}
type Z2 = (Z,Z)

z2 :: Encoder Z2
z2 = compose (Iso zpair zunpair) nat

zpair (x,y) = (nat2z . bitpair) (z2nat x,z2nat y)
zunpair z = (nat2z n,nat2z m) where (n,m)= (bitunpair . z2nat) z
\end{code}
working as follows:
\begin{codex}
*ISO> map zunpair [-5..5]
[(-1,1),(-2,-1),(-2,0),(-1,-1),(-1,0),(0,0),(0,-1),(1,0),(1,-1),(0,1),(0,-2)]
*ISO> map zpair it
[-5,-4,-3,-2,-1,0,1,2,3,4,5]

*ISO> as z2 z (-2008)
(63,-26)
*ISO> as z z2 it
-2008
\end{codex}
Figure \ref{isozunpair} shows the curve covering the lattice of integer
coordinates generated by the function {\tt zunpair}.

\FIG{isozunpair}
{Curve generated by unpairing function on signed integers}
{0.40}{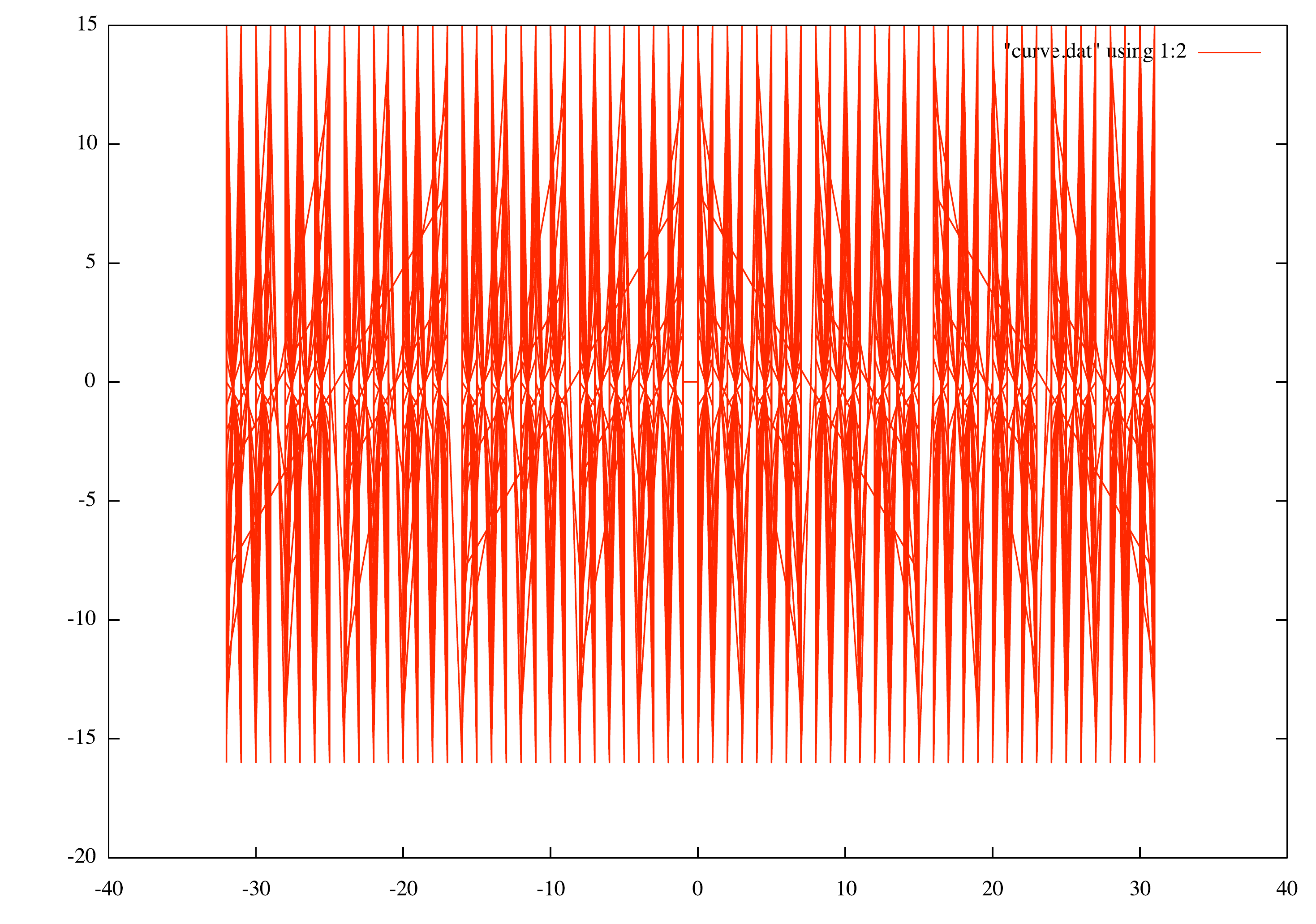}

The same construction can be extended to multiset pairing functions:
\begin{code}
mz2 :: Encoder Z2
mz2 = compose (Iso mzpair mzunpair) nat

mzpair (x,y) = (nat2z . mset_pair) (z2nat x,z2nat y)
mzunpair z = (nat2z n,nat2z m) where (n,m)= (mset_unpair . z2nat) z
\end{code}
working as follows:
\begin{codex}
*ISO> as mz2 z (-42)
(1,-8)
*ISO> as z mz2 it
-42
\end{codex}

\subsection{Gauss Integers and Pairing Functions}
Visualizing complex variable functions requires 4 dimensions even for
1-variable functions. This is usually handled by associating a color/hue value
to the {\em phase} while representing the {\em modulus} along the z-axis.
However, for 2-argument complex functions as simple as the sum, difference and
the product 6 dimensions would be needed.
Let us start shapeshifting operations on Gauss Integers
(pairs of integers with a real and imaginary part) in combination with a
mapping to ordinary integers using the  (commutative!) multiset
pairing/unpairing isomorphism provided by the Encoder {\tt mz2}:
\begin{code}
gauss_sum (ab,cd) = mzpair (a+b,c+d) where
  (a,b)=mzunpair ab
  (c,d)=mzunpair cd

gauss_dif (ab,cd) = mzpair (a-b,c-d) where
  (a,b)=mzunpair ab
  (c,d)=mzunpair cd
  
gauss_prod (ab,cd) = mzpair (a*c-b*d,b*c+a*d) where
  (a,b)=mzunpair ab
  (c,d)=mzunpair cd
\end{code}
Clearly one can now fit these operations in 3-dimensions as shown
in Figures \ref{isogsum}, \ref{isogdif}, \ref{isogprod} visualizing
sums, differences and products of Gauss Integers obtained by
unpairing integers in $[-2^4..2^4-1]$.

\FIG{isogsum}
{Sums of Gauss Integers visualized with Pairing functions}
{0.40}{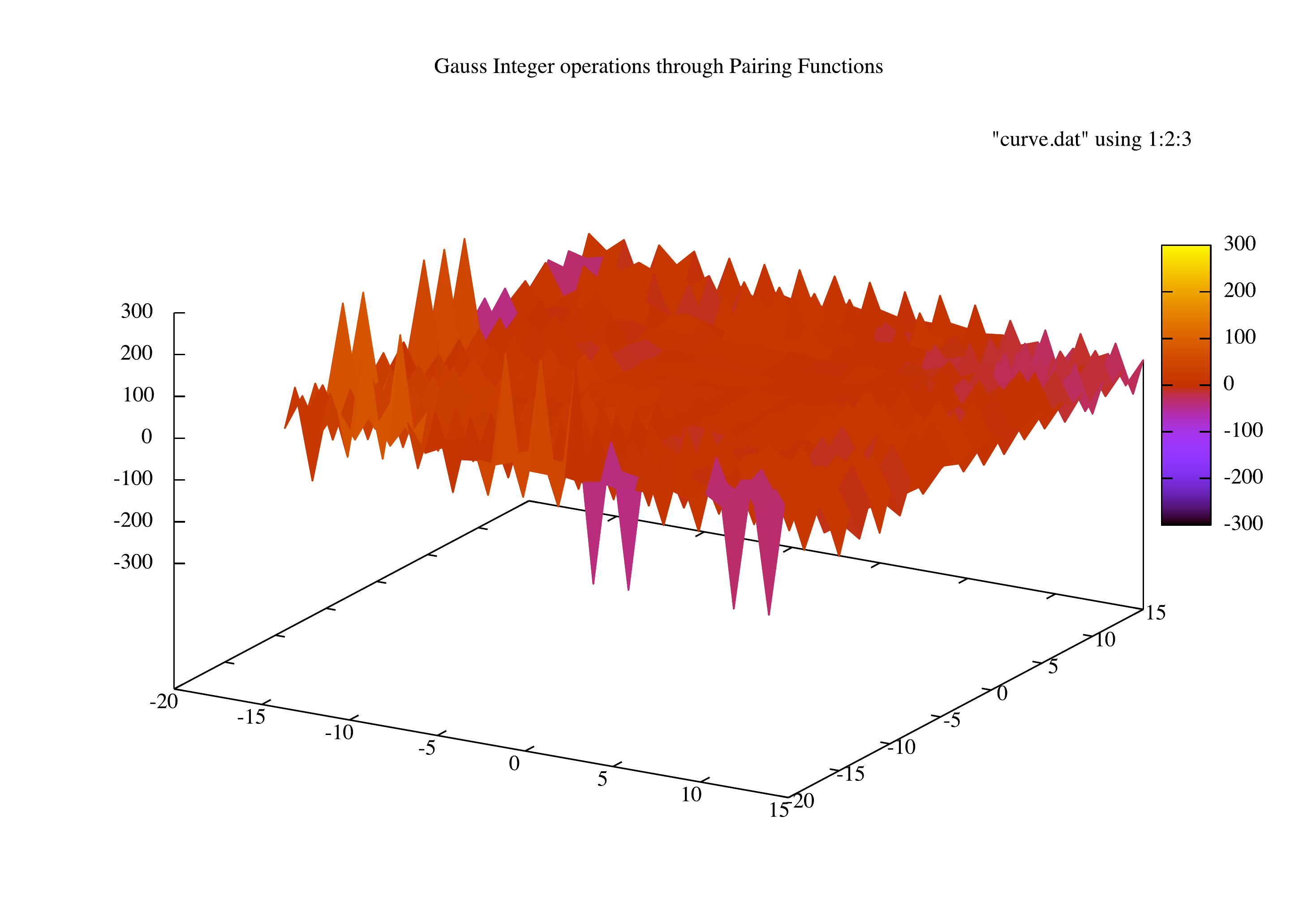}

\FIG{isogdif}
{Differences of Gauss Integers visualized with Pairing functions}
{0.40}{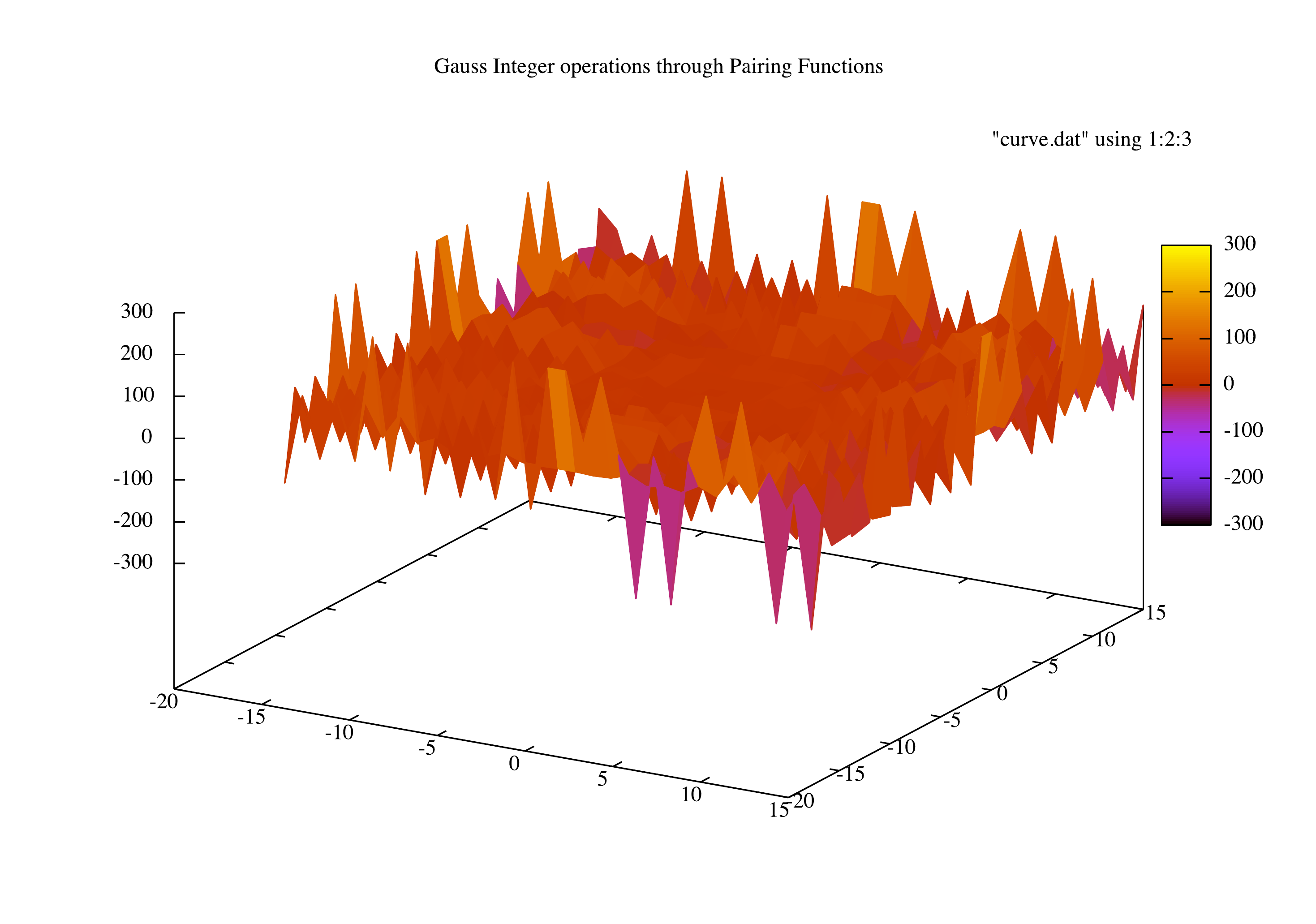}

\FIG{isogprod}
{Products of Gauss Integers visualized with Pairing functions}
{0.40}{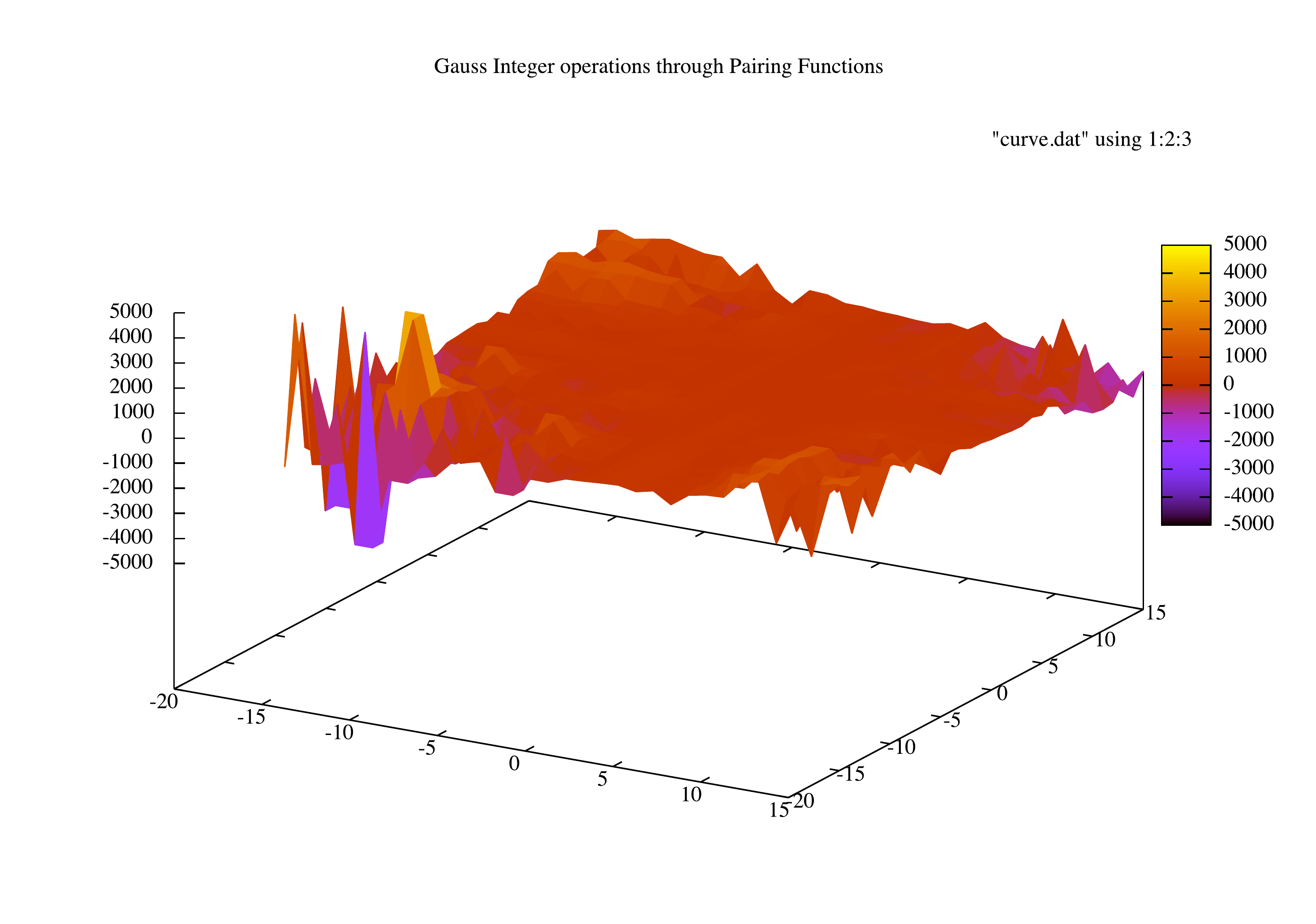}

\subsection{Some algebraic properties of pairing functions}
The following propositions state some simple
algebraic identities between pairing operations acting on ordered, unordered and multiset pairs.

\begin{prop}
Given the function definitions:
\begin{code}
bitlift x = bitpair (x,0)
bitlift' = (from_base 4) . (to_base 2)

bitclip = fst . bitunpair
bitclip' = (from_base 2) . (map (`div` 2)) . (to_base 4) . (*2)

bitpair' (x,y) = (bitpair (x,0))   +   (bitpair(0,y))
xbitpair (x,y) = (bitpair (x,0)) `xor` (bitpair (0,y))
obitpair (x,y) = (bitpair (x,0))  .|.  (bitpair (0,y))

pair_product (x,y) = a+b where
  x'=bitpair (x,0)
  y'=bitpair (0,y)
  ab=x'*y'
  (a,b)=bitunpair ab
\end{code}
the following identities hold:
\begin{equation}
bitlift \equiv bitlift'
\end{equation}
\begin{equation}
bitclip \equiv bitclip'
\end{equation}
\begin{equation}
bitclip \circ bitlift \equiv id 
\end{equation}
\begin{equation}
bitpair (0,n) \equiv 2*bitpair(n,0)
\end{equation}
\begin{equation}
bitpair (0,n) \equiv 2*(bitlift~n)
\end{equation}
\begin{equation}
bitpair (n,n) \equiv 3*(bitlift~n)
\end{equation}
\begin{equation}  \label{bitpow}
bitpair (2^n,0) \equiv  ({2^n})^2
\end{equation}
\begin{equation}  \label{biteq}
bitpair (2^{2^n}+1,0) \equiv 2^{2^{n+1}}+1
\end{equation}
\begin{equation}
bitpair' \equiv bitpair \equiv xbitpair \equiv obitpair
\end{equation}
\begin{equation}
bitpair (x,y) \equiv (bitlift~x)+2*(bitlift~y) 
\end{equation}
\begin{equation}
pair\_product \equiv *
\end{equation}
\end{prop}

\begin{prop}
Given the function definitions
\begin{code}
bitpair'' (x,y) = mset_pair (min x y,x+y) 

bitpair''' (x,y) = unord_pair [min x y,x+y+1]

mset_pair' (a,b) = bitpair (min a b, (max a b) - (min a b)) 

mset_pair'' (a,b) = unord_pair [min a b, (max a b)+1]

unord_pair' [a,b] = bitpair (min a b, (max a b) - (min a b) -1) 

unord_pair'' [a,b] = mset_pair (min a b, (max a b)-1)
\end{code}
the following identities hold:
\begin{equation}
bitpair \equiv bitpair'' \equiv bitpair '''
\end{equation}
\begin{equation} \label{mseteq}
mset\_pair \equiv mset\_pair' \equiv mset\_pair ''
\end{equation}
\begin{equation}
unord\_pair \equiv unord\_pair' \equiv unord\_pair ''
\end{equation}
\end{prop}

\section{Cons-Lists with Pairing/Unpairing}

The simplest application of pairing/unpairing operations is encoding
of cons-lists of natural numbers, defined as the data type:
\begin{code}
data CList = Atom Nat | Cons CList CList 
  deriving (Eq,Ord,Show,Read)
\end{code}

First, to provide an infinite supply of atoms, we encode them
as even numbers:
\begin{code}
to_atom n = 2*n
from_atom a | is_atom a = a `div` 2
is_atom n = even n  && n>=0
\end{code}
Next, as we want atoms and cons cell disjoint, we will encode the later as
odd numbers:
\begin{code}
is_cons n = odd n && n>0
decons z | is_cons z = pepis_unpair ((z-1) `div` 2)
cons x y = 2*(pepis_pair (x,y))+1
\end{code}
We can deconstruct a natural number by recursing over
applications of the unpairing-based {\tt decons} combinator:
\begin{code}
nat2cons n | is_atom n = Atom (from_atom n)
nat2cons n | is_cons n = 
  Cons (nat2cons hd) 
       (nat2cons tl) where
    (hd,tl) = decons n     
\end{code}
We can reverse this process by recursing with the {\tt cons} combinator
on the CList data type:
\begin{code}
cons2nat (Atom a) =  to_atom a
cons2nat (Cons h t) = cons (cons2nat h) (cons2nat t)
\end{code}
The following example shows both transformations as inverses.
\begin{codex}
*ISO> cons2nat (Cons (Atom 0) (Cons (Atom 1) (Cons (Atom 2) (Atom 3))))
26589
*ISO> nat2cons 26589
Cons (Atom 0) (Cons (Atom 1) (Cons (Atom 2) (Atom 3)))
\end{codex}
We obtain the Encoder:
\begin{code}
clist :: Encoder CList
clist = compose (Iso cons2nat nat2cons) nat
\end{code}
The Encoder works as follows:
\begin{codex}
*ISO> as clist nat 101
Cons (Atom 0) (Cons (Atom 0) (Atom 3))
\end{codex}
and can be used to generate random LISP-like data and code skeletons
from natural numbers.

\section{Revisiting Multiset Encodings}
We will now use pairing/unpairing functions, in combination with mappings to
sequences and sets to design an efficient encoding of multisets.

The function {\tt fmset2nat} starts by grouping the elements of a multiset. The
lengths of the groups (decremented by 1), as well as an element of each are then
collected in 2 lists. Then the second list is morphed
from a set to a sequence, as this provides a more compact representation without
changing the length of the list. The first list, seen as a sequence is then
paired element by element with the second list. 
Finally, the resulting numbers, seen as a sequence, are then fused together. 
\begin{code}
fmset2nat pairingf ms = m where  
  mss= group (sort ms) 
  xs=map (pred . genericLength) mss
  zs=map head mss
  ys=set2fun zs
  ps=zip xs ys
  ns=map pairingf ps 
  m=fun2nat ns
\end{code}
The function {\tt fnat2mset} reverses the process step by step:
\begin{code}
fnat2mset unpairingf m = rs where
   ns=nat2fun m
   ps=map unpairingf ns
   (xs,ys)=unzip ps
   xs'=map succ xs
   zs=fun2set ys
   f k x = genericTake k (repeat x) 
   rs = concat (zipWith f xs' zs)
\end{code}
After instantiating these generic functions to interesting pairing/unpairing
functions
\begin{code}
bmset2nat = fmset2nat bitpair
nat2bmset = fnat2mset bitunpair

bmset2nat' = fmset2nat pepis_pair
nat2bmset' = fnat2mset pepis_unpair
\end{code}
We obtain the Encoders:
\begin{code}
bmset :: Encoder [Nat]
bmset = compose (Iso bmset2nat nat2bmset) nat

bmset' :: Encoder [Nat]
bmset' = compose (Iso bmset2nat' nat2bmset') nat
\end{code}
working as follows:
\begin{codex}
ISO> as bmset nat 2008
[1,1,2,3,3,4,5,6,7]
*ISO> as nat bmset it
2008
*ISO> map (as bmset nat) [0..7]
[[],[0],[0,0],[0,1],[1],[0,1,1],[0,0,1],[0,1,2]]
*ISO> as bmset' nat 2008
[0,0,0,1,2,2,3,4,5,6]
\end{codex}
Note that, in contrast to the intractable prime number based multiset encoding
{\tt pmset}, this time we obtain an encoding, linear in the
bitsize of the natural numbers involved, as in the case of {\tt mset}.
Note also that the construction is generic in the sense that it works with any
pairing / unpairing function.
Like in the case of {\tt mset} and {\tt pmset} multiset encodings we can extend
these encodings to a hylomorphism {\tt hfbm}:
\begin{code}
nat_bmset = Iso nat2bmset bmset2nat

hfbm :: Encoder T
hfbm = compose (hylo nat_bmset) nat

nat_bmset' = Iso nat2bmset' bmset2nat'

hfbm' :: Encoder T
hfbm' = compose (hylo nat_bmset') nat
\end{code}
working as follows:
\begin{codex}
*ISO> as hfbm nat 42
H [H [],H [],H [H []],H [H []],H [H [],H []],H [H [],H []]]
*ISO> as nat hfbm it
42
*ISO> as hfbm' nat 2008
H [H [],H [],H [],H [H []],H [H [],H []],
   H [H [],H []],H [H [],H [H []]],H [H [H []]],
   H [H [],H [H []],H [H []]],H [H [],H [],H [H []]]]
*ISO> as nat hfbm' it
2008   
\end{codex}

\section{Pairing Functions and Encodings of Binary Decision Diagrams}
\label{encbdd}
As a variation on the theme of pairing/unpairing functions, we will show in this
section that a Binary Decision Diagram ($BDD$) 
representing the same logic function as an $n$-variable $2^n$ bit truth
table can be obtained by applying {\tt bitunpair} recursively to {\tt tt}.
More precisely, we will show that applying this {\em unfolding} operation
results in a complete binary tree of depth $n$ representing
a $BDD$ that returns {\tt tt} when evaluated applying
its boolean operations.

The binary tree type {\tt BT} has the constants {\tt
B0} and {\tt B1} as leaves representing the boolean values $0$ and $1$.
Internal nodes (that will represent {\tt if-then-else} decision points), 
will be marked with the constructor {\tt D}. 
We will also add integers to represent logic
variables, ordered identically in each
branch, as first arguments of {\tt D}. 
The two other arguments will be subtrees 
that represent {\tt THEN} 
and {\tt ELSE} branches:
\begin{code}
data BT a = B0 | B1 | D a (BT a) (BT a) 
             deriving (Eq,Ord,Read,Show)
\end{code}

The constructor {\tt BDD} wraps together a 
tree of type {\tt BT} and the number of logic
variables occurring in it.
\begin{code}
data BDD a = BDD a (BT a) deriving (Eq,Ord,Read,Show)
\end{code}

\subsection{Unfolding natural numbers to binary trees}
The following functions apply {\tt bitunpair} recursively, 
on a Natural Number {\tt tt}, 
seen as an $n$-variable $2^n$ bit truth table, 
to build a complete binary tree of depth $n$, 
that we will represent using the {\tt BDD} data type. 
\begin{code}
unfold_bdd :: Nat2 -> BDD Nat
unfold_bdd (n,tt) = BDD n bt where 
  bt=if tt<max then split_with bitunpair n tt
     else error 
       ("unfold_bdd: last arg "++ (show tt)++
       " should be < " ++ (show max))
     where max = 2^2^n

  split_with _ n 0 | n<1 =  B0
  split_with _ n 1 | n<1 =  B1
  split_with f n tt = D k (split_with f k tt1) 
                          (split_with f k tt2) where
    k=pred n
    (tt1,tt2)=f tt
\end{code}
The following examples 
show results returned by {\tt unfold\_bdd} 
for the $2^{2^n}$ truth tables associated to $n$ variables,
for $n=2$:
\begin{codex}
 BDD 2 (D 1 (D 0 B0 B0) (D 0 B0 B0))
 BDD 2 (D 1 (D 0 B1 B0) (D 0 B0 B0))
 BDD 2 (D 1 (D 0 B0 B0) (D 0 B1 B0))
 ...
 BDD 2 (D 1 (D 0 B1 B1) (D 0 B1 B1))
\end{codex}
Note that no boolean operations have been performed so far
and that we still have to prove that such
trees actually represent BDDs associated to truth tables.

\subsection{Folding binary trees to natural numbers}
One can ``evaluate back'' the binary tree of data type BDD,
by using the pairing function {\tt bitpair}.  
The inverse of {\tt unfold\_bdd} is implemented as follows:
\begin{code}
fold_bdd :: BDD Nat -> Nat2
fold_bdd (BDD n bt) = 
  (n,fuse_with bitpair bt) where
    fuse_with rf B0 = 0
    fuse_with rf B1 = 1
    fuse_with rf (D _ l r) = 
      rf (fuse_with rf l,fuse_with rf r)
\end{code}
Note that this is a purely structural operation
and that integers in first argument position
of the constructor {\tt D} are actually ignored.

The two bijections work as follows:
\begin{codex}
*ISO>unfold_bdd (3,42)
  BDD 3 
    (D 2 
      (D 1 (D 0 B0 B0) 
           (D 0 B0 B0)) 
      (D 1 (D 0 B1 B1) 
           (D 0 B1 B0)))

*ISO>fold_bdd it
  42
\end{codex}

\subsection{Boolean Evaluation of BDDs}
Practical uses of BDDs involve reducing them by
sharing nodes and eliminating identical
branches \cite{bryant86graphbased}.
Note that in this case {\tt bdd2nat} 
might give a different result as it computes
different pairing operations. 
Fortunately,
we can try to fold the binary tree 
back to a natural number
by evaluating it as a boolean
function.

The function {\tt eval\_bdd} describes the $BDD$ evaluator:
\begin{code}
eval_bdd (BDD n bt) = eval_with_mask (bigone n) n bt
 
eval_with_mask m _ B0 = 0 
eval_with_mask m _ B1 = m
eval_with_mask m n (D x l r) = 
  ite_ (var_mn m n x) 
         (eval_with_mask m n l) 
         (eval_with_mask m n r)
         
var_mn mask n k = mask `div` (2^(2^(n-k-1))+1)
bigone nvars = 2^2^nvars - 1         
\end{code}

The {\em projection functions} {\tt var\_mn}
can be combined with the usual bitwise integer operators, 
to obtain new bitstring truth tables, 
encoding all possible value combinations of their arguments, 
as shown in \cite{knuth06draft}.
Note that the constant $0$ evaluates to $0$ while the constant $1$
is evaluated as $2^{2^n}-1$ by the function {\tt bigone}.

The function {\tt ite\_} used in {\tt eval\_with\_mask} 
implements the boolean function  {\tt if x then t else e}
using arbitrary length bitvector operations:
\begin{code}
ite_ x t e = ((t `xor` e).&.x) `xor` e
\end{code}

\noindent {\em 
As the following example shows, it turns out that
boolean evaluation {\tt eval\_bdd}
faithfully emulates {\tt fold\_bdd}!
}

\begin{codex}
*ISO> unfold_bdd (3,42)
BDD 3 (D 2 (D 1 (D 0 B0 B0) (D 0 B0 B0)) 
           (D 1 (D 0 B1 B1) (D 0 B1 B0)))
*ISO> eval_bdd it
42
\end{codex}

\subsection{The Equivalence} \label{equiv}
We will now state the surprising (and new!) result
that boolean evaluation and structural transformation with
repeated application of
{\em pairing}
produce the same result:

\begin{prop} \label{tt}
The complete binary tree of depth $n$, obtained by recursive 
applications of {\tt bitunpair} on a truth table $tt$
computes an (unreduced) BDD, that, when evaluated, 
returns the truth table, i.e.
\begin{equation}
fold\_bdd \circ unfold\_bdd \equiv id
\end{equation}

\begin{equation}
eval\_bdd \circ unfold\_bdd \equiv id
\end{equation}
\end{prop}
\begin{proof} The function {\tt unfold\_bdd} builds a
binary tree by splitting the bitstring $tt \in [0..2^n-1]$ up to depth $n$. 
Observe that this corresponds to the Shannon expansion \cite{shannon_all} of the
formula associated to the truth table, using variable order $[n-1,...,0]$.
Observe that the effect of {\tt bitunpair} is the same as
\begin{itemize}
  \item the effect of {\tt var\_mn m n (n-1)} 
     acting as a mask selecting the left branch, and
\item 
     the effect of its complement, acting as a mask selecting the right
     branch.
\end{itemize}
Given that $2^n$ is the double of $2^{n-1}$, the same invariant holds at each
step, as the bitstring length of the truth table reduces to half. 
\end{proof}
We can thus assume from now on, that the BDD data type defined in
section \ref{encbdd} actually represents BDDs mapped one-to-one to truth tables
given as natural numbers.
An interesting application of this result would
be to investigate practical uses of 
{\tt bitpair}/{\tt bitunpair} operations in actual circuit design.

\section{Ranking and Unranking of BDDs} \label{rank}

One more step is needed to extend the mapping between $BDDs$ with $n$
variables to a bijective mapping from/to $Nat$: 
we will have to ``shift towards infinity'' 
the starting point of each new block\footnote{defined by the same number of
variables} of BDDs in $Nat$ as BDDs of larger and larger sizes are enumerated.

First, we need to know by how much - so we will count the number
of boolean functions with up to $n$ variables.
\begin{code}
bsum 0 = 0
bsum n | n>0 = bsum1 (n-1)

bsum1 0 = 2
bsum1 n | n>0 = bsum1 (n-1)+ 2^2^n
\end{code}
The stream of all such sums can now be generated as usual\footnote{{\tt bsums}
is sequence A060803 in The On-Line Encyclopedia of Integer
Sequences, \url{http://www.research.att.com/~njas/sequences}}:
\begin{code}
bsums = map bsum [0..]
\end{code}
\begin{codex}
*ISO> genericTake 7 bsums
  [0,2,6,22,278,65814,4295033110]
\end{codex}

What we are really interested into, is decomposing {\tt n} into
the distance {\tt n-m} to the
last {\tt bsum} {\tt m} smaller than {\tt n},
and the index that generates the sum, {\tt k}.
\begin{code}
to_bsum n = (k,n-m) where 
  k=pred (head [x|x<-[0..],bsum x>n])
  m=bsum k
\end{code}
{\em Unranking} of an arbitrary BDD is now easy - the index {\tt k}
determines the number of variables and {\tt n-m} determines
the rank. Together they select the right BDD
with {\tt unfold\_bdd} and {\tt bdd}.
\begin{code}
nat2bdd n = unfold_bdd (k,n_m) where (k,n_m)=to_bsum n
\end{code}
{\em Ranking} of a BDD is even easier: we shift its rank
within the set of BDDs with {\tt nv} 
variables, by the value {\tt (bsum nv)} that
counts the ranks previously assigned.
\begin{code}
bdd2nat bdd@(BDD nv _) = (bsum nv)+tt where
  (_,tt) =fold_bdd bdd
\end{code}
As the following example shows
{\tt bdd2nat}
implements the inverse of
{\tt nat2bdd}.
\begin{codex}
*ISO> nat2bdd 42
BDD 3 (D 2 (D 1 (D 0 B0 B1) (D 0 B1 B0)) 
           (D 1 (D 0 B0 B0) (D 0 B0 B0)))
*ISO> bdd2nat it
42
\end{codex}

This provides the Encoder:
\begin{code}
pbdd :: Encoder (BDD Nat)
pbdd = compose (Iso bdd2nat nat2bdd) nat
\end{code}
working as follows:
\begin{codex}
*ISO> as pbdd nat 2008
BDD 4 (D 3 (D 2 B0 (D 1 (D 0 B0 B1) B1)) 
      (D 2 (D 1 (D 0 B1 B1) B0) (D 1 B0 B1)))
*ISO> as nat pbdd it
2008
\end{codex}

We can now repeat the {\em ranking} function construction for {\tt eval\_bdd}:
\begin{code}
ev_bdd2nat bdd@(BDD nv _) = (bsum nv)+(eval_bdd bdd)
\end{code}
We can confirm that {\tt ev\_bdd2nat} also acts as an inverse to
{\tt nat2bdd}:
\begin{codex}
*ISO> ev_bdd2nat (nat2bdd 2008)
2008
\end{codex}

We obtain the Encoder:
\begin{code}
bdd :: Encoder (BDD Nat)
bdd = compose (Iso ev_bdd2nat nat2bdd) nat
\end{code}
working as follows:
\begin{codex}
*ISO> as bdd nat 2008
BDD 4 (D 3 (D 2 (D 1 (D 0 B0 B0) (D 0 B0 B0)) 
                (D 1 (D 0 B0 B1) (D 0 B1 B0))) 
           (D 2 (D 1 (D 0 B1 B1) (D 0 B0 B0)) 
                (D 1 (D 0 B0 B0) (D0 B1 B0))))
*ISO> as nat bdd it
2008
\end{codex}
This result can be seen as an intriguing isomorphism between
boolean, arithmetic and symbolic computations.

\subsection{Reducing the $BDDs$}
We will sketch here a simplified reduction mechanism for BDDs
eliminating identical branches. As nodes of a BDD are mapped
bijectively to unique natural numbers we will omit
the (trivial) implementation of node sharing, with the
implicit assumption that subtrees having the same encoding
are shared.

The function {\tt bdd\_reduce} reduces a $BDD$ by collapsing identical 
left and right subtrees, and the function {\tt bdd} 
associates this reduced form to $n \in Nat$.
\begin{code}
bdd_reduce (BDD n bt) = BDD n (reduce bt) where
  reduce B0 = B0
  reduce B1 = B1
  reduce (D _ l r) | l == r = reduce l
  reduce (D v l r) = D v (reduce l) (reduce r)

unfold_rbdd = bdd_reduce . unfold_bdd  
\end{code}

The results returned by {\tt unfold\_rbdd} for {\tt n=2} are:
\begin{codex}
  BDD 2 (C 0)
  BDD 2 (D 1 (D 0 (C 1) (C 0)) (C 0))
  BDD 2 (D 1 (C 0) (D 0 (C 1) (C 0)))
  BDD 2 (D 0 (C 1) (C 0))
  ...
  BDD 2 (D 1 (D 0 (C 0) (C 1)) (C 1))
  BDD 2 (C 1)
\end{codex}
We can now define the {\em unranking} operation on reduced BDDs
\begin{code}
nat2rbdd = bdd_reduce . nat2bdd 
\end{code}
and obtain the Encoder
\begin{code}
rbdd :: Encoder (BDD Nat)
rbdd = compose (Iso ev_bdd2nat nat2rbdd) nat
\end{code}
working as follows
\begin{codex}
*ISO> as rbdd nat 2008
BDD 4 (D 3 (D 2 B0 (D 1 (D 0 B0 B1) (D 0 B1 B0))) 
           (D 2 (D 1 B1 B0) (D 1 B0 (D 0 B1 B0))))
*ISO> as nat rbdd it
2008
\end{codex}

To be able to compare its space complexity
with other representations we will define 
a size operation on a BDD as follows:
\begin{code}
bdd_size (BDD _ t) = 1+(size t) where
  size B0 = 1
  size B1 = 1
  size (D _ l r) = 1+(size l)+(size r)
\end{code}
This measures the size of the BDD or reduced BDD as an expression tree.
To take into account sharing (as present in a standard ROBDD implementation)
one can simply eliminate duplicated subtrees:
\begin{code}
robdd_size (BDD _ t) = 1+(rsize t) where
  rsize = genericLength . nub . rbdd_nodes
  rbdd_nodes B0 = [B0]
  rbdd_nodes B1 = [B1]
  rbdd_nodes (D v l r) = 
    [(D v l r)] ++ (rbdd_nodes l) ++ (rbdd_nodes r)
\end{code}

\section{Generalizing BDD ranking/unranking functions}

\subsection{Encoding BDDs with Arbitrary Variable Order}
While the encoding built around the equivalence described in Prop. \ref{tt}
between bitwise pairing/unpairing operations and boolean decomposition
is arguably as simple and elegant as possible, it is useful
to parametrize BDD generation with respect to an arbitrary
variable order. This is of particular importance when using
BDDs for circuit minimization, as different variable orders
can make circuit sizes flip from linear to exponential in
the number of variables \cite{bryant86graphbased}.

Given a permutation of $n$ variables represented as
natural numbers in $[0..n-1]$ and a truth table
$tt \in [0..2^{2^n}-1]$ we can define: 
\begin{code}
to_bdd vs tt | 0<=tt && tt <= m = 
  BDD n (to_bdd_mn vs tt m n) where
    n=genericLength vs
    m=bigone n
to_bdd _ tt = error 
   ("bad arg in to_bdd=>" ++ (show tt)) 
\end{code}
where the function {\tt to\_bdd\_mn} recurses over
the list of variables {\tt vs} and applies
Shannon expansion \cite{shannon_all},
expressed as bitvector operations. This computes
branches $f1$ and $f0$, to be used as {\tt then} and {\tt else}
parts, when evaluating back the BDD to a truth table
with if-the-else functions.
\begin{code}
to_bdd_mn []      0 _ _ = B0
to_bdd_mn []      _ _ _ = B1
to_bdd_mn (v:vs) tt m n = D v l r where
  cond=var_mn m n v
  f0= (m `xor` cond) .&. tt
  f1= cond .&. tt 
  l=to_bdd_mn vs f1 m n
  r=to_bdd_mn vs f0 m n
\end{code}
\begin{prop}
The function {\tt to\_bdd} builds an (unreduced) BDD corresponding
to a truth table {\tt tt} for variable order {\tt vs} that returns
{\tt tt} when evaluated as a boolean function.
\end{prop}
We can reduce the resulting BDDs, and convert back from BDDs and reduced BDDs to
truth tables with boolean evaluation: 
\begin{code}
to_rbdd vs tt = bdd_reduce (to_bdd vs tt)
from_bdd bdd = eval_bdd bdd
\end{code}
We can obtain BDDs and reduced BDDs of various sizes as follows:
\begin{codex}
*ISO> as perm nat 5
[0,2,1]
*ISO> to_bdd (as perm nat 5) 42
BDD 3 (D 0 (D 2 (D 1 B0 B0) (D 1 B1 B1)) 
           (D 2 (D 1 B0 B0) (D 1 B1 B0)))
*ISO> to_rbdd (as perm nat 5) 42
BDD 3 (D 0 (D 2 B0 B1) (D 2 B0 (D 1 B1 B0)))
*ISO> to_rbdd (as perm nat 8) 42
BDD 3 (D 2 B0 (D 0 B1 (D 1 B1 B0)))
ISO> from_bdd it
42
\end{codex}
Finally, we can, obtain a minimal BDD expressing a logic function of $n$
variables given as a truth table as follows:
\begin{code}
to_min_bdd n t = search_bdd min n t

search_bdd f n tt = snd $ foldl1 f 
 (map (sized_rbdd tt) (all_permutations n)) where
    sized_rbdd tt vs = (robdd_size b,b) where 
      b=to_rbdd vs tt
 
all_permutations n = if n==0 then [[]] else
  [nth2perm (n,i)|i<-[0..(factorial n)-1]] where
     factorial n=foldl1 (*) [1..n]
\end{code}
As the following examples show, this can provide an effective 
multilevel boolean formula minimization up to functions with
6-7 arguments.
\begin{codex}
*ISO> to_min_bdd 3 42
BDD 3 (D 0 (D 2 B0 B1) (D 1 (D 2 B0 B1) B0))
*ISO> to_min_bdd 4 2008
BDD 4 (D 3 (D 1 (D 0 B0 B1) (D 0 B1 B0)) 
      (D 2 (D 1 (D 0 B0 B1) B0) (D 0 B1 B0)))
*ISO> to_min_bdd 7 2008
BDD 7 (D 0 (D 1 (D 2 (D 6 
                     (D 4 (D 3 B0 B1) (D 3 B1 B0)) 
      (D 5 (D 4 (D 3 B0 B1) B0) 
                (D 3 B1 B0))) B0) B0) B0)
*ISO> robdd_size it
12
\end{codex}

\subsection{Multi-Terminal Binary Decision Diagrams (MTBDD)} \label{multi}
MTBDDs \cite{DBLP:journals/fmsd/FujitaMY97,CBGP08} are a natural generalization
of BDDs allowing non-binary values as leaves.
Such values are typically 
bitstrings representing the outputs
of a multi-terminal boolean function,
encoded as unsigned integers.

We shall now describe an encoding of $MTBDDs$
that can be extended to ranking/unranking functions,
in a way similar to $BDDs$ as shown in section \ref{rank}.

Our {\tt MTBDD} data type is a binary tree like the one used for $BDDs$,
parameterized by two integers {\tt m} and {\tt n}, indicating
that an MTBDD represents a function from $[0..n-1]$ to $[0..m-1]$,
or equivalently, an $n$-input/$m$-output boolean function.

\begin{code}   
data MT a = L a | M a (MT a) (MT a) deriving (Eq,Ord,Read,Show)
data MTBDD a = MTBDD a a (MT a) deriving (Show,Eq)
\end{code}

The function  {\tt to\_mtbdd} creates,
from a natural number tt representing a truth table,
an MTBDD representing
functions of type $N \rightarrow M$ with $M=[0..2^m-1], N=[0..2^n-1]$.
Similarly to a BDD, it is represented as binary tree 
of $n$ levels, except that its leaves are in $[0..{2^m}-1]$.
\begin{code}
to_mtbdd m n tt = MTBDD m n r where 
  mlimit=2^m
  nlimit=2^n
  ttlimit=mlimit^nlimit
  r=if tt<ttlimit 
    then (to_mtbdd_ mlimit n tt)
    else error 
      ("bt: last arg "++ (show tt)++
      " should be < " ++ (show ttlimit))
\end{code}
Given that correctness of the range of
{\tt tt} has been checked, the function {\tt to\_mtbdd\_} 
applies {\tt bitmerge\_unpair} 
recursively up to depth $n$, where
leaves in range $[0..mlimit-1]$ are created.
\begin{code}  
to_mtbdd_ mlimit n tt|(n<1)&&(tt<mlimit) = L tt
to_mtbdd_ mlimit n tt = (M k l r) where 
   (x,y)=bitunpair tt
   k=pred n
   l=to_mtbdd_ mlimit k x
   r=to_mtbdd_ mlimit k y
\end{code}
Converting back from $MTBDDs$ to natural numbers is
basically the same thing as for $BDDs$, except that
assertions about the range of leaf data are enforced.
\begin{code}
from_mtbdd (MTBDD m n b) = from_mtbdd_ (2^m) n b

from_mtbdd_ mlimit n (L tt)|(n<1)&&(tt<mlimit)=tt
from_mtbdd_ mlimit n (M _ l r) = tt where 
   k=pred n
   x=from_mtbdd_ mlimit k l
   y=from_mtbdd_ mlimit k r
   tt=bitpair (x,y)
\end{code}
The following examples show that {\tt to\_mtbdd} and {\tt from\_mtbdd}
are indeed inverses values in $[0..2^n-1] \times [0..2^m-1]$. 
\begin{codex}
>to_mtbdd 3 3 2008
  MTBDD 3 3 
    (M 2 
      (M 1 
         (M 0 (L 2) (L 1)) 
         (M 0 (L 2) (L 1))) 
      (M 1 
         (M 0 (L 2) (L 0)) 
         (M 0 (L 1) (L 1))))

>from_mtbdd it
2008

>mprint (to_mtbdd 2 2) [0..3]
  MTBDD 2 2 
    (M 1 (M 0 (L 0) (L 0)) (M 0 (L 0) (L 0)))
  MTBDD 2 2 
    (M 1 (M 0 (L 1) (L 0)) (M 0 (L 0) (L 0)))
  MTBDD 2 2 
    (M 1 (M 0 (L 0) (L 0)) (M 0 (L 1) (L 0)))
  MTBDD 2 2 
    (M 1 (M 0 (L 1) (L 0)) (M 0 (L 1) (L 0)))
\end{codex}

\section{Revisiting Encodings of Finite Functions} \label{tfun}
We will now generalize the {\tt bitpair} pairing function to $k$-tuples and then we will 
derive an alternative encoding for finite functions. 

\subsection{Tuple Encodings as Generalized Bitpair} \label{tuple}
The function {\tt to\_tuple:} $Nat \rightarrow Nat^k$ converts a natural 
number to a $k$-tuple by splitting its bit representation into $k$ groups, 
from which the $k$ members in the tuple are finally rebuilt. This operation 
can be seen as a transposition of a bit matrix obtained by expanding 
the number in base $2^k$:
\begin{code}  
to_tuple k n = map (from_base 2) (
    transpose (
      map (to_maxbits k) (
        to_base (2^k) n
      )
    )
  )
\end{code}
To convert a $k$-tuple back to a natural number we will merge their 
bits, $k$ at a time. This operation uses the transposition of a bit 
matrix obtained from the tuple, seen as a number in base $2^k$, 
with help from bit crunching functions given in APPENDIX:
\begin{code}
from_tuple ns = from_base (2^k) (
    map (from_base 2) (
      transpose (
        map (to_maxbits l) ns
      )
    )
  ) where 
      k=genericLength ns
      l=max_bitcount ns
\end{code}
The following example shows the decoding of {\tt 42}, its decomposition 
in bits (right to left), the formation of a $3$-tuple and the encoding 
of the tuple back to {\tt 42}.
\begin{codex}
*ISO> to_base 2 42
[0,1,0,1,0,1]
*ISO> to_tuple 3 42
[2,1,2]
*ISO> to_base 2 2
[0,1]
*ISO> to_base 2 1
[1]
*ISO> from_tuple [2,1,2]
42
\end{codex}
Fig. \ref{3tuple} shows multiple steps of the same decomposition, 
with shared nodes collected in a DAG. 
\VFIGS{3tuple}{Repeated 3-tuple expansions}{42}{2008}{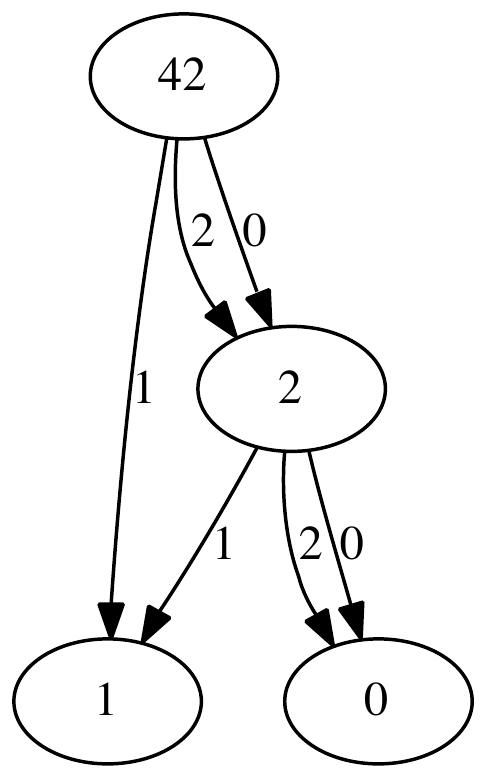}{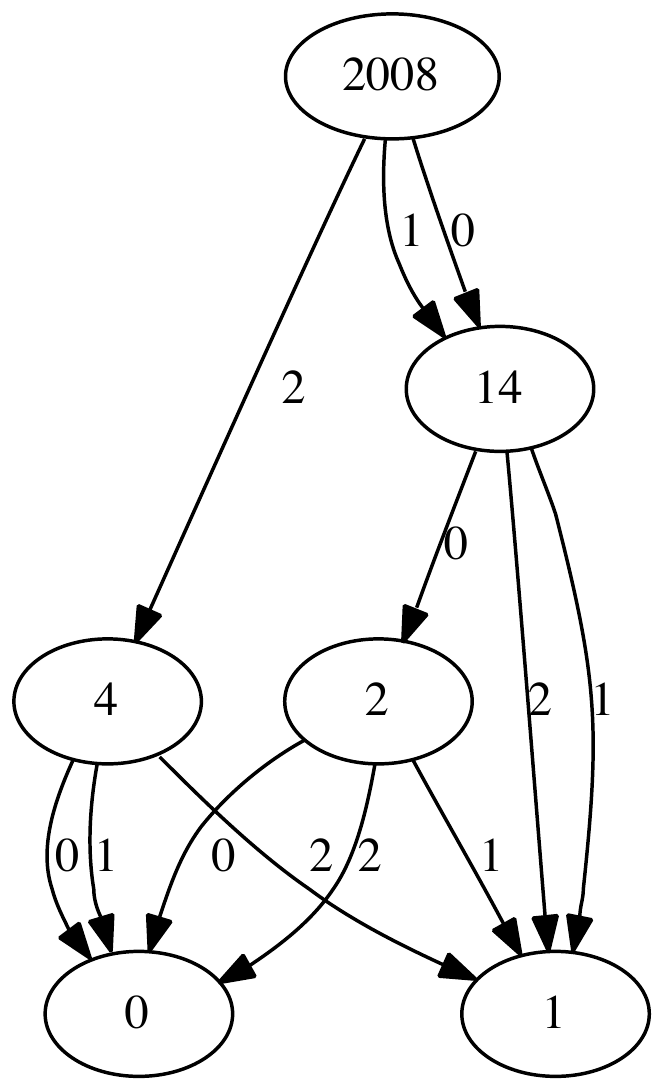}

\subsection{Encoding Finite Functions as Tuples}

As finite sets can be put in a bijection with an initial segment 
of $Nat$,
a {\tt finite function} can be seen as a function defined from an
initial segment of $Nat$ to $Nat$.
We can encode and decode a finite function from $[0..k-1]$ to $Nat$ 
(seen as the list of its values), as a natural number:
\begin{code} 
ftuple2nat [] = 0
ftuple2nat ns = succ (pepis_pair (pred k,t)) where
  k=genericLength ns 
  t=from_tuple ns

nat2ftuple 0 = []
nat2ftuple kf = to_tuple (succ k) f where 
  (k,f)=pepis_unpair (pred kf)
\end{code}
As the length of the tuple, {\tt k}, is usually smaller than the number 
obtained by merging the bits of the {\tt k}-tuple, we have picked the 
Pepis pairing function, exponential in its first argument and linear 
in its second, to embed the length of the tuple needed for the decoding. 
This suggest the following alternative Encoder for finite functions:
\begin{code}
fun' :: Encoder [Nat]
fun' = compose (Iso ftuple2nat nat2ftuple) nat
\end{code}
as well as the related alternative hylomorphism:
\begin{code}
nat_fun' = Iso nat2ftuple ftuple2nat

hff' :: Encoder T
hff' = compose (hylo nat_fun') nat
\end{code}

The encoding/decoding and the hylomorphism work as follows:
\begin{codex}
*ISO> as fun' nat 2008
[3,2,3,1]
*ISO> as nat fun' it
2008

*ISO> as hff' nat 2008
H [H [H [H []]],H [H [],H []],H [H [H []]],H [H []]]
*ISO> as nat hff' it
2008
\end{codex}

\section{Directed Graphs, Undirected graphs, Multigraphs and Hypergraphs}
We will now show that more complex data types like digraphs and hypergraphs
have extremely simple encoders. This shows once more the importance
of compositionality in the design of our embedded transformation
language.

\subsection{Encoding Directed Graphs} \label{digraphs}
We can find a bijection from 
directed graphs (with no isolated vertices, corresponding to their
 view as binary relations), to finite sets by fusing their list
 of ordered pair representation into finite sets with a pairing
 function:
\begin{code}
digraph2set ps = map bitpair ps
set2digraph ns = map bitunpair ns
\end{code}
The resulting Encoder is:
\begin{code}
digraph :: Encoder [Nat2]
digraph = compose (Iso digraph2set set2digraph) set
\end{code}
working as follows:
\begin{codex}
*ISO> as digraph nat 2008
[(1,1),(2,0),(2,1),(3,1),(0,2),(1,2),(0,3)]
*ISO> as nat digraph it
2008
*ISO> as rbdd digraph [(1,1),(2,0),(2,1),
                        (3,1),(0,2),(1,2),(0,3)]
BDD 4 (D 3 (D 2 B0 (D 1 (D 0 B0 B1) (D 0 B1 B0))) 
           (D 2 (D 1 B1 B0) (D 1 B0 (D 0 B1 B0))))
\end{codex}
Fig. \ref{f5} shows the digraph associated to {\tt 2008}.

\FIG{f5}{2008 as a digraph}{0.60}{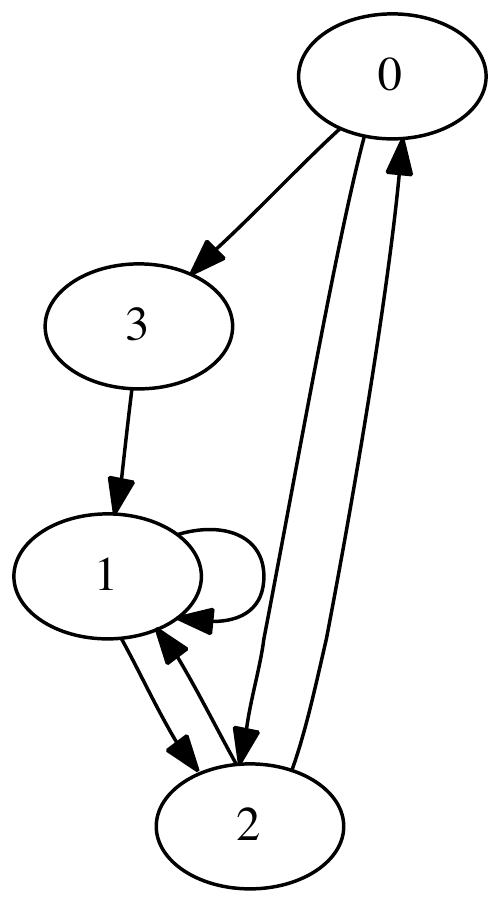}

\subsection{Encoding Undirected Graphs} \label{ugraphs}
We can find a bijection from 
undirected graphs to finite sets by fusing their list
 of unordered pair representation into finite sets with a pairing
 function on unordered pairs:
\begin{code}
graph2set ps = map unord_pair ps
set2graph ns = map unord_unpair ns
\end{code}
The resulting Encoder is:
\begin{code}
graph :: Encoder [[Nat]]
graph = compose (Iso graph2set set2graph) set
\end{code}
working as follows:
\begin{codex}
*ISO> as graph nat 2008
[[1,3],[2,3],[2,4],[3,5],[0,3],[1,4],[0,4]]
*ISO> as nat graph it
2008
*ISO> as nat graph 
        [[1,3],[3,2],[2,4],[5,3],[0,3],[4,1],[0,4]]
2008
\end{codex}
Note that, as expected, the result is invariant to changing the order of
elements in pairs like {\tt [1,4]} and {\tt [3,5]} to {\tt [4,1]} and
{\tt [5,3]}.

\subsection{Encoding Directed Multigraphs} \label{mdigraphs}
We can find a bijection from 
directed multigraphs (directed graphs with multiple edges between pairs of
vertices) to finite sequences by fusing their list of ordered pair
representation into finite sequences with a pairing function:

The resulting Encoder is:
\begin{code}
mdigraph :: Encoder [Nat2]
mdigraph = compose (Iso digraph2set set2digraph) fun
\end{code}
working as follows:
\begin{codex}
*ISO> as mdigraph nat 2008
[(1,1),(0,0),(1,0),(0,0),(0,0),(0,0),(0,0)]
*ISO> as nat mdigraph it
2008
\end{codex}
Note that the only change to the {\tt digraph} Encoder is replacing
the composition with {\tt set} by a composition with {\tt fun}.
 
\subsection{Encoding Undirected Multigraphs} \label{mgraphs}
We can find a bijection from 
undirected multigraphs (undirected graphs with multiple edges between 
unordered pairs of
vertices) to finite sequences by fusing their list of pair
representation into finite sequences with a 
pairing function on unordered pairs:

The resulting Encoder is:
\begin{code}
mgraph :: Encoder [[Nat]]
mgraph = compose (Iso graph2set set2graph) fun
\end{code}
working as follows:
\begin{codex}
*ISO> as mgraph nat 2008
[[1,3],[0,1],[1,2],[0,1],[0,1],[0,1],[0,1]]
*ISO> as nat mgraph it
2008
\end{codex}
Note that the only change to the {\tt graph} Encoder is replacing
the composition with {\tt set} by a composition with {\tt fun}.
 
\subsection{Encoding Hypergraphs}
\begin{df}
A hypergraph (also called {\em set system}) is a pair $H=(X,E)$ where
$X$ is a set and $E$ is a set of non-empty subsets of $X$.
\end{df}
We can easily
derive a bijective encoding of {\em hypergraphs}, 
represented as sets of sets:
\begin{code}
set2hypergraph = map nat2set
hypergraph2set = map set2nat
\end{code}
The resulting Encoder is:
\begin{code}
hypergraph :: Encoder [[Nat]]
hypergraph = compose (Iso hypergraph2set set2hypergraph) set
\end{code}
working as follows
\begin{codex}
*ISO> as hypergraph nat 2008
[[0,1],[2],[1,2],[0,1,2],[3],[0,3],[1,3]]
*ISO> as nat hypergraph it
2008
\end{codex}

\section{Encoding SAT problems}
Boolean Satisfiability (SAT) problems are encoded as lists of lists representing
conjunctions of disjunctions of positive or negative propositional symbols.

After defining:
\begin{code}
set2sat = map (set2disj . nat2set) where
  shift0 z = if (z<0) then z else z+1
  set2disj = map (shift0. nat2z)
  
sat2set = map (set2nat . disj2set) where
  shiftback0 z = if(z<0) then z else z-1
  disj2set = map (z2nat . shiftback0)
\end{code}
we obtain the Encoder
\begin{code}
sat :: Encoder [[Z]]
sat = compose (Iso sat2set set2sat) set
\end{code}
working as follows:
\begin{codex}
*ISO> as sat nat 2008
[[1,-1],[2],[-1,2],[1,-1,2],[-2],[1,-2],[-1,-2]]
*ISO> as nat sat it
2008
\end{codex}
Clearly this encoding can be used to generate random SAT problems out of
easier to generate random natural numbers.

\section{An Encoder for Graph Models}
Graph models \cite{bucciarelli,berline} provide a semantics of
$\lambda$-calculus (Y-combinator included) in terms of sets of finite sets of natural numbers.
Following \cite{bucciarelli}  a {\em graph model} is a
pair $(D,p)$ where D is an infinite set and $p:D^{*}\times D \rightarrow D$ is
an injective total function. We will strengthen this to be a bijection, for the
case $D=Nat$ as follows.
\begin{code}
gmodel2nat (set,m) = pred (fun2nat (m : (set2fun set)))
nat2gmodel n = (fun2set xs,m) where (m:xs) = nat2fun (succ n)
\end{code}
This provides the Encoder:
\begin{code}
type Gdomain= ([Nat],Nat)
gmodel :: Encoder Gdomain
gmodel = compose (Iso gmodel2nat nat2gmodel) nat
\end{code}
working as follows:
\begin{codex}
*ISO> as gmodel nat 42
([0,2,4],0)
*ISO> as nat gmodel it
42
\end{codex}
The interests of such models is that they provide an accurate
set theoretic semantics for untyped lambda calculus describing
key computational mechanisms like $\beta$-conversion and fixpoint
combinators.

\section{A mapping to a dense set: Dyadic Rationals in $[0,1)$}
So far our isomorphisms have focused on natural numbers,
finite sets and other 
discrete data types.
Dyadic rationals are fractions with denominators 
restricted to be exponents of 2.
They are a {\em dense} set in $\mathcal{R}$ i.e. they provide arbitrarily
close approximations for any real number. An interesting isomorphism to
such a set would allow borrowing things like distance or average
functions that could have interesting interpretations in symbolic or boolean
domains. It also makes sense to pick a bounded subdomain of the dyadic
rationals that can be meaningful as the range of
probabilistic boolean functions or fuzzy sets.
We will build an Encoder for Dyadic Rationals in $[0,1)$ by
providing a bijection from finite sets of natural numbers
seen this time as {\em negative} exponents of 2.
\begin{code}
dyadic :: Encoder (Ratio Nat)
dyadic = compose (Iso dyadic2set set2dyadic) set
\end{code}
The function {\tt set2dyadic} mimics {\tt set2nat} defined in subsection
\ref{natset}, except for the use of negative exponents and computation on
rationals.
\begin{code}
set2dyadic :: [Nat] -> Ratio Nat
set2dyadic ns = rsum (map nexp2 ns) where
  nexp2 0 = 1%2
  nexp2 n = (nexp2 (n-1))*(1%2)

  rsum [] = 0%1
  rsum (x:xs) = x+(rsum xs)
\end{code}
The function {\tt dyadic2set} extracts negative exponents of two from
a dyadic rational and it is modeled after {\tt nat2set} defined in subsection
\ref{natset}.
\begin{code}
dyadic2set :: Ratio Nat -> [Nat]
dyadic2set n | good_dyadic n = dyadic2exps n 0 where
  dyadic2exps 0 _ = []
  dyadic2exps n x = 
    if (d<1) then xs else (x:xs) where
      d = 2*n
      m = if d<1 then d else (pred d)
      xs=dyadic2exps m (succ x)
dyadic2set _ =  
  error  "dyadic2set: argument not a dyadic rational"
\end{code}
As not all rational numbers are dyadics in $[0,1)$, the predicate {\tt
good\_dyadic} is needed validate the input of {\tt dyadic2set}. 
This also ensures that {\tt dyadic2set} always terminates returning
a finite set.
\begin{code}
good_dyadic kn = (k==0 && n==1) 
  || ((kn>0%1) && (kn<1%1) && (is_exp2 n)) where 
    k=numerator kn
    n=denominator kn

    is_exp2 1 = True
    is_exp2 n | even n = is_exp2 (n `div` 2)
    is_exp2 n = False
\end{code}

Some examples of borrow/lend operations are:
\begin{code}
dyadic_dist x y = abs (x-y)

dist_for t x y =  as dyadic t 
  (borrow2 (with dyadic t) dyadic_dist x y)
dsucc = borrow (with nat dyadic) succ
dplus = borrow2 (with nat dyadic) (+)

dconcat = lend2 dyadic (++)
\end{code}
\begin{codex}
*ISO> dist_for nat  6 7
1%2
*ISO> dist_for set [1,2,3] [3,4,5]
21%64
*ISO> dsucc (3%8)
7%8
\end{codex}
Fig. \ref{f7} shows the dyadic rationals associated
to natural numbers in [0..255].

\FIG{f7}{Dyadic rationals associated to n in [0..255]}{0.40}{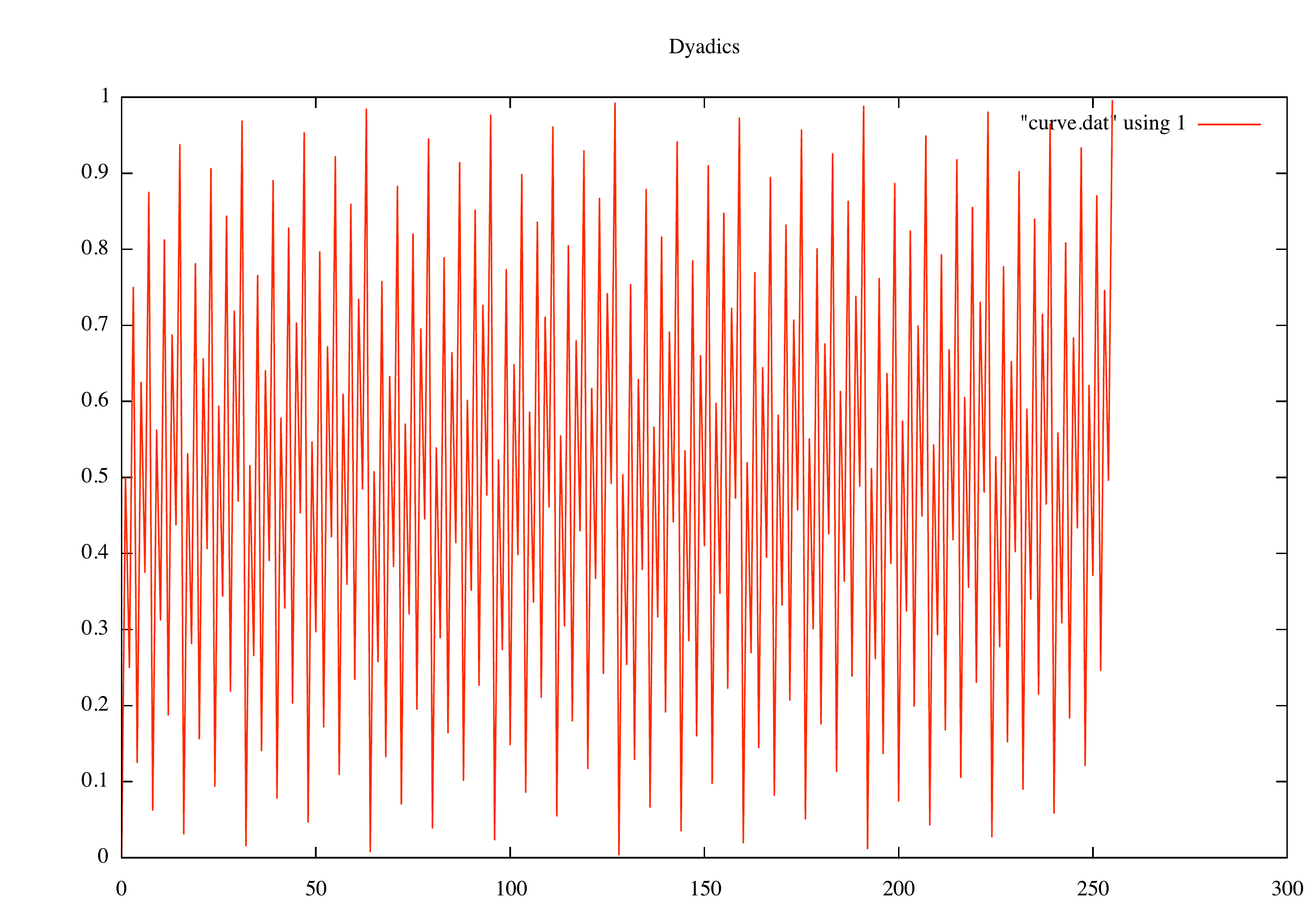}

\section{Strings and Parenthesis Languages}

\subsection{Encoding Strings}
As strings can be seen just as a notational equivalent
of lists of natural numbers 
we obtain an Encoder immediately as: 
\begin{code}
string :: Encoder String
string = Iso string2fun fun2string

string2fun cs = map (fromIntegral . ord) cs

fun2string ns = map (chr . fromIntegral) ns 
\end{code}
Note however that this is only an isomorphism
within the {\tt chr/ord} conversion range,
therefore we shall assume this constraint
as a {\em law} governing this Encoder.
\begin{codex}
*ISO> as set string "hello"
[104,206,315,424,536]
*ISO> as string set it
"hello"
\end{codex}

\subsection{Encoding a Parenthesis Language}

An encoder for a parenthesis language is obtained by
combining a parser and writer. As Hereditarily Finite Functions
naturally map one-to-one to a parenthesis expression
we will choose them as target of the transformers.
\begin{code}
pars :: Encoder [Char]
pars = compose (Iso pars2hff hff2pars) hff
\end{code}

The parser recurses over a string and builds a {\tt HFF} as follows:
\begin{code}
pars2hff cs = parse_pars '(' ')' cs

parse_pars l r cs | newcs == [] = t where
  (t,newcs)=pars_expr l r cs

  pars_expr l r (c:cs) | c==l = ((H ts),newcs) where 
     (ts,newcs) = pars_list l r cs
     
  pars_list l r (c:cs) | c==r = ([],cs)
  pars_list l r (c:cs) = ((t:ts),cs2) where 
    (t,cs1)=pars_expr l r (c:cs)
    (ts,cs2)=pars_list l r cs1
\end{code}
The writer recurses over a {\tt HFF} and collects
matching parenthesis pairs:
\begin{code}
hff2pars = collect_pars '(' ')'

collect_pars l r (H ns) =
  [l]++ 
    (concatMap (collect_pars l r) ns)
  ++[r] 
\end{code}
The transformations of {\tt 42} look as follows:
\begin{codex}
*ISO> as pars nat 42
"((())(())(()))"
*ISO> as hff pars it
H [H [H []],H [H []],H [H []]]
*ISO> as nat hff it
42
\end{codex}
Alternatively, by using a {\tt 0} and {\tt 1} as left and right parenthesis we
can define:
\begin{code}
bitpars2hff cs = parse_pars 0 1 cs
hff2bitpars = collect_pars 0 1

hff_pars :: Encoder [Nat]
hff_pars = compose (Iso bitpars2hff hff2bitpars) hff
\end{code}
working as follows:
\begin{codex}
*ISO> as hff_pars nat 2008
[0,0,0,1,0,1,1,0,1,0,0,1,1,0,1,0,1,0,1,0,1,1]
*ISO> as nat hff_pars it
2008
*ISO> as nat bits (as hff_pars nat 2008)
7690599
\end{codex}
As the last example shows, the information density of
a parenthesis representation is lower. This is expected,
given that order is constrained by balancing and content is
constrained by having the same number of {\tt 0s} and {\tt 1s}.
The following example
\begin{codex}
*ISO> map ((as nat bits) .  (as hff_pars nat)) [0..7]
[5,27,119,115,495,483,471,467]
\end{codex}
shows that this application is injective only.
Therefore a succinct representation of an abstract tree
structure can be obtained by encoding it as a
natural number as in:
\begin{codex}
*ISO> as nat pars "((()())()(())()()()())"
2008
\end{codex}
Note however, that
\begin{codex}
*ISO> as nat bits (as hff_pars nat (2^2^16))
32639
\end{codex}
while the conventional representation of the same number would
have a few thousand digits. This suggest defining:
\begin{code}
nat2parnat n = as nat bits (as hff_pars nat n)

parnat2nat n = as nat hff_pars (as bits nat n)
\end{code}
and find out that
\begin{codex}
*ISO> [x|x<-[0..2^16],nat2parnat x<x]
[8192,16384,32768,32769,49152,65536]
\end{codex}
One can see that more compact representations only happen
for a few numbers that are powers of two or ``sparse'' sums of
powers of two.
A good way to evaluate ``information density'' for an arbitrary
data type that is isomorphic to {\tt Nat} through one of our encoders
is to compute the total bitsize of its actual encoding over an
interval like $[0..2^{n-1}]$. For instance,
\begin{code}
hff_bitsize n= sum (map hff_bsize [0..2^n-1])

hff_bsize k=genericLength (as bits nat (nat2parnat k)) 
\end{code}
Knowing that the optimal bit representation of all numbers in $[0..2^{n-1}]$
totals $n*2^n$ ($2^n$ of them, $n$ bits each), we can define a measure of
information density for a bit-encoded parenthesis language seen as a
representation for {\tt HFF} as:
\begin{code}
info_density_hff n = (n*2^n)%(hff_bitsize n)
\end{code}
One can see that information density progressively increases to converge
to a value above half of the ``perfect'' value of {\tt 1}:
\begin{codex}
*ISO> map info_density_hff [0..12]
[0%1,1%3,4%9,12%25,1%2,80%157,16%31,112%215,
    32%61,48%91,1024%1933,2816%5297,2048%3841]
*ISO> map fromRational it
[0.0,0.3333333333333333,0.4444444444444444,0.48,0.5,
 0.5095541401273885,0.5161290322580645,0.5209302325581395,
 0.5245901639344263,0.5274725274725275,0.5297465080186239,
 0.5316216726448934,0.5331944806040094]    
\end{codex}
To compare this with the information density of hereditarily finite sets,
multisets and permutations, we can also map their structure to a bit-represented
parenthesis language by defining the encoder:
\begin{code}
pars_hf=Iso bitpars2hff hff2bitpars

hff_pars' :: Encoder [Nat]
hff_pars' = compose pars_hf hff'

hfs_pars :: Encoder [Nat]
hfs_pars = compose pars_hf hfs

hfpm_pars :: Encoder [Nat]
hfpm_pars = compose pars_hf hfpm

hfm_pars :: Encoder [Nat]
hfm_pars = compose pars_hf hfm

bhfm_pars :: Encoder [Nat]
bhfm_pars = compose pars_hf hfbm

bhfm_pars' :: Encoder [Nat]
bhfm_pars' = compose pars_hf hfbm'

hfp_pars :: Encoder [Nat]
hfp_pars = compose pars_hf hfp
\end{code}
and then defining:
\begin{code}
parsize_as t n = genericLength (hff2bitpars (as t nat n))

parsizes_to m t = map (parsize_as t) [0..2^m-1]
 
nat2hfsnat n = as nat bits (as hfs_pars nat n)

hfs_bitsize n= sum (map hfs_bsize [0..2^n-1])

hfs_bsize k=genericLength (as bits nat (nat2hfsnat k)) 
  
info_density_hfs n = (n*2^n)%(hfs_bitsize n)
\end{code}
The intuition that hereditarily finite functions have higher information
density than hereditarily finite sets can now be conjectured:
\begin{codex}
*ISO> map info_density_hfs [0..12]
[0%1,1%3,2%5,3%8,1%3,5%16,2%7,7%27,4%17,3%13,2%9,11%52,1%5]
*ISO> map fromRational it
[0.0,0.3333333333333333,0.4,0.375,0.3333333333333333,
 0.3125,0.2857142857142857,0.25925925925925924,0.23529411764705882,
 0.23076923076923078,0.2222222222222222,0.21153846153846154,0.2]
\end{codex}
Contrary to the case of bit-encoded HFFs, in this case
information density is decreasing for larger values - an observation
that can help with finding a simple proof for the conjecture.
More generally, such techniques suggest applications to
experimental mathematics.

\section{Self-delimiting codes} \label{selfdelim}
A more precise estimate of the actual size of various bitstring representations
requires also counting the overhead for ``delimiting'' their components.
An asymptotically optimal mechanism for this is the use of a {\em universal
self-delimiting code} for instance, the {\em Elias omega code}.
To implement it, the encoder proceeds by recursively encoding length of the
string, the length of the length of the strings etc.

\begin{code}
to_elias :: Nat -> [Nat]
to_elias n = (to_eliasx (succ n))++[0]

to_eliasx 1 = []
to_eliasx n = xs where
  bs=to_lbits n
  l=(genericLength bs)-1
  xs = if l<2 then bs else (to_eliasx l)++bs
\end{code}
The decoder first rebuilds recursively the
sequence of lengths and then the actual bitstring.
It makes sense to design the decoder to extract the number
represented by the self-delimiting code from a sequence/stream 
of bits and also return what is left after the extraction.
\begin{code}
from_elias :: [Nat] -> (Nat, [Nat])
from_elias bs = (pred n,cs) where (n,cs)=from_eliasx 1 bs

from_eliasx n (0:bs) = (n,bs)
from_eliasx n (1:bs) = r where 
  hs=genericTake n bs
  ts=genericDrop n bs
  n'=from_lbits (1:hs)
  r=from_eliasx n' ts 

to_lbits = reverse . (to_base 2)

from_lbits = (from_base 2) . reverse
\end{code}
We obtain the Encoder:
\begin{code}
elias :: Encoder [Nat]
elias = compose (Iso (fst . from_elias) to_elias) nat
\end{code}
working as follows:
\begin{codex}
*ISO> as elias nat 42
[1,0,1,0,1,1,0,1,0,1,1,0]
*ISO> as nat elias it
42
*ISO> as elias nat 2008
[1,1,1,0,1,0,1,1,1,1,1,0,1,1,0,0,1,0]
*ISO> as nat elias it
2008
\end{codex}
Note that self-delimiting codes are not {\em onto} the regular language
$\{0,1\}^*$, therefore this Encoder cannot be used to map arbitrary bitstrings
to numbers.

\section{Encoding DNA}
We have covered so far encodings for ``artificial entities'' used in various
fields. We will now add an encoding of ``natural origin'', DNA bases and
strands. While it is an (utterly) simplified model of the real thing, it captures
some essential algebraic properties of DNA bases and strands.

We start with a DNA data type, following \cite{algDNA,haskellDNA}:
\begin{code}
data Base = Adenine | Cytosine | Guanine | Thymine 
  deriving(Eq,Ord,Show,Read)

type DNA = [Base]
\end{code}
We will encode/decode the DNA base alphabet as follows:
\begin{code}
alphabet2code Adenine = 0
alphabet2code Cytosine = 1
alphabet2code Guanine = 2 
alphabet2code Thymine = 3

code2alphabet 0 = Adenine
code2alphabet 1 = Cytosine
code2alphabet 2 = Guanine 
code2alphabet 3 = Thymine
\end{code}
The mapping is simply a symbolic variant of conversion to/from base 4:
\begin{code}
dna2nat  = (from_base 4) . (map alphabet2code)

nat2dna = (map code2alphabet) . (to_base 4)
\end{code}
We can now define a decoder for base sequences as follows:
\begin{code}
dna :: Encoder DNA
dna = compose (Iso dna2nat nat2dna)  nat
\end{code}

A first set of DNA operations act on base sequences.
The transformation between complements looks as follows:  
\begin{code}
dna_complement :: DNA -> DNA
dna_complement = map to_compl where
  to_compl Adenine = Thymine
  to_compl Cytosine = Guanine
  to_compl Guanine = Cytosine
  to_compl Thymine = Adenine
\end{code}
Reversing is just list reversal.
\begin{code}
dna_reverse :: DNA -> DNA
dna_reverse = reverse
\end{code}
As reversal and complement are independent operations
their composition is commutative - we can pick
reversing first and then complementing:
\begin{code}
dna_comprev :: DNA -> DNA
dna_comprev = dna_complement . dna_reverse
\end{code}
The following examples show interaction of DNA codes
with other data types and their operations:
\begin{codex}
*ISO> as dna nat 2008
[Adenine,Guanine,Cytosine,Thymine,Thymine,Cytosine]
*ISO> borrow (with dna nat) dna_reverse 42
42
*ISO> borrow (with dna nat) dna_reverse 2008
637
*ISO> borrow (with dna nat) dna_complement 2008
2087
*ISO> borrow (with dna nat) dna_comprev 2008
3458
*ISO> borrow (with dna bits) 
        dna_comprev [1,0,1,0,1,1,0,1,0,1]
[1,1,1,0,1,0,0,0,0,1,1]
\end{codex}
Note that each of these DNA operations induces
a bijection $Nat \rightarrow Nat$.

Like signed integers, DNA strands have ``polarity'' - their direction matters:
\begin{code}
data Polarity =  P3x5 | P5x3 
  deriving(Eq,Ord,Show,Read)

data DNAstrand = DNAstrand Polarity DNA 
  deriving(Eq,Ord,Show,Read)
\end{code}
Polarity can be easily encoded as parity even/odd:
\begin{code}
strand2nat (DNAstrand polarity strand) = 
  add_polarity polarity (dna2nat strand) where 
    add_polarity P3x5 x = 2*x
    add_polarity P5x3 x = 2*x-1
    
nat2strand n =
  if even n 
     then DNAstrand P3x5 (nat2dna (n `div` 2))
     else DNAstrand P5x3 (nat2dna ((n+1) `div` 2))
\end{code}
We can now define an Encoder for DNA strands:
\begin{code}
dnaStrand :: Encoder DNAstrand
dnaStrand = compose (Iso strand2nat nat2strand) nat
\end{code}
Two additional operations lift DNA sequences to strands with polarities:
\begin{code}
dna_down :: DNA -> DNAstrand
dna_down = (DNAstrand P3x5) . dna_complement

dna_up :: DNA -> DNAstrand
dna_up = DNAstrand P5x3
\end{code}
We can now lend or borrow operations as follows:
\begin{codex}
*ISO> as dnaStrand nat 1234
DNA P3x5 [Cytosine,Guanine,Guanine,Cytosine,Guanine]
*ISO> lend (with dnaStrand nat) succ 
   (DNAstrand P5x3 [Adenine,Cytosine,Guanine,Thymine])
DNAstrand P5x3 [Cytosine,Cytosine,Guanine,Thymine]
\end{codex}

The DoubleHelix is a stable combination of two complementary
strands. This built-in redundancy protects against unwanted
mutations.
\begin{code}
data DoubleHelix = DoubleHelix DNAstrand DNAstrand
   deriving(Eq,Ord,Show,Read)

dna_double_helix :: DNA -> DoubleHelix
dna_double_helix s = 
  DoubleHelix (dna_up s) (dna_down s)
\end{code}
We can now generate a double helix from a natural number:
\begin{codex}
*ISO> dna_double_helix (nat2dna 33)
DoubleHelix 
  (DNAstrand P5x3 [Cytosine,Adenine,Guanine])
  (DNAstrand P3x5 [Guanine,Thymine,Cytosine])
\end{codex}
This can be used for generating random instances of double helixes
by reusing a random generator for natural numbers.

\section{Testing It All}
We will now describe
a random testing mechanism
to validate our Encoders.

While QuickCheck 
\cite{DBLP:journals/sigplan/ClaessenH02}
provides an elegant general purpose
random tester, it would require
writing a specific adaptor for each isomorphism.
We will describe here a shortcut through
a few higher order combinators.

First, we build a simple random generator for $nat$
\begin{code}
rannat = rand (2^50)

rand :: Nat->Nat->Nat
rand max seed = n where 
  (n,g)=randomR (0,max) (mkStdGen (fromIntegral seed))   
\end{code}

We can now design a generic random test for {\em any}
Encoder as follows:
\begin{code}
rantest :: Encoder t->Bool
rantest t = and (map (rantest1 t) [0..255])

rantest1 t n = x==(visit_as t x) where  x=rannat n

visit_as t = (to nat) . (from t) . (to t) . (from nat) 
\end{code}
Note that in {\tt rantest1}, {\tt visit\_at} starts
with a random natural number from which it generates
its test data of a given type. After testing the encoder,
the result is brought back as a natural number that
should be the same as the original random number.

We can now implement our tester {\tt isotest} that in a few
seconds goes over of thousands of test cases and aggregates the result
with a final {\tt and}:
\begin{code}
isotest = and (map rt [0..25])

rt 0 = rantest nat
rt 1 = rantest fun
rt 2 = rantest set
rt 3 = rantest bits
rt 4 = rantest funbits
rt 5 = rantest hfs
rt 6 = rantest hff
rt 7 = rantest uhfs
rt 8 = rantest uhff
rt 9 = rantest perm
rt 10 = rantest hfp
rt 11 = rantest nat2
rt 12 = rantest set2
rt 13 = rantest clist
rt 14 = rantest pbdd
rt 15 = rantest bdd
rt 16 = rantest rbdd
rt 17 = rantest digraph
rt 18 = rantest graph
rt 19 = rantest mdigraph
rt 20 = rantest mgraph
rt 21 = rantest hypergraph
rt 22 = rantest dyadic
rt 23 = rantest string
rt 24 = rantest pars
rt 25 = rantest dna
\end{code}
The empirical correctness test of the ``whole enchilada'' follows:
\begin{codex}
*ISO> isotest
True
\end{codex}
suggesting that the probability of having errors in
the code described so far is extremely small.

\section{Applications}
Besides their utility as a uniform basis for a general purpose
data conversion library, let us point out
some specific applications of our isomorphisms.

\subsection{Combinatorial Generation}
A free combinatorial generation algorithm (providing
a constructive proof of recursive enumerability)
for a given structure is obtained simply through
an isomorphism from $nat$:
\begin{code}
nth thing = as thing nat
nths thing = map (nth thing)
stream_of thing = nths thing [0..]
\end{code}
\begin{codex}
*ISO> nth set 42
[1,3,5]

*ISO> nth bits 42
[1,1,0,1,0]

*ISO> take 3 (stream_of hfs)
[H [],H [H []],H [H [H []]]]

*ISO> take 3 (stream_of bdd)
[BDD 0 B0,BDD 0 B1,BDD 1 (D 0 B0 B0)]
\end{codex}

\subsection{Random Generation}
Combining {\tt nth} with a random generator for $nat$
provides free algorithms for random generation of
complex objects of customizable size:
\begin{code}
ran thing seed largest = head (random_gen thing seed largest 1)

random_gen thing seed largest n = genericTake n
  (nths thing (rans seed largest))
  
rans seed largest = 
    randomRs (0,largest) (mkStdGen seed)
\end{code}
For instance
\begin{codex}
*ISO> random_gen set 11 999 3
[[0,2,5],[0,5,9],[0,1,5,6]]
\end{codex}
generates a list of 3 random sets.

For instance
\begin{codex}
*ISO>ran digraph 5 (2^31)
[(1,0),(0,1),(2,1),(1,3),(2,2),(3,2),(4,0),(4,1),
 (5,1),(6,0),(6,1),(7,1),(5,3),(6,2),(6,3)]
 
*ISO> ran hfs 7 30
H [H [],H [H [],H [H []]],H [H [H [H []]]]]
*ISO> ran dnaStrand 1 123456789

DNAstrand P5x3 [Guanine,Thymine,Guanine,Cytosine,
  Cytosine,Thymine,Thymine,Thymine,Thymine,
  Adenine,Thymine,Cytosine,Cytosine]
\end{codex}
generate a random digraph, a hereditarily finite set and a DNA strand.

Random generator for various data types are useful for further automating
test generators in tools like QuickCheck 
\cite{DBLP:journals/sigplan/ClaessenH02}
by generating customized random tests.

An interesting other application is generating random problems or programs of a
given type and size.
For instance
\begin{codex}
*ISO> ran sat 8 (2^31)
[[-1],[1,-1],[-1,2],[1,-1,2],[-2],[1,-2],[-1,-2],[1,-1,-2],
[2,-2],[1,2,-2],[-1,2,-2],[3],[1,-1,3],[1,-1,2,3],[1,-2,3],
[-1,-2,3],[2,-2,3],[1,2,-2,3],[-1,2,-2,3]]
 
*ISO> ran clist 8 12345
Cons (Atom 0) (Cons (Cons (Atom 0) (Atom 0)) (Atom 100))
\end{codex}
generate, respectively, a random SAT-problem and a random {\tt Cons}-list.

\subsection{Succinct Representations}
Depending on the information theoretical density of
various data representations as well as on the
constant factors involved in various data structures,
significant data compression can be achieved by
choosing an alternate isomorphic representation,
as shown in the following examples:

\begin{codex}
*ISO> as hff hfs (H [H [H []],H [H [],
        H [H []]],H [H [],H [H [H []]]]])
H [H [H []],H [H []],H [H []]]
*ISO> as nat hff (H [H [H []],H [H []],H [H []]])
42
*ISO> as fun bits [0,1,0,0,0,0,0,0,0,0,0]
[0,10]
*ISO> as rbdd hfs (H [H [],H [H [],H [H []]],
                      H [H [H []],H [H [H []]]]])
BDD 3 (D 1 B1 B0)
*ISO> as hff bdd (BDD 3 (D 2 
        (D 1 (D 0 B1 B0) (D 0 B0 B1)) 
        (D 1 (D 0 B1 B1) (D 0 B1 B1))))
H [H [],H [H [],H [],H []]]

\end{codex}

In particular, mapping to efficient arbitrary length
integer implementations (usually C-based
libraries), can provide more compact
representations or improved performance
for isomorphic higher level 
data representations. Alternatively,
lazy representations as provided by
functional binary numbers or BDDs,
for very large
integers encapsulating results of some
computations might turn out to be more
effective space-wise or time-wise.

We can compare representations sharing
a common datatype to conjecture about their
asymptotic information density.

\subsection{Experimental Mathematics}

\subsubsection{Comparing compactness of representations}

For instance, after defining:
\begin{code}
length_as t = fit genericLength (with nat t)
sum_as t = fit sum (with nat t)
size_as t = fit tsize (with nat t)
\end{code}
one can conjecture that finite functions are more compact than
permutations which are more compact than sets asymptotically
\begin{codex}
*ISO> length_as set 123456789012345678901234567890
54
*ISO> length_as perm 123456789012345678901234567890
28
*ISO> length_as fun 123456789012345678901234567890
54
*ISO> sum_as set 123456789012345678901234567890
2690
*ISO> sum_as perm 123456789012345678901234567890
378
*ISO> sum_as fun 123456789012345678901234567890
43
\end{codex}
One might observe that the same trend applies also
to their hereditarily finite derivatives:
\begin{codex}
*ISO> size_as hfs 123456789012345678901234567890
627
*ISO> size_as hfp 123456789012345678901234567890
276
*ISO> size_as hff 123456789012345678901234567890
91
\end{codex}
While confirming or refuting this conjecture is beyond the
scope of this paper, the affirmative case would imply,
interestingly, that ``order'' (permutations) has
asymptotically higher information density than ``content'' (sets),
and explain why finite functions (that involve both) dominate
data representations in various computing fields. 

Based on the same experiment, reduced BDDs
(especially if one implements sharing, as computed by {\tt robdd\_size})
also provide relatively compact representations:
\begin{codex}
*ISO> bdd_size $ as bdd 
        nat 123456789012345678901234567890
256
*ISO> bdd_size $ as rbdd 
        nat 123456789012345678901234567890
144
*ISO> robdd_size $ as rbdd 
         nat 123456789012345678901234567890
39
\end{codex}
Figures \ref{f8}, \ref{f9}, \ref{f10} compare the sizes of bitstring,
BDD, HFF, HFS, HFP representations, 
first with the most succinct ones  (bitstring, BDDs, HFF) grouped
together in Fig. \ref{f8}, 
then the less succinct ones (HFS and HFP) in Fig. \ref{f9} 
and finally all representations together for
{\tt n} in the larger interval $[0..2^{16}-1]$.

\FIG{f8}{Comparison of curve1=Bit, curve2=BDD and curve3=HFF sizes}
{0.40}{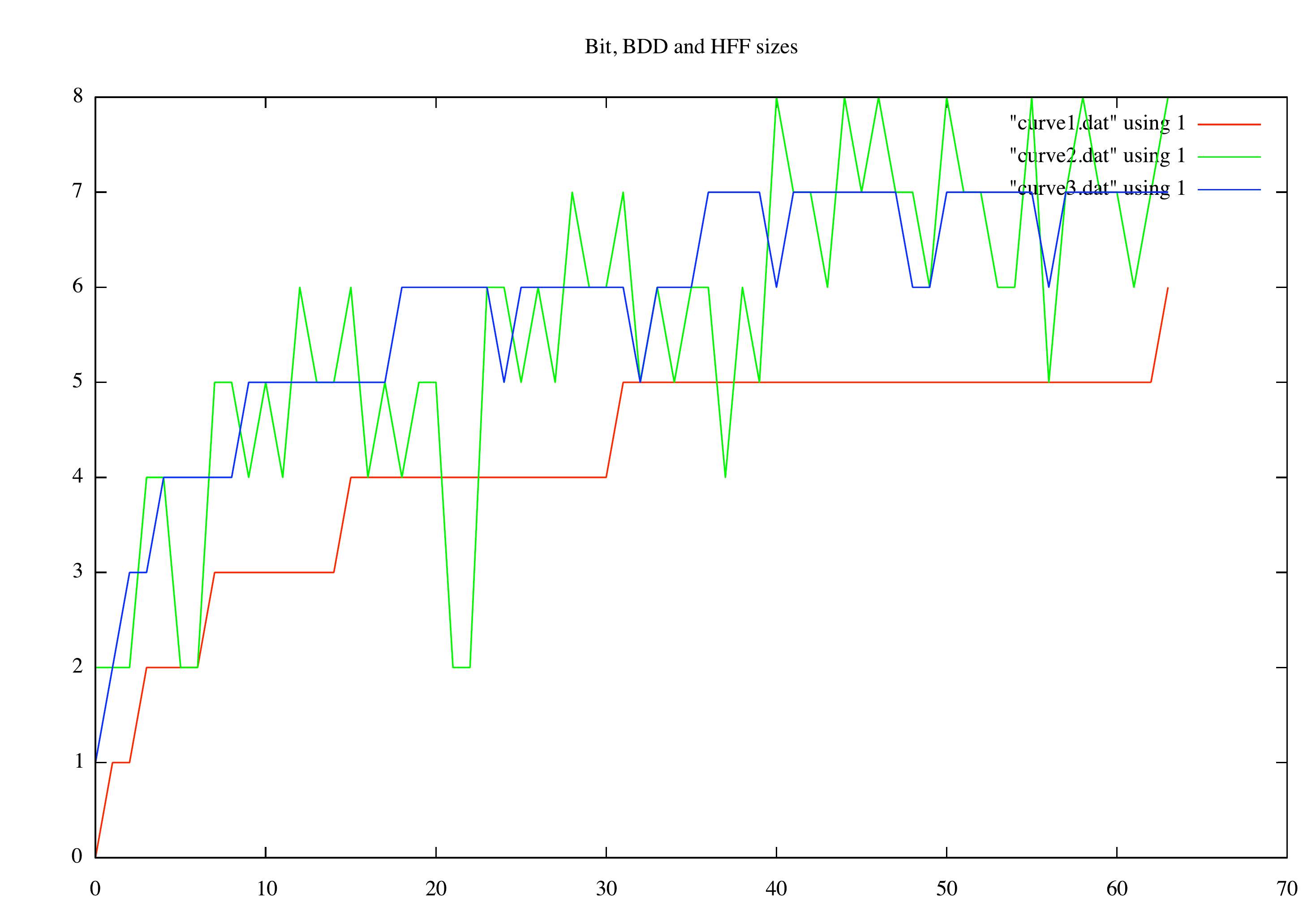}

\FIG{f9}{Comparison of curve1=HFS and curve1=HFP sizes}{0.40}{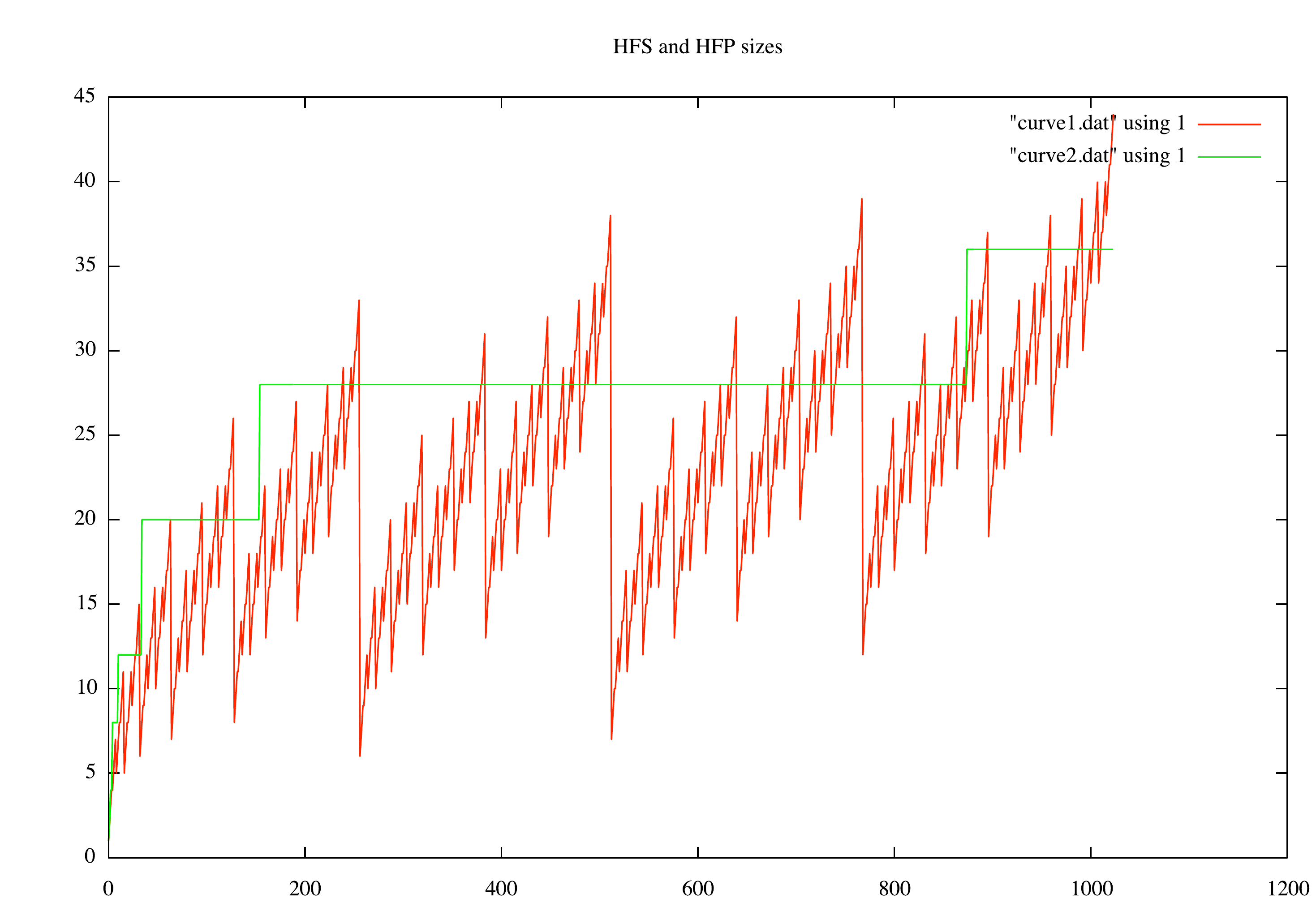}

\FIG{f10}
{Comparison of all representation sizes at a larger scale}
{0.40}{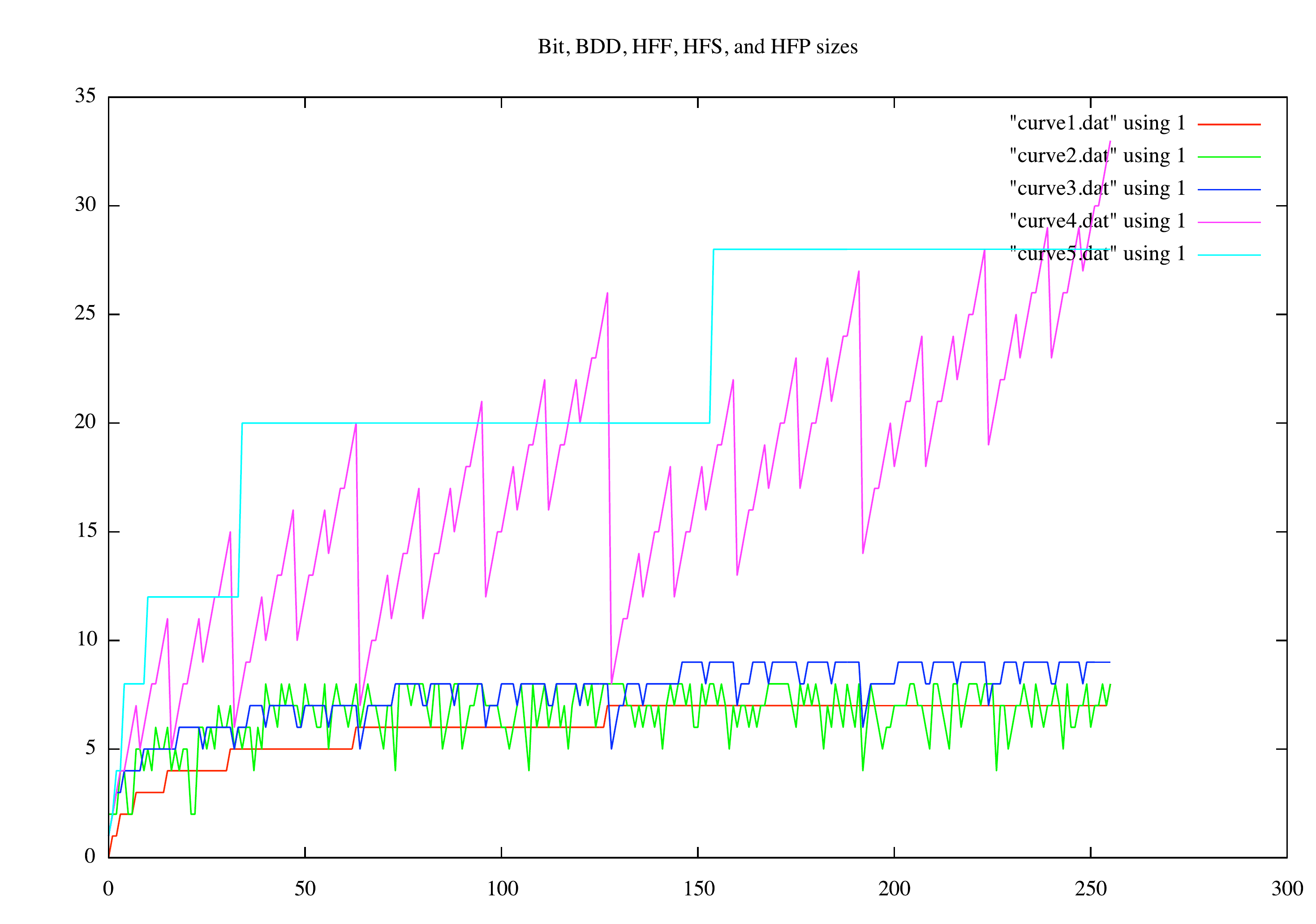}

It is also interesting to observe the ability of some representations to
express huge numbers that normally overflow computer memory 
but which are genuinely ``low complexity'' as a result
of a small numbers of simple computational steps 
that generate them. 

For instance,
\begin{codex}
 *ISO> map (as nat pars) 
     [ "()","(())","((()))","(((())))", "((((()))))","(((((())))))"]
 [0,1,2,4,16,65536]
*ISO> as hff pars "((()))"
H [H [H []]]
\end{codex}
shows that parenthesis sequences (structurally isomorphic to hereditarily finite
functions) can represent succinctly the fast growing but low complexity series
$a_n=2^{2^n}$. Clearly, terms of the series would exhaust computer memory quite
quickly using a conventional bitvector based arbitrary size integer
representation! This suggest the usefulness of a {\em universal}
possibly lazy ``shapeshifting'' algorithm, that can decide on 
the most efficient data representation
automatically, using size estimates, at the time when
data is actually constructed.

\subsubsection{Sparseness criteria}

As a first step, one can introduce a ``sparseness criteria'' by comparing
the size of a representation {\tt f} with the size of the self-delimiting
Elias omega code.

One can obtain an encoding of such sequences by encoding its length and then
encoding each term, parametrized by a function $f:Nat \rightarrow [Nat]$:
\begin{code}
nat2self f n = (to_elias l) ++ concatMap to_elias ns where
  ns = f n
  l=genericLength ns
  
nat2sfun n = nat2self (as fun nat) n   
\end{code}
This function is injective (but not onto!) and its action can be reversed
by first decoding the length $l$ and then extracting self delimited sequences 
$l$ times.
\begin{code}
self2nat g ts = (g xs,ts') where 
  (l,ns) = from_elias ts
  (xs,ts')=take_from_elias l ns

  take_from_elias 0 ns = ([],ns) 
  take_from_elias k ns = ((x:xs),ns'') where
     (x,ns')=from_elias ns
     (xs,ns'')=take_from_elias (k-1) ns'
  
sfun2nat ns = xs where
  (xs,[])=self2nat (as nat fun) ns
\end{code}
We obtain the Encoder:
\begin{code}
sfun :: Encoder [Nat]
sfun = compose (Iso sfun2nat nat2sfun) nat
\end{code}
working as follows:
\begin{codex}
*ISO> as sfun nat 42
[1,0,1,0,0,0,1,0,0,1,0,0,1,0,0]
*ISO> as nat sfun it
42
\end{codex}

A simple concept of sparseness is derived by comparing the size of a
self-delimiting code for a number {\tt n} vs. the size of its
self-delimiting representation as a finite sequence, finite set or  finite
permutation as shown in Fig. \ref{f11a}, computed as follows:
\begin{code}
linear_sparseness_pair t n = 
  (genericLength (to_elias n),genericLength (nat2self (as t nat) n))

linear_sparseness f n = x/y where (x,y)=linear_sparseness_pair f n 
\end{code}

\FIG{f11a}
{Sparseness measures with curve1=fun, curve2=set, curve3=perm up to $2^7$} 
{0.40}{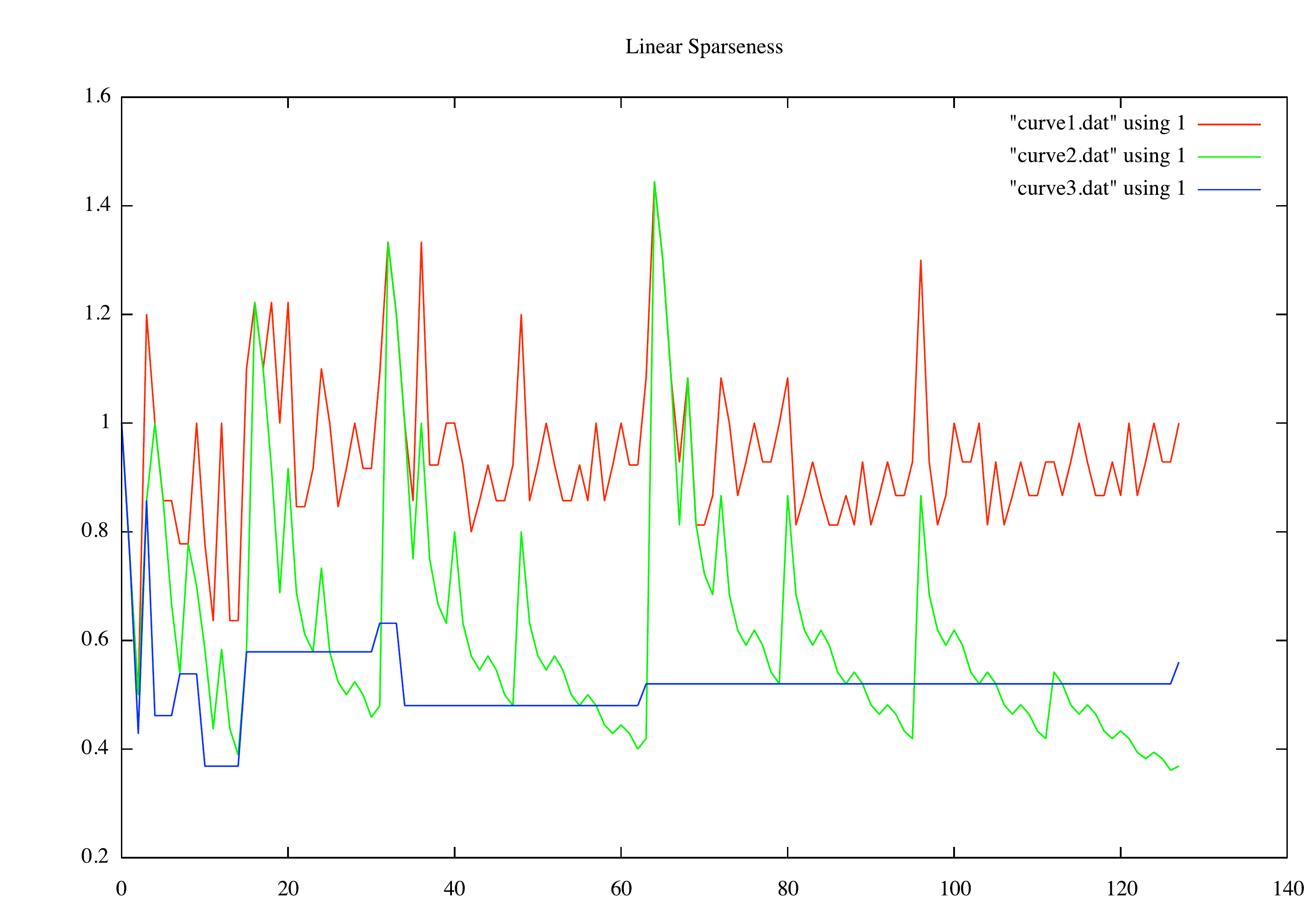}

We can also extend this comparison the hereditarily finite representations,
which, as a pleasant surprise,  turn out to provide self-delimiting codes.
\begin{code}
sparseness_pair f n = 
  (genericLength (to_elias n),genericLength (as f nat n))

sparseness f n = x/y where (x,y)=sparseness_pair f n 
\end{code}

One can then compare (self-delimiting) parenthesis language representations for
hereditarily finite encoders provided by HFF, HFS, HFP and discover the
``peaks'' of sparseness as shown in Fig. \ref{f11} and \ref{f12}.

\FIG{f11}
{Sparseness measures with curve1=HFF, curve2=HFS, curve3=HFP up to $2^8$} 
{0.40}{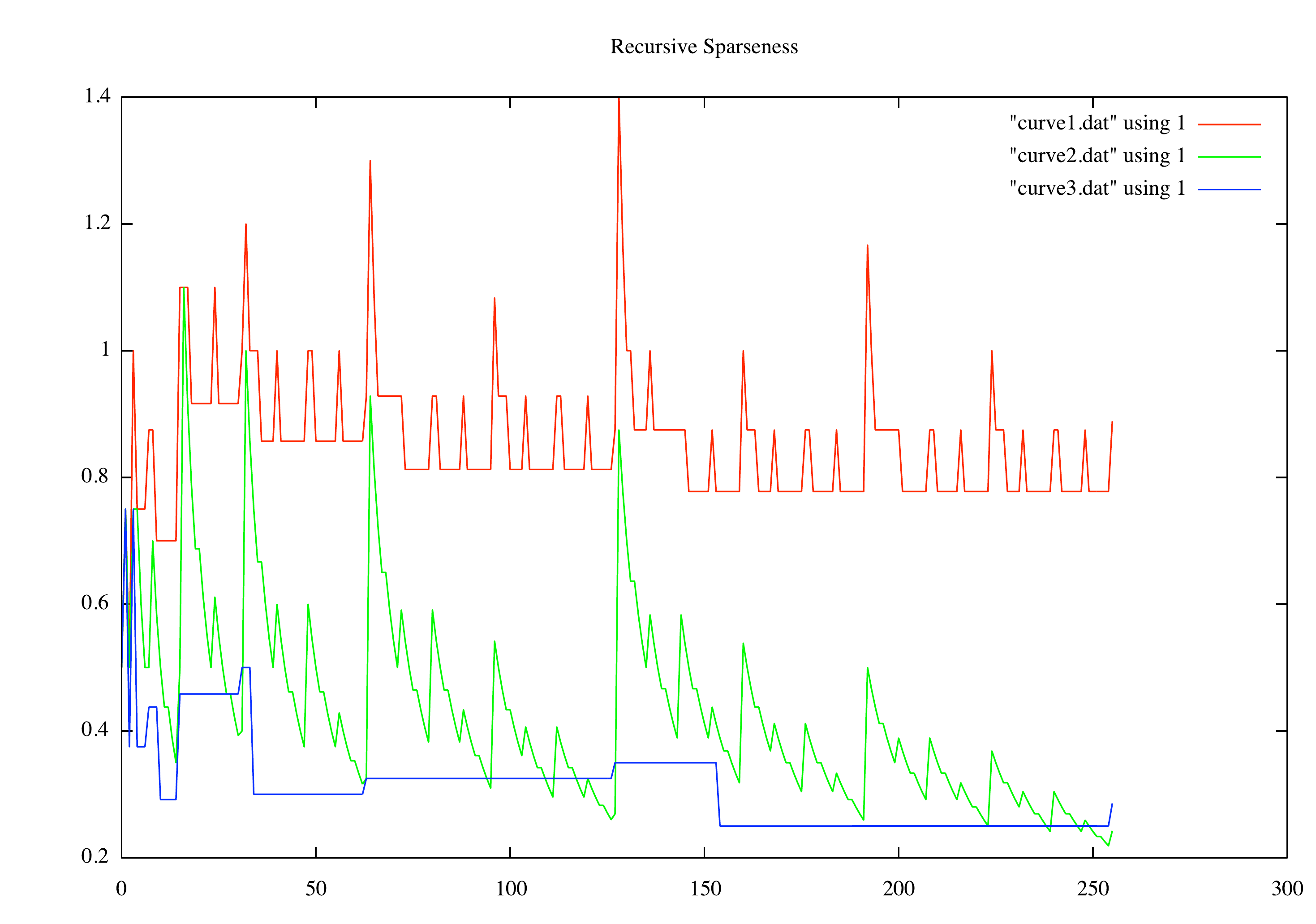}

\FIG{f12}
{Sparseness measures with curve1=HFF curve2=HFS, curve3=HFP up to $2^{14}$} 
{0.40}{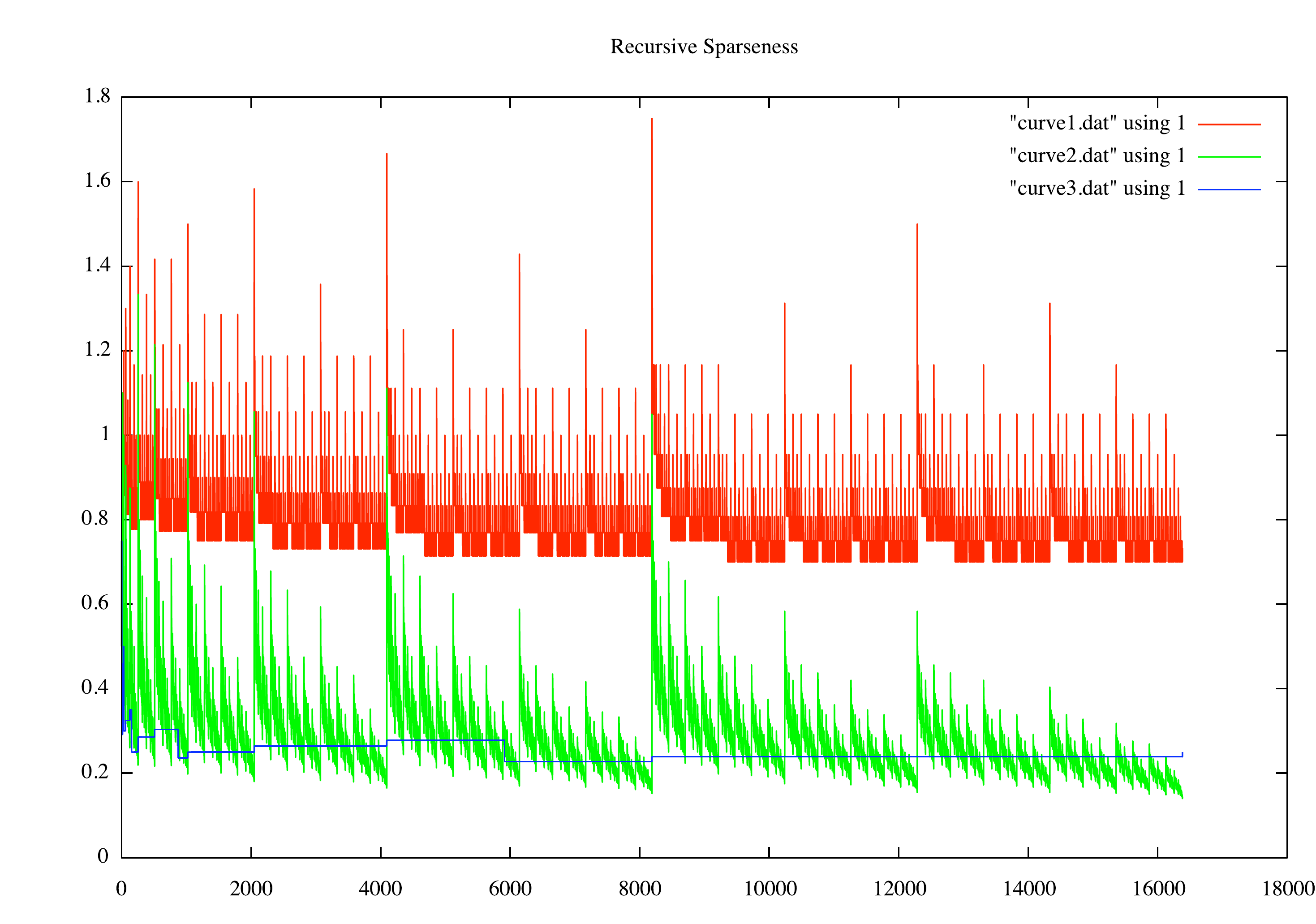}
 
\subsubsection{A new self-delimiting code}

While the HFF representation is generally less compact than Elias omega code,
its simplicity suggest it as a possibly useful self-delimiting code,
especially interesting for streams of ``sparse'' values, as shown in Fig.
\ref{f14}.

\FIG{f14}
{Comparison of codes: curve1=Undelimited curve2=Elias, curve3=HFF up to $2^{7}$} 
{0.40}{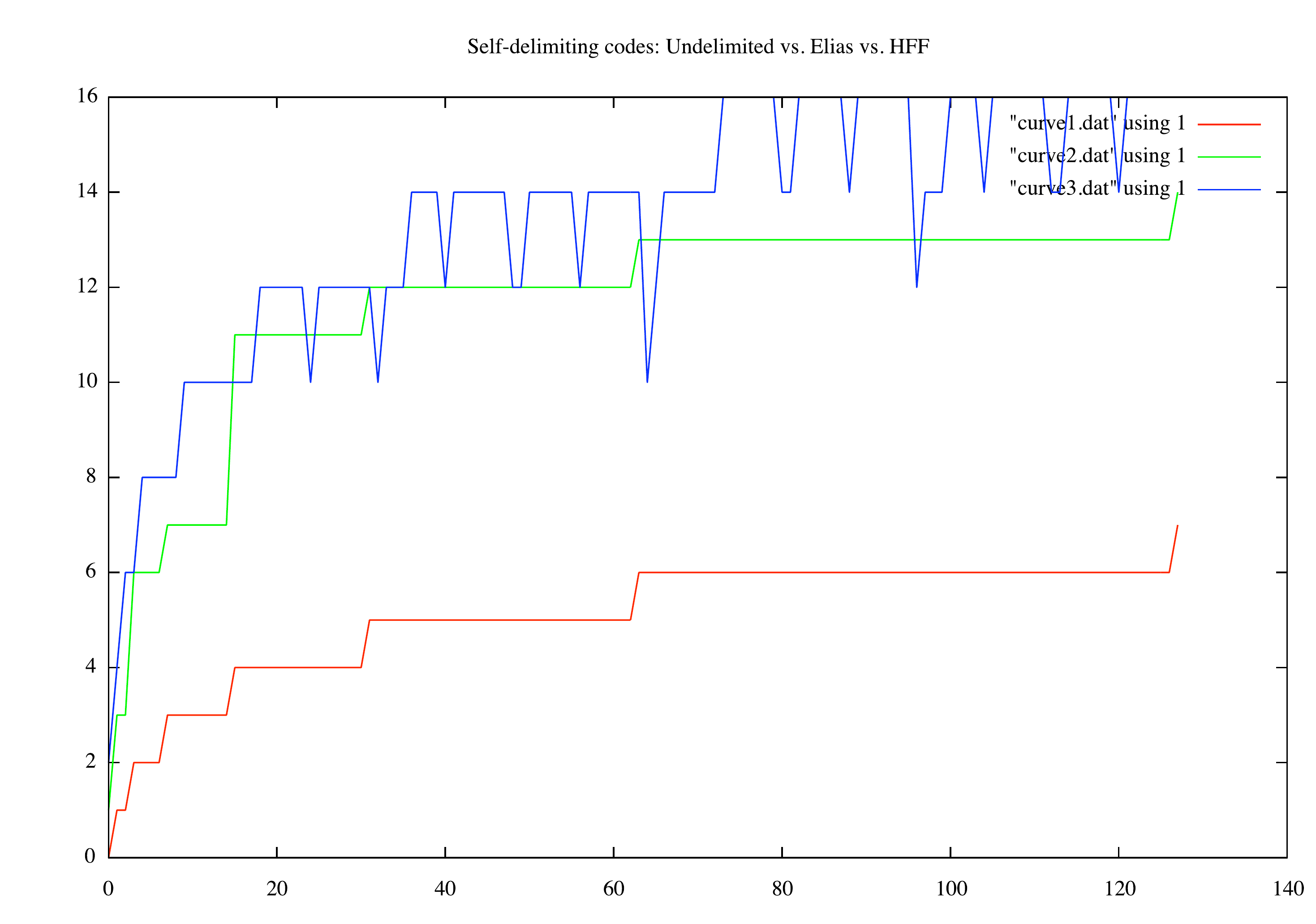}
One can collect values that have smaller HFF codes than Elias omega codes i.e.
``sparse numbers'' with:
\begin{code}
sparses_to m = [n|n<-[0..m-1],
  (genericLength (as hff_pars nat n)) 
  <
  (genericLength (as elias nat n))]
\end{code}
working as follows
\begin{codex}
ISO> sparses_to (2^11)
[15,16,17,24,32,64,65,96,128,129,192,256,257,258,259,320,384,
 385,448,512,513,514,515,516,517,518,519,520,544,576,640,641,704,768,
 769,770,771,832,896,897,960,1024,1025,1026,1027,1028,1029,1030,1031,
 1032,1088,1152,1280,1281,1408,1536,1537,1538,1539,1664,1792,1793,1920]
\end{codex}
and notice that the list collects an unusually large number of various popular
memory chip and computer screen sizes. Figure \ref{f15} shows distribution of
``sparse numbers'' in $[0..2^{18}]$.

\FIG{f15}
{Sparse numbers in $[0..2^{18}]$, x=nth sparse number, y=its value} 
{0.40}{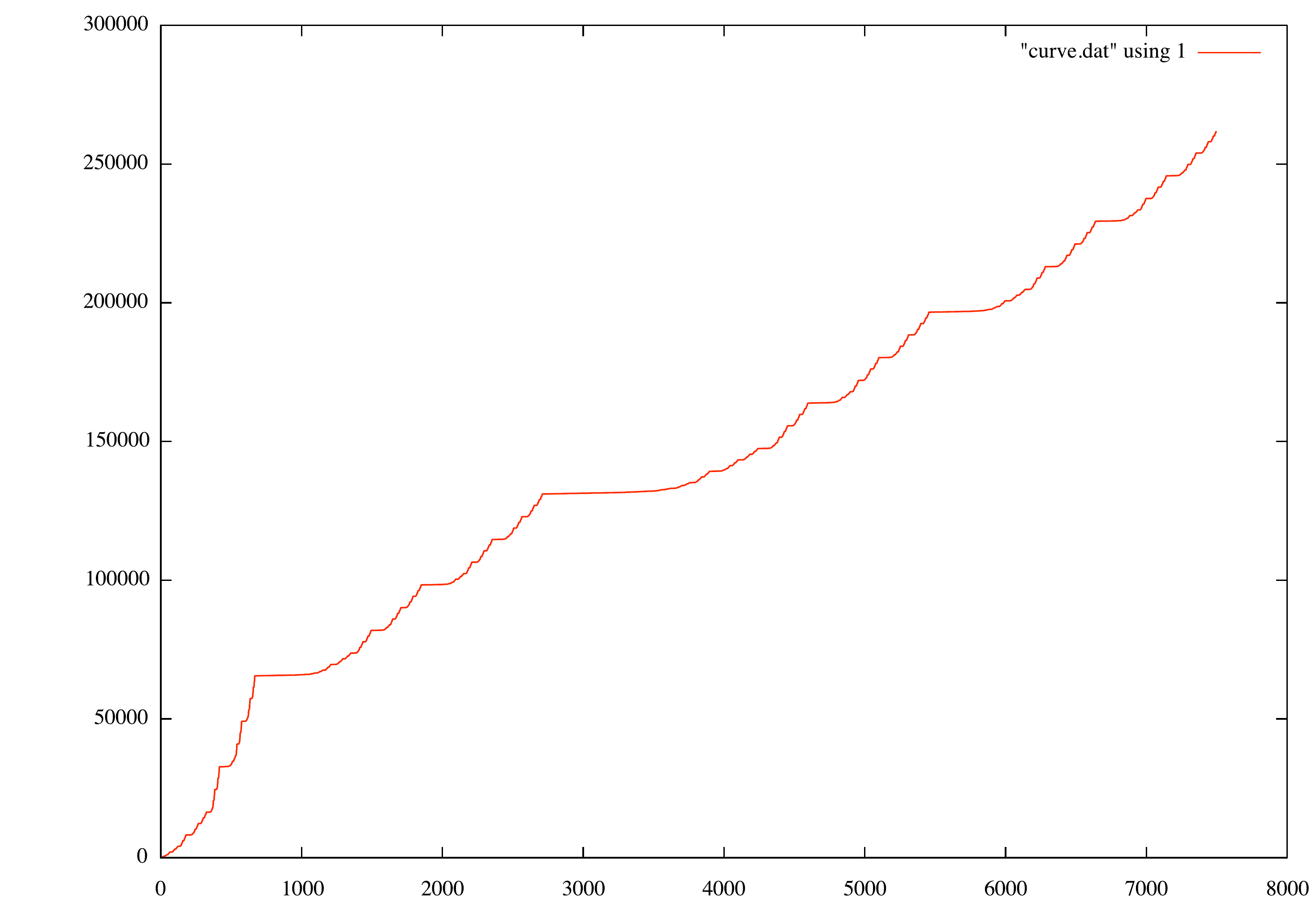}

\subsubsection{Primes and Pairing Functions}
Products of two prime numbers have the interesting property that they are
special a case where no information is lost by multiplication in the sense of
\cite{DBLP:journals/tit/Pippenger05}. Indeed, in this case multiplication is
reversible, i.e. the two factors can be recovered given the product. 
As the product is comparatively easy to compute, while in case of large primes
factoring is believed intractable, this property has well-known uses in
cryptography.
Given the isomorphism between natural numbers and primes mapping a prime to its
position in the sequence of primes, one can transport pairing/unpairing
operations to prime numbers
\begin{code}
ppair pairingf (p1,p2) | is_prime p1 && is_prime p2 = 
  from_pos_in ps (pairingf (to_pos_in ps p1,to_pos_in ps p2)) where 
    ps = primes
 
punpair unpairingf p | is_prime p = (from_pos_in ps n1,from_pos_in ps n2) where 
  ps=primes
  (n1,n2)=unpairingf (to_pos_in ps p)
\end{code}
working as follows:
\begin{codex}
*ISO> ppair bitpair (11,17)
269
*ISO> punpair bitunpair it
(11,17)
\end{codex}
Clearly, this defines a bijection $f : Primes \times Primes \rightarrow Primes$
that is tempting to compare with the product of two primes. 
Figs. \ref{isoppairs} and \ref{isomsetprimes} shows the surfaces
generated by products and multiset pairings of primes. While both commutative
operations are reversible and likely to be asymptotically equivalent in
terms of information density, one can notice the much smoother transition in the case
of lossless multiplication.
\FIG{isoppairs}
{Lossless multiplication of primes}
{0.40}{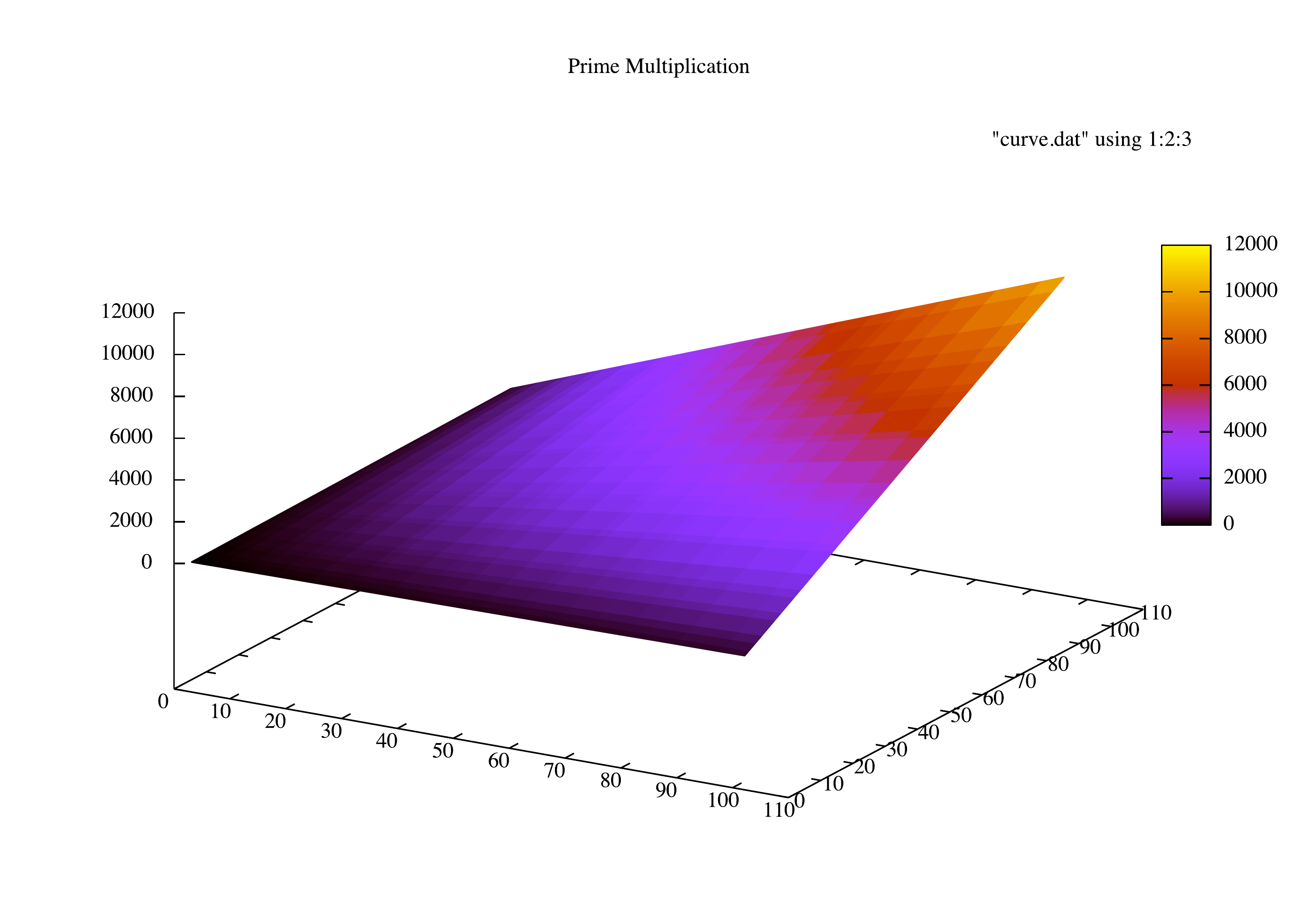}

\FIG{isomsetprimes}
{Lossless multiset pairing of primes}
{0.40}{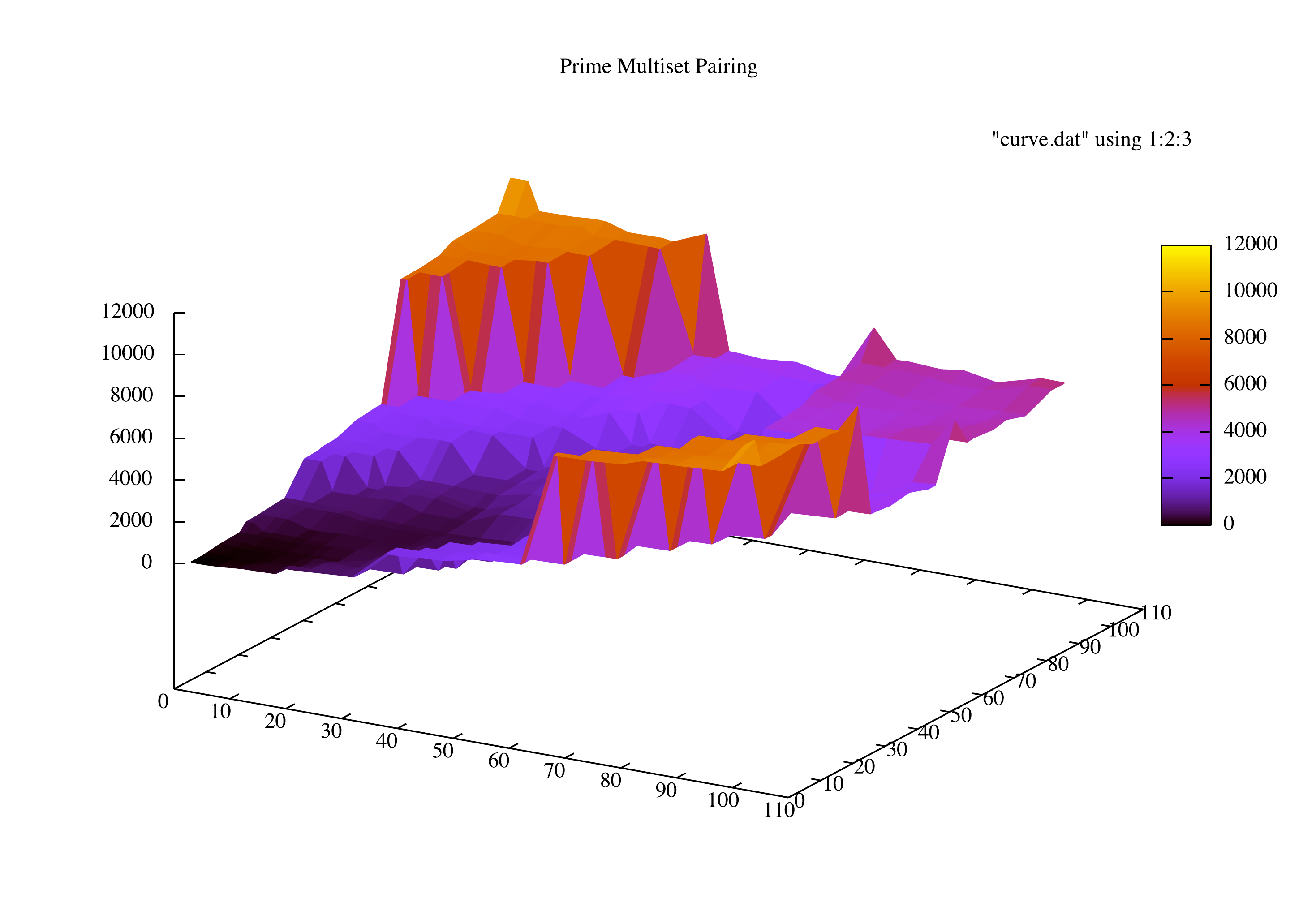}

We have seen that recursive application of the unpairing function {\tt
bitunpair} provided an isomorphisms between natural numbers and BDDs. 
Given an {\em unpairing function} {$u:Nat \rightarrow Nat \times Nat$} and a
predicate {\tt p(n)} over the set of natural numbers, it makes sense to
investigate subsets of $Nat$ such that if {\tt p} holds for {\tt n} then it also
holds after applying the unpairing function {\tt u} to {\tt n}. More
interestingly, one can look at subsets for which this property holds recursively.

Assuming a prime recognizer {\tt is\_prime} and a generator {\tt primes} 
for the stream of prime numbers (see Appendix), we can define:
\begin{code}
hyper_primes u = [n|n<-primes, all_are_primes (uparts u n)] where
  all_are_primes ns = and (map is_prime ns)
  
uparts u = sort . nub . tail . (split_with u) where
    split_with _ 0 = []
    split_with _ 1 = []
    split_with u n = n:(split_with u n0)++(split_with u n1) where
      (n0,n1)=u n  
\end{code}
working as follows:
\begin{codex}
*ISO> take 20 (hyper_primes bitunpair)
[2,3,5,7,11,13,17,19,23,29,31,43,47,59,71,79,83,89,103,139]
*ISO> take 20 (hyper_primes pepis_unpair)
[2,3,5,7,11,13,19,23,29,31,43,53,59,107,127,173,223,251,311,347]
\end{codex}
This leads to the following conjectures, in increasing order of
generality:
\begin{conj}
The sets generated by (hyper\_primes bitpair) and (hyper\_primes pepis\_unpair)
are infinite.
\end{conj}

\begin{conj}
If {\tt u} is a bijection from  $u:Nat \rightarrow Nat \times Nat$ such that:
\begin{enumerate}
\item if $n>1$ and $u~n = (n_0,n_1)$ then $n_0 < n$ and $n_1<n$
\item  {\tt p} is a
predicate on $Nat$ such that $P=\{n:p(n)\}$ is infinite
\end{enumerate}
then the set $P \cap \{n : {uparts}~u~n\}$ is also infinite.
\end{conj}

Figure \ref{isop12} 
shows the complete unpairing graph for two
hyper-primes obtained with {\tt bitunpair}.


\VFIGS{isop12}
{{\tt mset\_unpair} hyper-primes}
{1783}{2109167}{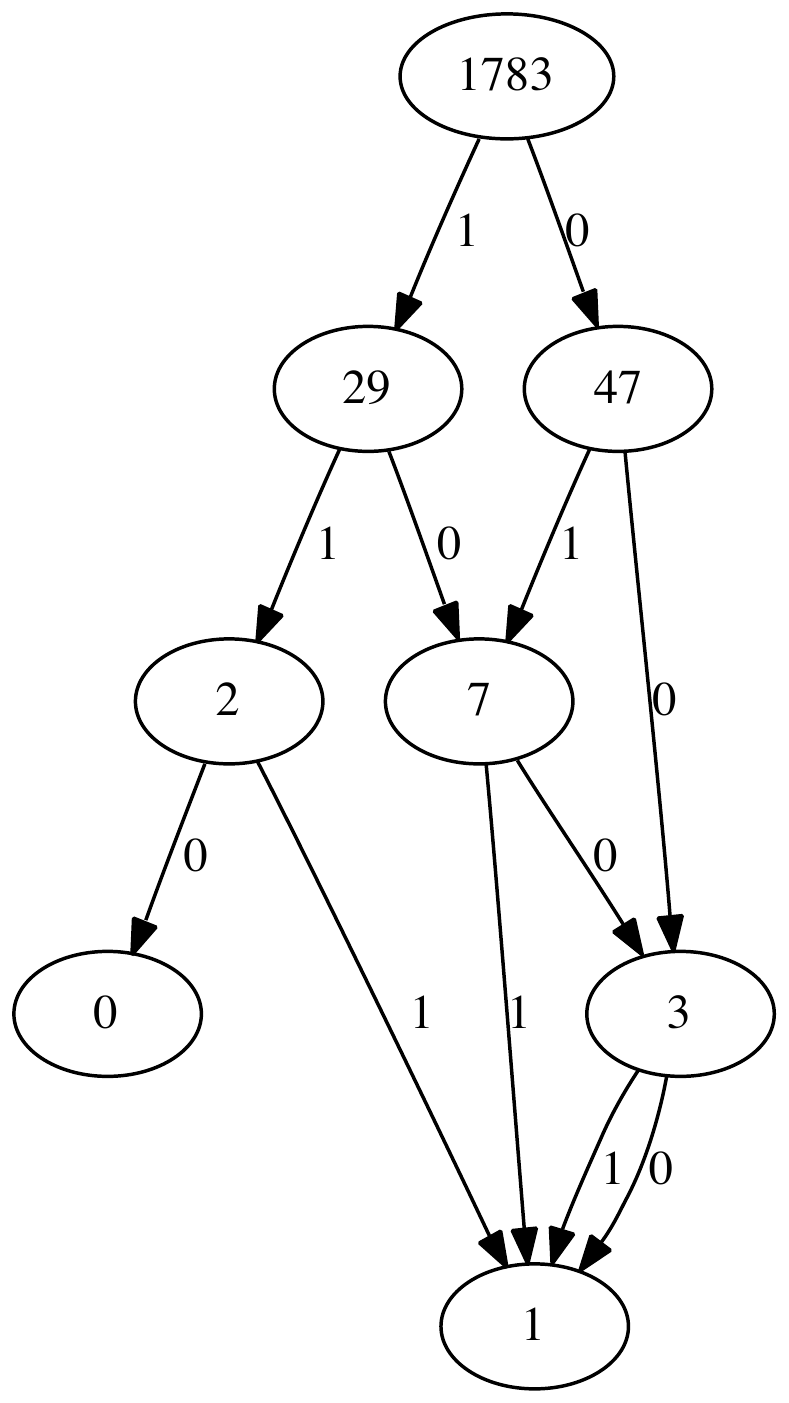}{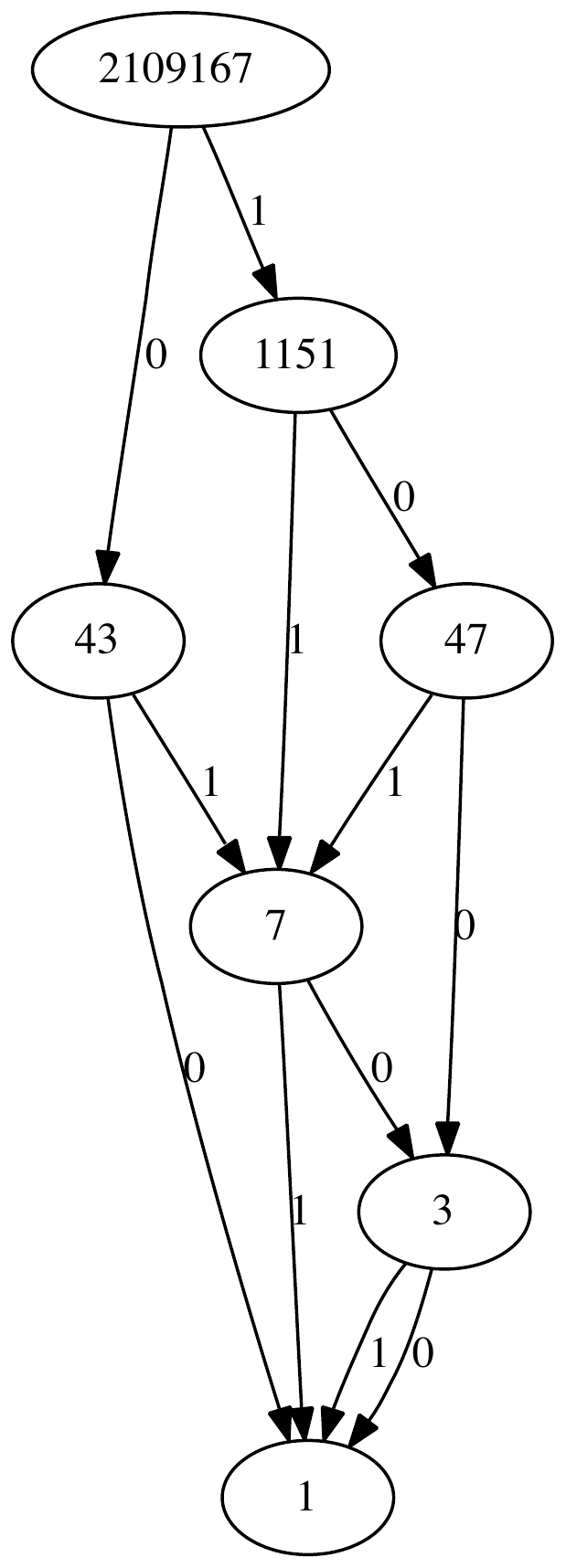}

It is interesting to compare the action of pairing of natural
numbers with their action on functions on primes and hyper-primes with
products. Clearly products are not reversible, except when numbers are primes,
while pairing functions are always reversible. To factor in the fact that
products commute while pairing functions do not, we have considered $2xy$
instead of $xy$.

Figures 
\ref{isoprimes} and \ref{isohypers} show this
comparison.

\FIG{isoprimes}{Pairing of primes vs. 2xy}{0.40}{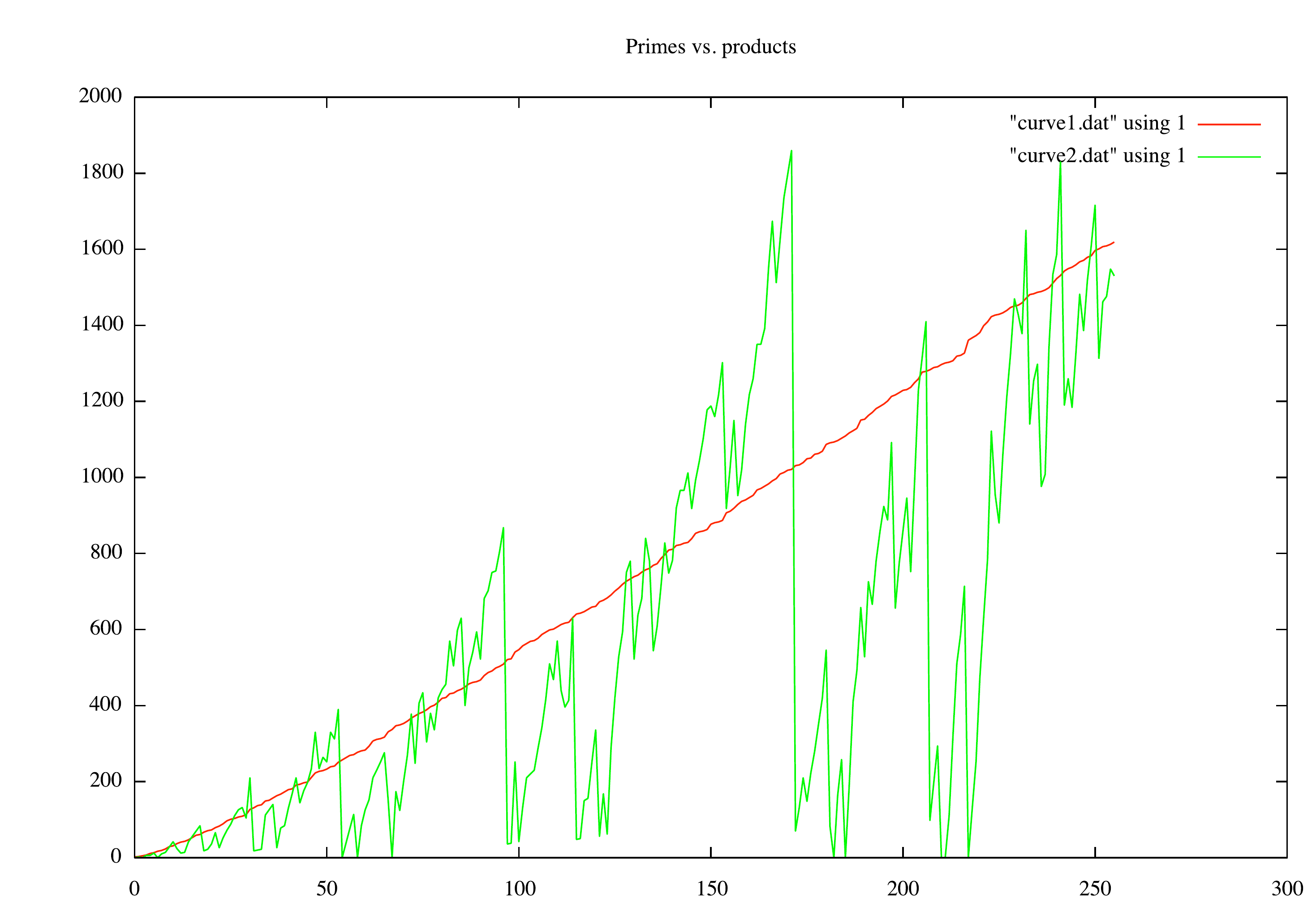}
\FIG{isohypers}{Pairing of hyper-primes vs. 2xy}{0.40}{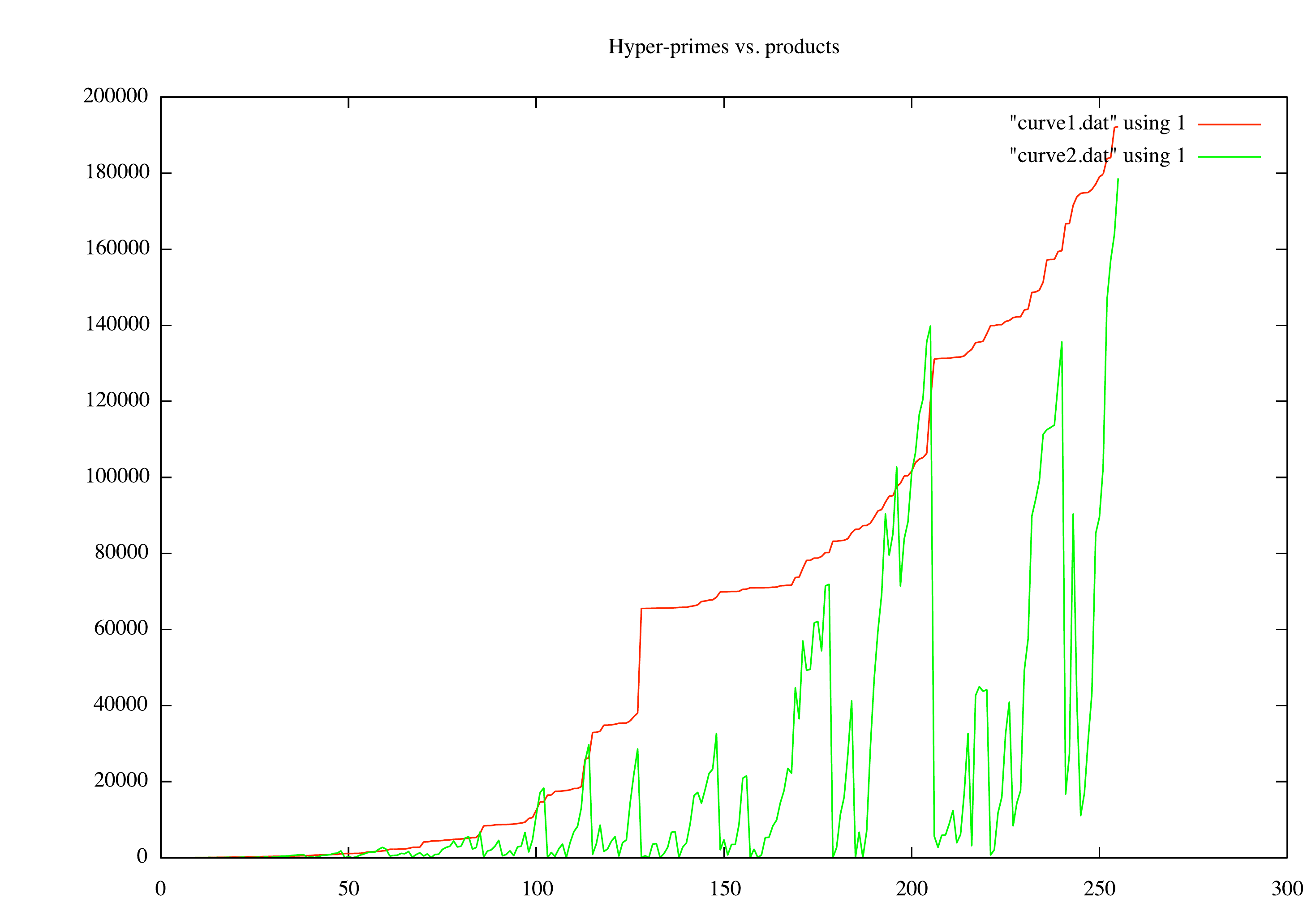}

\clearpage

\subsubsection{Hyper-primes and Fermat primes}
One could expect to model more closely the behavior of primes and products by
focusing on commutative functions like the multiset pairing function
{\tt mset\_pair}:
\begin{codex}
*ISO> take 16 (hyper_primes mset_unpair)
[2,3,5,13,17,113,173,257,10753,17489,34897,34961,43633,43777,65537,142781101]
\end{codex}
We remind that:
\begin{df}
A Fermat-prime is a prime of the form $2^{2^n}+1$ with $n>0$.
\end{df}
Fig. \ref{fermat} shows a hyper-prime that is also a Fermat prime
and a hyper-prime that is not a Fermat prime.

\VFIGS{fermat}
{{\tt mset\_unpair} hyper-primes}
{Fermat prime}{Non-Fermat prime}{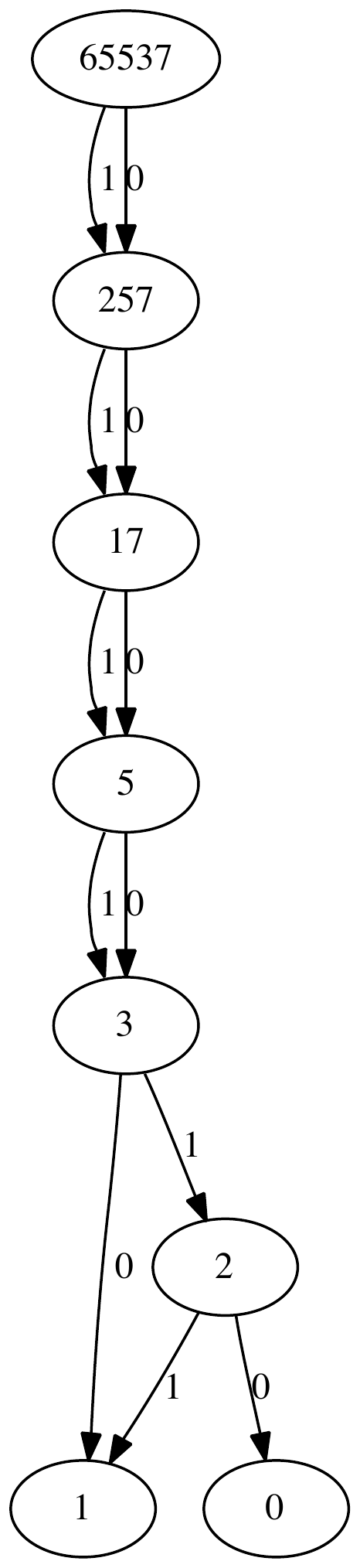}{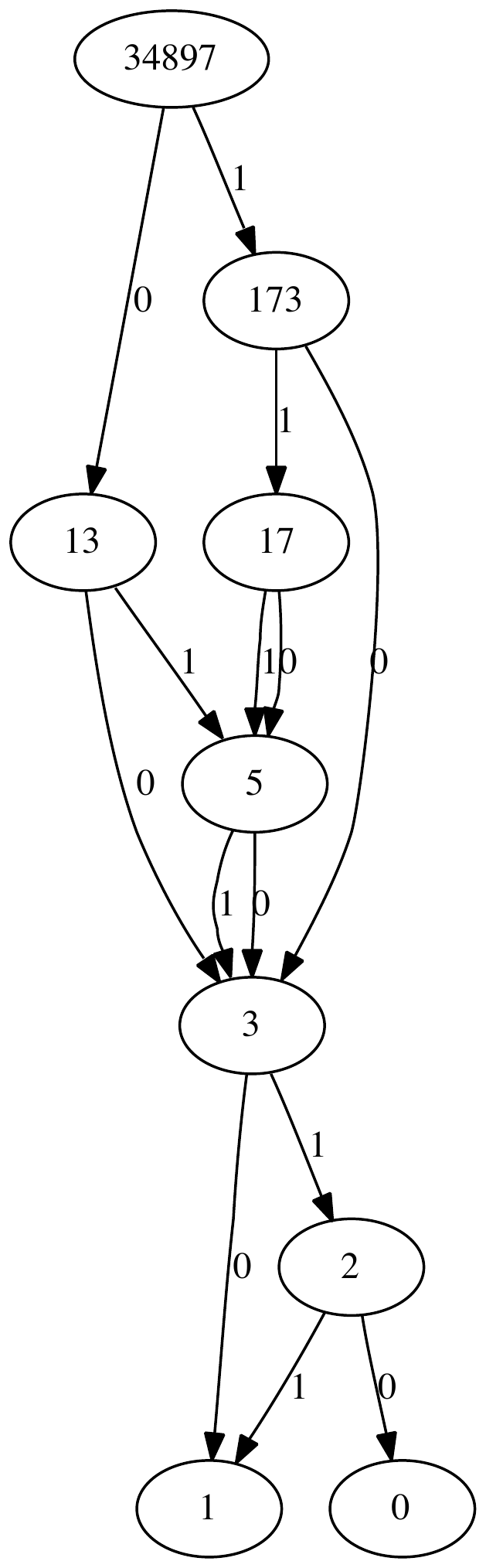}


This time a more interesting conjecture emerges.
We can now state that:
\begin{conj}
All Fermat primes are {\tt mset\_unpair} induced hyper-primes.
\end{conj}
We will just observe that this would follow from
the widely believed conjecture that there the only Fermat primes are
[3,5,17,257,65537] as these 5 primes are indeed on our list of
{\tt mset\_unpair} hyperprimes.

In the event of the alternative, we will now state:
\begin{prop} \label{pfermat}
If there are Fermat primes other than [3,5,17,257,65537] then there are
Fermat primes that are not  {\tt mset\_unpair} hyper-primes.
\end{prop}
To prove Prop. \ref{pfermat} we need a few additional results.
First, the following known fact, implying that we only need to prove that there
are primes of the form $2^{2^n}+1$ that are not hyper-primes.
\begin{lem}
If $n>0$ and $2^n+1$ is prime then $n$ is a power of $2$. 
\end{lem}
It is easy to prove, from the definition of {\tt mset\_pair} that:
\begin{lem}
\begin{equation}
mset\_pair~(2^{2^n}+1,2^{2^n}+1) \equiv 2^{2^{n+1}}+1
\end{equation}
\end{lem}
Indeed, from the identity \ref{mseteq} we obtain
\begin{equation}
mset\_pair (a,a) \equiv bitpair (a,0)
\end{equation}
and then observe that from \ref{biteq} it follows that
\begin{equation}
bitpair (2^{2^n}+1,0) \equiv 2^{2^{n+1}}+1
\end{equation}
We can now prove Prop. \ref{fermat}.
If $2^{2^{n+1}}+1$ is a Fermat prime that is also a
hyper-prime, then $2^{2^{n}}+1$ would be also a Fermat prime that is hyper-prime.
This would form a descending sequence of consecutive Fermat primes - a
contradiction, given that it has been proven (by Leonhard Euler in 1732) that
for instance, $2^{32}+1 = 641 * 6,700,417$ is not a prime.

\subsection{A surprising ``free algorithm'': strange\_sort}
A simple isomorphism like
{\tt nat\_set} can exhibit
interesting properties as a building block of more intricate mappings
like Ackermann's encoding, but let's also note a (surprising to us)
``free algorithm'' -- sorting a list of distinct elements without
explicit use of comparison operations:
\begin{code}
strange_sort = (from nat_set) . (to nat_set)
\end{code}
\begin{codex}
*ISO> strange_sort [2,9,3,1,5,0,7,4,8,6]
[0,1,2,3,4,5,6,7,8,9]
\end{codex}
This algorithm emerges as a consequence of the commutativity of
addition and the unicity of the decomposition of
a natural number as a sum of powers of $2$.
The cognoscenti might notice that
such surprises are not totally unexpected
in the world of functional programming.
In a different context, they
go back as early as 
Wadler's Free Theorems \cite{wadler:theor}.
In a similar way, to sort sequences with repeated elements one can write
\begin{code}
strange_sort' = (to mset) . (from mset)
strange_sort'' = (as mset nat) . (as nat mset)
\end{code}
\begin{codex}
*ISO> strange_sort' [2,4,1,1,0,3,17,1.4]
[0,1,1,1,2,3,4,4,17]
*ISO> strange_sort'' [2,4,1,1,0,3,17,1,4]
[0,1,1,1,2,3,4,4,17]
\end{codex}

\subsection{Circuit Minimization}
Let us consider the classic problem of synthesizing a half adder, composed
of an XOR (\verb~^~) and an AND  (\verb~*~) function. We can combine
the two functions with an if-then-else with selector variable A to
obtain: \verb~ITE(A,B^C,B*C)~ with the following truth table:
\begin{codex}
[0,0,0]:0
[0,0,1]:0
[0,1,0]:0
[0,1,1]:1
[1,0,0]:0
[1,0,1]:1
[1,1,0]:1
[1,1,1]:0
\end{codex}
Note that this {\tt 3} argument single output function (encoded as
the natural number {\tt 22} by reading its value column in binary), fuses the
two operations with the upper half of the truth table representing 
the {\tt AND} and the lower half representing the {\tt XOR}.
When running {\tt to\_min\_bdd} on this function we obtain:
\begin{codex}
ISO> from_base 2 [0,1,1,0, 1,0,0,0]
22
*ISO> to_min_bdd 3 22
BDD 3 (D 0 
  (D 1 (D 2 B0 B1) (D 2 B1 B0)) 
              (D 1 (D 2 B1 B0) B0))
\end{codex}

\subsection{Other Applications}
A fairly large number of useful algorithms in fields 
ranging from data compression, coding theory and cryptography 
to compilers, circuit design and computational complexity
involve bijective functions between heterogeneous data
types. Their systematic encapsulation in a generic API
that coexists well with strong typing can bring 
significant simplifications to various software modules
with the added benefits of reliability and easier 
maintenance.
In a Genetic Programming context \cite{koza92} the use 
of isomorphisms between bitvectors/natural numbers 
on one side, and trees/graphs representing HFSs, HFFs on the other side, 
looks like a promising phenotype-genotype connection.
Mutations and crossovers in a data type close to the problem
domain are transparently mapped to numerical domains
where evaluation functions can be computed easily.
In particular, ``biological proven'' encodings like 
DNA strands are likely to provide interesting 
genotypes implementations. 
In the context of Software Transaction Memory implementations
(like Haskell's STM \cite{DBLP:journals/cacm/HarrisMJH08}),
encodings through isomorphisms are subject to efficient
shortcuts, as undo operations in case of transaction failure
can be performed by applying inverse transformations without
the need to save the intermediate chain of
data structures involved.

\section{Related work} \label{related}
The closest reference on encapsulating bijections
as a Haskell data type is \cite{bijarrows} 
and Conal Elliott's composable
bijections module \cite{bijeliot},
where, in a more complex setting,
Arrows \cite{hughes:arrows} are used 
as the underlying abstractions.
While our {\tt Iso} data type is similar
to the {\em Bij} data type in \cite{bijeliot} and
BiArrow concept of \cite{bijarrows},
the techniques for using such isomorphisms
as building blocks of an embedded composition
language centered around encodings
as Natural Numbers are new.

As the domains between which we define our
isomorphisms can be organized as categories,
it is likely that some of our constructs would benefit
from {\em natural transformation} \cite{matcat} and {\em n-category}
formulations \cite{Baez97anintroduction}. 

{\em Ranking} functions can be traced back to G\"{o}del numberings
\cite{Goedel:31,conf/icalp/HartmanisB74} associated to formulae. 
Together with their inverse {\em unranking} functions they are also 
used in combinatorial generation
algorithms
\cite{conf/mfcs/MartinezM03,knuth06draft,Ruskey90generatingbinary,Myrvold01rankingand}.
However the generic view of such transformations as hylomorphisms obtained compositionally
from simpler isomorphisms, as described in this paper,
is new.

Natural Number encodings of Hereditarily Finite Sets have 
triggered the interest of researchers in fields ranging from 
Axiomatic Set Theory and Foundations of Logic to 
Complexity Theory and Combinatorics
\cite{finitemath,kaye07,abian78,avigad97,DBLP:journals/mlq/Kirby07,DBLP:conf/foiks/LeontjevS00}.
Computational and Data Representation aspects of Finite Set Theory 
have been described in logic programming and theorem proving contexts 
in \cite{DBLP:journals/tplp/PiazzaP04,DBLP:conf/types/Paulson94}. 

Pairing functions have been used in work on decision problems as early as
\cite{robinson50,robinsons68b}. A
typical use in the foundations of mathematics is
\cite{DBLP:journals/tcs/CegielskiR01}.
An extensive study of various pairing functions and their 
computational properties is presented in 
\cite{DBLP:conf/ipps/Rosenberg02a}.

Various mappings from natural numbers to rational numbers are described 
in \cite{rationals}, also in a functional programming framework.

We have learned from Knuth's recent work on combinatorial
algorithms \cite{knuth06draft} the techniques related to
bitvector encodings of projection functions and boolean operations
and about BDDs and reduced ordered BDDs from Bryant's
seminal paper on the topic \cite{bryant86graphbased}.
However, the connection with pairing/unpairing functions
and the equivalence results of subsection \ref{equiv} are new.

The concepts of hereditarily finite functions and
permutations as well as their encodings, are
likely to be new, given that our sustained 
search efforts have not lead so far to anything
similar. 

Some other techniques, ranging from
factoradics to cons-lists and
functional binary numbers
to DNA encodings and dyadic rationals are
for sure part of the scientific commons. 
In that
case our focus was to express them as
elegantly ans possible in a uniform framework.
In these cases as well, most of the time
it was faster to ``just do it'', by implementing
them from scratch in a functional programming 
framework, rather than adapting procedural 
algorithms found elsewhere.

\section{Conclusion} \label{concl}
We have shown the expressiveness of Haskell as a
metalanguage for executable mathematics, by describing
encodings
for functions and finite sets
in a uniform framework
as data type isomorphisms with a groupoid structure.
Haskell's higher order functions and recursion patterns
have helped the design of an embedded data transformation
language.
Using higher order combinators a 
simplified QuickCheck style random testing
mechanism has been implemented as
an empirical correctness test.
The framework has been extended
with hylomorphisms providing
generic mechanisms for encoding
Hereditarily Finite Sets and 
Hereditarily Finite Functions.
In the process, a few surprising
``free algorithms'' have emerged
as well as a generalization of
Ackermann's encoding to Hereditarily Finite 
Sets with Urelements. We plan to explore in
depth in the near future, some of the results
that are likely to be of interest in fields
ranging from combinatorics and boolean logic
to data compression and arbitrary precision
numerical computations.

\bibliographystyle{INCLUDES/splncs}
\bibliography{INCLUDES/theory,tarau,INCLUDES/proglang,INCLUDES/biblio,INCLUDES/syn}

\begin{thebibliography}{10}

\bibitem{sac09fISO}
Tarau, P.:
\newblock {Isomorphisms, Hylomorphisms and Hereditarily Finite Data Types in
  Haskell}.
\newblock In: {Proceedings of ACM SAC'09}, Honolulu, Hawaii, ACM (March 2009)

\bibitem{lakoff}
Lakoff, G., Johnson, M.:
\newblock Metaphors We Live By.
\newblock University of Chicago Press, Chicago, IL, USA (1980)

\bibitem{Cook04theoriesfor}
Cook, S.:
\newblock Theories for complexity classes and their propositional translations.
\newblock In: Complexity of computations and proofs. (2004)  1--36

\bibitem{Cook93functionalinterpretations}
Cook, S., Urquhart, A.:
\newblock Functional interpretations of feasibly constructive arithmetic.
\newblock Annals of Pure and Applied Logic \textbf{63} (1993)  103--200

\bibitem{Goedel:31}
G\"{o}del, K.:
\newblock \"{U}ber formal unentscheidbare {S\"{a}tze der Principia Mathematica
  und verwandter Systeme I}.
\newblock Monatshefte f\"{u}r Mathematik und Physik \textbf{38} (1931)
  173--198

\bibitem{conf/icalp/HartmanisB74}
Hartmanis, J., Baker, T.P.:
\newblock On simple goedel numberings and translations.
\newblock In Loeckx, J., ed.: ICALP. Volume~14 of Lecture Notes in Computer
  Science., Springer (1974)  301--316

\bibitem{DBLP:journals/sigplan/ClaessenH02}
Claessen, K., Hughes, J.:
\newblock Testing monadic code with quickcheck.
\newblock SIGPLAN Notices \textbf{37}(12) (2002)  47--59

\bibitem{multisetOver}
Singh, D., Ibrahim, A.M., Yohanna, T., Singh, J.N.:
\newblock An overview of the applications of multisets.
\newblock Novi Sad J. Math \textbf{52}(2) (2007)  73--92

\bibitem{DBLP:journals/jfp/Hutton99}
Hutton, G.:
\newblock {A Tutorial on the Universality and Expressiveness of Fold}.
\newblock J. Funct. Program. \textbf{9}(4) (1999)  355--372

\bibitem{DBLP:conf/fpca/MeijerH95}
Meijer, E., Hutton, G.:
\newblock {Bananas in Space: Extending Fold and Unfold to Exponential Types}.
\newblock In: FPCA. (1995)  324--333

\bibitem{ackencoding}
Ackermann, W.F.:
\newblock {Die Widerspruchsfreiheit der allgemeinen Mengenlhere}.
\newblock Mathematische Annalen (114) (1937)  305--315

\bibitem{DBLP:journals/tplp/PiazzaP04}
Piazza, C., Policriti, A.:
\newblock {Ackermann Encoding, Bisimulations, and OBDDs}.
\newblock TPLP \textbf{4}(5-6) (2004)  695--718

\bibitem{DBLP:journals/jct/Gobel80}
G{\"o}bel, F.:
\newblock On a 1-1-correspondence between rooted trees and natural numbers.
\newblock J. Comb. Theory, Ser. B \textbf{29}(1) (1980)  141--143

\bibitem{knuth_art_1997-1}
Knuth, D.E.:
\newblock The art of computer programming, volume 2 (3rd ed.): seminumerical
  algorithms.
\newblock Addison-Wesley Longman Publishing Co., Inc., Boston, MA, USA (1997)

\bibitem{DBLP:journals/dmtcs/MantaciR01}
Mantaci, R., Rakotondrajao, F.:
\newblock A permutations representation that knows what "eulerian" means.
\newblock Discrete Mathematics {\&} Theoretical Computer Science \textbf{4}(2)
  (2001)  101--108

\bibitem{pepis}
Pepis, J.:
\newblock Ein verfahren der mathematischen logik.
\newblock The Journal of Symbolic Logic \textbf{3}(2) (jun 1938)  61--76

\bibitem{kalmar1}
Kalmar, L.:
\newblock On the reduction of the decision problem. first paper. ackermann
  prefix, a single binary predicate.
\newblock The Journal of Symbolic Logic \textbf{4}(1) (mar 1939)  1--9

\bibitem{kalmar2}
{Kalmar, Laszlo}, {Suranyi, Janos}:
\newblock On the reduction of the decision problem.
\newblock The Journal of Symbolic Logic \textbf{12}(3) (sep 1947)  65--73

\bibitem{kalmar3}
{Kalmar, Laszlo}, {Suranyi, Janos}:
\newblock On the reduction of the decision problem: Third paper. pepis prefix,
  a single binary predicate.
\newblock The Journal of Symbolic Logic \textbf{15}(3) (sep 1950)  161--173

\bibitem{robinson50}
Robinson, J.:
\newblock General recursive functions.
\newblock Proceedings of the American Mathematical Society \textbf{1}(6) (dec
  1950)  703--718

\bibitem{robinson55}
Robinson, J.:
\newblock A note on primitive recursive functions.
\newblock Proceedings of the American Mathematical Society \textbf{6}(4) (aug
  1955)  667--670

\bibitem{robinson68a}
Robinson, J.:
\newblock Recursive functions of one variable.
\newblock Proceedings of the American Mathematical Society \textbf{19}(4) (aug
  1968)  815--820

\bibitem{robinsons68b}
Robinson, J.:
\newblock Finite generation of recursively enumerable sets.
\newblock Proceedings of the American Mathematical Society \textbf{19}(6) (dec
  1968)  1480--1486

\bibitem{robinson67}
Robinson, J.:
\newblock An introduction to hyperarithmetical functions.
\newblock The Journal of Symbolic Logic \textbf{32}(3) (sep 1967)  325--342

\bibitem{pigeon}
Pigeon, S.:
\newblock Contributions \`{a} la compression de donn\'{e}es.
\newblock Ph.d. thesis, Universit\'{e} de Montr\'{e}al, Montr\'{e}al (2001)

\bibitem{bryant86graphbased}
Bryant, R.E.:
\newblock Graph-based algorithms for boolean function manipulation.
\newblock {IEEE} Transactions on Computers \textbf{35}(8) (1986)  677--691

\bibitem{knuth06draft}
Knuth, D.:
\newblock {The Art of Computer Programming, Volume 4, draft} (2006)
  http://www-cs-faculty.stanford.edu/$\sim$knuth/taocp.html.

\bibitem{shannon_all}
Shannon, C.E.:
\newblock {Claude Elwood Shannon: collected papers}.
\newblock IEEE Press, Piscataway, NJ, USA (1993)

\bibitem{DBLP:journals/fmsd/FujitaMY97}
Fujita, M., McGeer, P.C., Yang, J.C.Y.:
\newblock Multi-terminal binary decision diagrams: An efficient data structure
  for matrix representation.
\newblock Formal Methods in System Design \textbf{10}(2/3) (1997)  149--169

\bibitem{CBGP08}
Ciesinski, F., Baier, C., Groesser, M., Parker, D.:
\newblock Generating compact {MTBDD}-representations from {Probmela}
  specifications.
\newblock In: Proc. 15th International SPIN Workshop on Model Checking of
  Software (SPIN'08). (2008)

\bibitem{bucciarelli}
Bucciarelli, A., Salibra, A.:
\newblock The sensible graph theories of lambda calculus.
\newblock In: LICS '04: Proceedings of the 19th Annual IEEE Symposium on Logic
  in Computer Science, Washington, DC, USA, IEEE Computer Society (2004)
  276--285

\bibitem{berline}
Berline, C.:
\newblock Graph models of $\lambda$-calculus at work, and variations.
\newblock Mathematical. Structures in Comp. Sci. \textbf{16}(2) (2006)
  185--221

\bibitem{algDNA}
Li, Z.:
\newblock Algebraic properties of dna operations.
\newblock Biosystems \textbf{52} (October 1999)  55--61(7)

\bibitem{haskellDNA}
Hinze, T., Sturm, M.:
\newblock A universal functional approach to dna computing and its experimental
  practicability.
\newblock In: Proceedings 6th DIMACS Workshop on DNA Based Computers, held at
  the University of Leiden, Leiden, The Netherlands, 13 - 17. (2000)  257--266

\bibitem{DBLP:journals/tit/Pippenger05}
Pippenger, N.:
\newblock The average amount of information lost in multiplication.
\newblock IEEE Transactions on Information Theory \textbf{51}(2) (2005)
  684--687

\bibitem{wadler:theor}
Wadler, P.:
\newblock Theorems for free!
\newblock In: FPCA '89: Proceedings of the fourth international conference on
  Functional programming languages and computer architecture, New York, NY,
  USA, ACM (1989)  347--359

\bibitem{koza92}
Koza, J.R.:
\newblock Genetic Programming: On the Programming of Computers by Means of
  Natural Selection.
\newblock MIT Press, Cambridge, MA, USA (1992)

\bibitem{DBLP:journals/cacm/HarrisMJH08}
Harris, T., Marlow, S., Jones, S.L.P., Herlihy, M.:
\newblock Composable memory transactions.
\newblock Commun. ACM \textbf{51}(8) (2008)  91--100

\bibitem{bijarrows}
Alimarine, A., Smetsers, S., van Weelden, A., van Eekelen, M., Plasmeijer, R.:
\newblock There and back again: arrows for invertible programming.
\newblock In: Haskell '05: Proceedings of the 2005 ACM SIGPLAN workshop on
  Haskell, New York, NY, USA, ACM Press (2005)  86--97

\bibitem{bijeliot}
{Conal Elliott}:
\newblock {Data.Bijections Haskell Module}.
\newblock http://haskell.org/haskellwiki/TypeCompose.

\bibitem{hughes:arrows}
Hughes, J.:
\newblock {Generalizing Monads to Arrows} Science of Computer Programming 37,
  pp. 67-111, May 2000.

\bibitem{matcat}
Mac~Lane, S.:
\newblock Categories for the Working Mathematician.
\newblock Springer-Verlag, New York, NY, USA (1998)

\bibitem{Baez97anintroduction}
Baez, J.C.:
\newblock An introduction to n-categories.
\newblock In: In 7th Conference on Category Theory and Computer Science,
  Springer-Verlag (1997)  1--33

\bibitem{conf/mfcs/MartinezM03}
Martinez, C., Molinero, X.:
\newblock Generic algorithms for the generation of combinatorial objects.
\newblock In Rovan, B., Vojtas, P., eds.: MFCS. Volume 2747 of Lecture Notes in
  Computer Science., Springer (2003)  572--581

\bibitem{Ruskey90generatingbinary}
Ruskey, F., Proskurowski, A.:
\newblock Generating binary trees by transpositions.
\newblock J. Algorithms \textbf{11} (1990)  68--84

\bibitem{Myrvold01rankingand}
Myrvold, W., Ruskey, F.:
\newblock Ranking and unranking permutations in linear time.
\newblock Information Processing Letters \textbf{79} (2001)  281--284

\bibitem{finitemath}
Takahashi, M.o.:
\newblock {A Foundation of Finite Mathematics}.
\newblock Publ. Res. Inst. Math. Sci. \textbf{12}(3) (1976)  577--708

\bibitem{kaye07}
Kaye, R., Wong, T.L.:
\newblock {On Interpretations of Arithmetic and Set Theory}.
\newblock Notre Dame J. Formal Logic Volume \textbf{48}(4) (2007)  497--510

\bibitem{abian78}
Abian, A., Lamacchia, S.:
\newblock On the consistency and independence of some set-theoretical
  constructs.
\newblock Notre Dame Journal of Formal Logic \textbf{X1X}(1) (1978)  155--158

\bibitem{avigad97}
Avigad, J.:
\newblock {The Combinatorics of Propositional Provability}.
\newblock In: ASL Winter Meeting, San Diego (January 1997)

\bibitem{DBLP:journals/mlq/Kirby07}
Kirby, L.:
\newblock {Addition and multiplication of sets}.
\newblock Math. Log. Q. \textbf{53}(1) (2007)  52--65

\bibitem{DBLP:conf/foiks/LeontjevS00}
Leontjev, A., Sazonov, V.Y.:
\newblock {Capturing LOGSPACE over Hereditarily-Finite Sets}.
\newblock In Schewe, K.D., Thalheim, B., eds.: FoIKS. Volume 1762 of Lecture
  Notes in Computer Science., Springer (2000)  156--175

\bibitem{DBLP:conf/types/Paulson94}
Paulson, L.C.:
\newblock {A Concrete Final Coalgebra Theorem for ZF Set Theory}.
\newblock In Dybjer, P., Nordstr{\"o}m, B., Smith, J.M., eds.: TYPES. Volume
  996 of Lecture Notes in Computer Science., Springer (1994)  120--139

\bibitem{DBLP:journals/tcs/CegielskiR01}
C{\'e}gielski, P., Richard, D.:
\newblock Decidability of the theory of the natural integers with the cantor
  pairing function and the successor.
\newblock Theor. Comput. Sci. \textbf{257}(1-2) (2001)  51--77

\bibitem{DBLP:conf/ipps/Rosenberg02a}
Rosenberg, A.L.:
\newblock Efficient pairing functions - and why you should care.
\newblock International Journal of Foundations of Computer Science
  \textbf{14}(1) (2003)  3--17

\bibitem{rationals}
Gibbons, J., Lester, D., Bird, R.:
\newblock Enumerating the rationals.
\newblock Journal of Functional Programming \textbf{16}(4) (2006)

\end{thebibliography}

\section*{Appendix}

The code in the paper is organized in a module with the following dependencies:

\begin{codex}
module ISO where
import Data.List
import Data.Bits
import Data.Graph
import Data.Graph.Inductive
import Graphics.Gnuplot.Simple
import Data.Char
import Ratio
import Random
\end{codex}

\subsection*{Bit crunching functions} 

The function
bitcount computes the number of bits needed to represent an integer and
max\_bitcount computes the maximum bitcount for a list of integers.
\begin{code}
bitcount n = head [x|x<-[1..],(2^x)>n]
max_bitcount ns = foldl max 0 (map bitcount ns)
\end{code}

The following function convert a number to to binary, padded with 0s, up to maxbits.
\begin{code}
to_maxbits maxbits n = 
  bs ++ (genericTake (maxbits-l)) (repeat 0) where 
    bs=to_base 2 n
    l=genericLength bs
\end{code}

\subsection*{Primes}
The following code implements factoring function {\tt to\_primes} a primality
test ({\tt is\_prime}) and a generator for the infinite stream of prime numbers
{\tt primes}.

\begin{code}
primes = 2 : filter is_prime [3,5..]

is_prime p = [p]==to_primes p

to_primes n | n>1 = to_factors n p ps where 
  (p:ps) = primes

to_factors n p ps | p*p > n = [n]
to_factors n p ps | 0==n `mod` p = p : to_factors (n `div` p)  p ps 
to_factors n p ps@(hd:tl) = to_factors n hd tl
\end{code}

We will briefly describe here the functions used to visualize various data
types with the help of Haskell libraries providing interfaces to {\tt graphviz}
and {\tt gnuplot}.

\subsection*{Multiset Operations}
The following functions provide multiset analogues of the usual set operations,
under the assumption that multisets are represented as non-decreasing sequences.
\begin{code}
msetInter [] _ = []
msetInter _ [] = []
msetInter (x:xs) (y:ys) | x==y = (x:zs) where zs=msetInter xs ys
msetInter (x:xs) (y:ys) | x<y = msetInter xs (y:ys)
msetInter (x:xs) (y:ys) | x>y = msetInter (x:xs) ys

msetDif [] _ = []
msetDif xs [] = xs
msetDif (x:xs) (y:ys) | x==y = zs where zs=msetDif xs ys
msetDif (x:xs) (y:ys) | x<y = (x:zs) where zs=msetDif xs (y:ys)
msetDif (x:xs) (y:ys) | x>y = zs where zs=msetDif (x:xs) ys

msetSymDif xs ys = sort ((msetDif xs ys) ++ (msetDif ys xs))

msetUnion xs ys = sort ((msetDif xs ys) ++ (msetInter xs ys) ++ (msetDif ys xs))
\end{code}

\subsection*{Building a multigraph from a natural number using a function
associating to each natural number a sequence or set of natural numbers.}

\begin{code}
fun2g ns = nat2fgs nat2fun ns
set2g ns = nat2sgs nat2set ns
perm2g ns = nat2fgs nat2perm ns
pmset2g ns = nat2fgs nat2pmset ns
bmset2g ns = nat2fgs nat2bmset ns

nat2fg f n = nat2gx fun_edge f nat2pftree n :: Gr Nat Int

nat2fgs f ns = nat2gsx fun_edge f nat2pftree ns :: Gr Nat Int

nat2sg f n = nat2gx set_edge f nat2pftree n :: Gr Nat ()

nat2sgs f ns = nat2gsx set_edge f nat2pftree ns :: Gr Nat ()

set_edge xs (a,b,i) = (lookUp a xs,lookUp b xs,())

fun_edge xs (a,b,i) = (lookUp a xs,lookUp b xs,i)

nat2gx e f g n = mkGraph vs  (map (e xs) es) where 
  es=g f n
  (xs,vs)=labeledVertices es

nat2gsx e f g ns = mkGraph vs  (map (e xs) es)  where 
  es=nub (concatMap (g f) ns)
  (xs,vs)=labeledVertices es
  
labeledVertices es= (xs,vs) where  
  xs=fvertices es
  is=[0..(length xs)-1]
  vs = zip is xs
     
nat2pftree f n = nub (nat2pftreex f (n,n,0))

nat2pftreex f (_,n,_) = ps ++ (concatMap (nat2pftreex f) ps) where
  ps = nat2pfun f (n,n,0)

nat2pfun _ (_,0,_) = []
nat2pfun f (_,n,_) | n> 0 = ps where 
  ps = zipWith (\x i->(n,x,i)) (f n) [0..]

fvertices ps = (sort . nub) (concatMap f ps) where
  f (a,b,_) = [a,b]

lookUp n ns = i where Just i=elemIndex n ns
\end{code}

\subsection*{Building Inductive Graphs from Lists of Pairs}

\begin{code}
pairs2gr ::  [(Nat,Nat)] -> Gr Nat ()

pairs2gr ps = mkGraph lvs les where 
  vs=to_vertices ps
  lvs=zip [0..] vs
  es=to_edges vs ps
  les=map f es
  f (x,y) = (x,y,())

to_vertices es = sort $ nub $ concat [[fst p,snd p]|p<-es]

to_edges vs ps = map (f vs) ps where
  f vs (x,y) = (lookUp x vs,lookUp y vs)
\end{code}

\subsection*{Generating labeled edge triplets by recursing over unpairing
functions} 

The following function represents a number as a set of triplets expressing
branches of decomposition with an unpairing function {\tt f}, for
instance, in the case of BDDs with function {\tt bitunpair}.

\begin{code}
unpairing_edges f tt = nub  (h f tt) where
  h _ tt | tt<2 = []
  h f n  = ys where
     (n0,n1)=f n
     ys= (n,n0,0):(n,n1,1):
           (h f n0) ++ 
           (h f n1)
\end{code}
The function works as follows:
\begin{codex}
*ISO> unpairing_edges bitunpair 42
[(42,0,0),(42,7,1),(7,3,0),(7,1,1),(3,1,0),(3,1,1)]
*JFISO> unpairing_edges pepis_unpair 42
[(42,0,0),(42,21,1),(21,1,0),(21,5,1),(5,1,0),(5,1,1)]
*ISO> 
\end{codex}

\subsection*{Generating labeled edge triplets by recursing over untupling
functions} 

The following function represents a number as a set of triplets expressing
branches of decomposition with an untupling function {\tt fk}, for
instance {\tt to\_tuple k}.

\begin{code}
untupling_edges f k tt = nub  (h f k tt) where
  h _ _ tt | tt<2 = []
  h f k n  = ys where
     ns = f k n
     ys = (zip3 (repeat n) ns [0..]) ++
          (concatMap (h f k) ns) 
  
\end{code}
The function works as follows:
\begin{codex}
*ISO> untupling_edges to_tuple 3 2008
[(2008,14,0),(2008,14,1),(2008,4,2),(14,2,0),(14,1,1),(14,1,2),
 (2,0,0),(2,1,1),(2,0,2),(4,0,0),(4,0,1),(4,1,2)]
\end{codex}

\subsection*{Building Inductive Graphs from Unpairing and Untupling Trees}

We can now turn a BDD as well as any other unpairing function generated tree
into an inductive graph, as follows:
\begin{code}
to_unpair_graph f tt = nat2fun_graph (unpairing_edges f) tt

to_untuple_graph f k tt = nat2fun_graph (untupling_edges f k) tt

nat2fun_graph f n = mkGraph vs fs :: Gr Nat Int where
  es=f n
  (xs,vs)=labeledVertices es
  fs=map (fun_edge xs) es
\end{code}
The functions work as follows:
\begin{codex}
*ISO> to_unpair_graph bitunpair 42
0:0->[]
1:1->[]
2:3->[(1,1),(0,1)]
3:7->[(0,2),(1,1)]
4:42->[(1,3),(0,0)]

*ISO> to_unpair_graph pepis_unpair 42
0:0->[]
1:1->[]
2:5->[(1,1),(0,1)]
3:21->[(1,2),(0,1)]
4:42->[(1,3),(0,0)]

*ISO> to_untuple_graph to_tuple 3 2008
0:0->[]
1:1->[]
2:2->[(2,0),(1,1),(0,0)]
3:4->[(2,1),(1,0),(0,0)]
4:14->[(0,2),(2,1),(1,1)]
5:2008->[(2,3),(1,4),(0,4)]
\end{codex}

\subsection*{Visualization with graphviz}

\begin{code}   
gviz g = writeFile "iso.gv" 
  ((graphviz g "" (0.0,0.0) (2,2) Portrait)++"\n")

funviz f n = gviz (nat2fg f n)  

setviz f n = gviz (nat2sg f n)

pviz t n = gviz (pairs2gr (as t nat n))

uviz f tt = gviz (to_unpair_graph f tt)

tviz f k tt = gviz (to_untuple_graph f k tt)
\end{code}

\subsection*{Plotting with gnuplot}

\begin{code}
plot3d f xs ys = plotFunc3d [Title ""] [] xs ys f

cplot3d f = plot3d (curry f)

plotpairs m | m<=2^8 = cplot3d bitpair ls ls where ls=[0..m-1]

plotdyadics m = plotList 
  [Title "Dyadics"] 
  (map (fromRational . (as dyadic nat)) [0..m-1])

sizes_to m t = map (size_as t) [0..m-1]

plot_hf m = plotLists [Title "Bit, BDD, HFF, HFS, and HFP sizes"] 
  ( 
    [bits_to m,bsizes_to m] ++ 
    (map (sizes_to m) [hff,hfs,hfm,hfp])
  )


plot_best m = plotLists [Title "Bit, BDD and HFF and HFF' sizes"] 
  ( 
    [bits_to m,bsizes_to m] ++ 
    (map (sizes_to m) [hff,hff'])
  )

plot_worse m = plotLists [Title "HFM, HFS and HFP sizes"] 
  ( 
    (map (sizes_to m) [hfm,hfs,hfp])
  )

plot hf m = plotx [hf] m

plotx hfx m = plotLists [Title "HF tree size"] 
  ( 
    (map (sizes_to (2^m-1)) hfx)
  )

-- plots pairs
pplot f m = plotPath [] (map (to_ints . f) [0..2^m-1]) 

zplot f m = plotPath [] (map (to_ints . f) [-(2^m)..2^m-1]) 

to_ints (i,j)=(fromIntegral i,fromIntegral j)

diplot n = plotPath [] (map to_ints (as digraph nat n))

bsize_of n = robdd_size (as rbdd nat n)

bsizes_to m = map bsize_of [0..m-1]

bits_to m = map s [0..m-1] where s n = genericLength (as bits nat n)

plot_linear_sparseness m = plotLists [Title "Linear Sparseness"] 
  [(map (linear_sparseness fun) [0..m-1]),
   (map (linear_sparseness pmset) [0..m-1]),
   (map (linear_sparseness mset) [0..m-1]),
   (map (linear_sparseness set) [0..m-1]),
   (map (linear_sparseness perm) [0..m-1])]


plot_sparseness m = plotLists [Title "Recursive Sparseness"] 
  [(map (sparseness hff_pars) [0..m-1]),
   (map (sparseness hfpm_pars) [0..m-1]),
   (map (sparseness hfm_pars) [0..m-1]),
   (map (sparseness hfs_pars) [0..m-1]),
   (map (sparseness hfp_pars) [0..m-1])]

plot_sparseness1 m = plotLists 
  [Title "Recursive Sequence vs. Multiset Sparseness"] 
  [
   (map (sparseness hff_pars) [0..m-1]),
   (map (sparseness hfpm_pars) [0..m-1])
  ]
  
plot_sparseness2 m = plotLists [Title "Recursive Multiset Sparseness"] 
  [
   (map (sparseness bhfm_pars) [0..m-1]),
   (map (sparseness hfm_pars) [0..m-1])
  ]

plot_sparseness3 m = plotLists [Title "Recursive Multiset Sparseness"] 
  [
   (map (sparseness hff_pars) [0..m-1]),
   (map (sparseness hff_pars') [0..m-1])
  ]


plot_sparseness4 m = plotLists 
  [Title "Recursive Multiset vs Multiset with Primes Sparseness"] [
   (map (sparseness hfm_pars) [0..m-1]),
   (map (sparseness hfpm_pars) [0..m-1])
  ]

plot_sparseness5 m = plotLists 
  [Title "Recursive Multisets vs. Sequences"] [
   (map (sparseness hff_pars) [0..m-1]),
   (map (sparseness hfm_pars) [0..m-1])
  ]
        
plot_selfdels m = plotLists 
   [Title "Self-delimiting codes: Undelimited vs. Elias vs. HFF"] 
   [(map (genericLength . (as bits nat)) [0..m-1]),
    (map (genericLength . (as elias nat)) [0..m-1]),
    (map (genericLength . (as hff_pars nat)) [0..m-1])]

plot_pairs_prods m = plotLists [Title "Pairs vs. products"]  
   [ms,prods] where
     ms=[1..m]
     pairs=map bitunpair ms
     prods=map prod pairs where prod (x,y)=2*x*y
 
plot_lifted_pairs m = 
   plotLists [Title "Lifted pairs"]  [us0,us1] where 
     ms=[0..m-1]
     pairs=map bitunpair ms
     us0=map fst pairs
     us1=map snd pairs
     
plot_lifted_pairs1 m = 
   plotLists [Title "Lifted pairs and products"]  [ps,s0,s1,xys] where 
     ms=[0..m-1]
     pairs=map bitunpair ms
     us0=map fst pairs
     us1=map snd pairs
     ps=zipWith (*) us0 us1
     s0=map (^2) us0
     s1=map (^2) us1
     xys=map f pairs where
       f (x,y) = x*y
 
plot_primes_prods m = plotLists [Title "Primes vs. products"]  
  [ps,prods] where 
     ms=[0..m]
     ps=genericTake m primes
     pairs=map bitunpair ps
     prods=map prod pairs where prod (x,y)=2*x*y     
   
plot_hypers_prods m = plotLists [Title "Hyper-primes vs. products"]
   [ps,prods] where 
     ms=[0..m]
     ps=genericTake m (hyper_primes bitunpair)
     pairs=map bitunpair ps
     prods=map prod pairs where prod (x,y)=2*x*y 
       
\end{code}

\subsection*{Generated Figures}

\begin{code}
f1=gviz  (nat2sg nat2set 2008)
f2=gviz (nat2fg nat2fun 2008)
f2a=gviz (nat2fg nat2mset 2008)
f3=gviz (nat2fg nat2perm 2008)

f4=gviz (nat2fg nat2perm 2009)
f5=pviz digraph 2008

f6=plotpairs 64
f7=plotdyadics 256

f8=plot_best (2^6)
f9=plot_worse (2^10)
f10=plot_hf (2^8)

f11a=plot_linear_sparseness (2^7)
f11=plot_sparseness (2^8)
f11b=plot_sparseness1 (2^8)
f11c=plot_sparseness2 (2^10)

f12=plot_sparseness (2^14)

f13=plot_sparseness (2^17)

f14=plot_selfdels (2^7)

f15=plotList [] (sparses_to (2^18))
\end{code}

\begin{code}

f16=gviz (nat2fgs nat2fun [0..7])

arp24 i =468395662504823 + 205619*23*i

arps24 = map arp24 [0..23]

arp25 i = 6171054912832631 + 366384*23*i

arps25 = map arp25 [0..24]

f17 = gviz (fun2g arps24)
f17a = gviz (fun2g arps25)

f18 = gviz (fun2g [2^65+1,2^131+3])

f18a = gviz (set2g [2^65+1,2^131+3])


f19 = gviz (fun2g [0..7])

f20 = gviz (pmset2g [0..7])

f20a = gviz (bmset2g [0..7])

f21 = gviz (set2g [0..7])

f22 = gviz (perm2g [0..7])

g1 tt= uviz bitunpair tt
g2 tt= uviz pepis_unpair tt
g2' tt= uviz pepis_unpair' tt
g3 tt= uviz rpepis_unpair tt
isofermat=uviz mset_unpair 65537
isofermat1=uviz mset_unpair 142781101
isonfermat=uviz mset_unpair 34897

isopairs = plot_pairs_prods 256
isoprimes = plot_primes_prods 256
isohypers = plot_hypers_prods 256

isounpair1=pplot bitunpair 10
isounpair2=pplot pepis_unpair 10
isounpair3=pplot mset_unpair 10

isozunpair n=zplot zunpair n

ms2pms n = as nat pmset (as mset nat n)

pms2ms n = as nat mset (as pmset nat n)

kms2pms 0 n = n
kms2pms k n = ms2pms (kms2pms (k-1) n) 

kpms2ms 0 n = n
kpms2ms k n = pms2ms (kpms2ms (k-1) n) 

lms k m = [x|x<-[0..2^m-1], kms2pms k x < kpms2ms k x]

xms k m = [x|x<-[0..2^m-1],kms2pms k x < x]

eqms k m = [x|x<-[0..2^m-1],kms2pms k x == x]

xpms k m = [x|x<-[0..2^m-1],kpms2ms k x < x]

eqpms k m = [x|x<-[0..2^m-1],kpms2ms k x == x]

qms k m = 
 [(toRational (kpms2ms k x)) - (toRational (kms2pms k x))|x<-[1..2^m-1]]

q1 k m = plotList []  (qms k m)

q2 k m = plotLists []  
  [map (kms2pms k) xs,map (kpms2ms k) xs] where 
    xs = [0..2^m-1]

mult_vs_pairing p1 p2 = (p1*p2) % (ppair bitpair (p1,p2))
mult_vs_mset_pairing p1 p2 = (p1*p2) % (ppair mset_pair (p1,p2))

q3 n = plotFunc3d 
        [Title "Prime Multiplication vs. Prime Pairing"] [] 
          ps ps mult_vs_pairing where
            ps=genericTake n primes

q4 n = plotFunc3d 
  [Title "Prime Multiplication vs. Prime Multiset Pairing"] [] 
        ps ps mult_vs_mset_pairing where
        ps=genericTake n primes
       
n4a n = plotFunc3d [Title "Multiplication"] [] 
        ps ps (*) where
          ps=[0..2^n-1]

n4b n = plotFunc3d [Title "Multiset Pairing"] [] 
        ps ps (curry mset_pair) where
          ps=[0..2^n-1]


n4c n = plotFunc3d [Title "mprod operation"] [] 
        ps ps (mprod) where
          ps=[0..2^n-1]

n4d n = plotFunc3d [Title "pmprod' operation"] [] 
        ps ps (pmprod') where
          ps=[0..2^n-1]

n4e n = plotFunc3d [Title "mprod' operation"] [] 
        ps ps (mprod') where
          ps=[0..2^n-1]


n4f n = plotFunc3d [Title "mprod' x y/ x * y"] [] 
        ps ps (\x y->(mprod' x y) % (x*y)) where
          ps=[1..2^n]
          
expMexp k m = plotLists []  
   [map (\x->x^k) xs, map (\x->mexp' x k) xs] where 
   xs = [0..2^m]
    

p4a n = plotFunc3d [Title "Prime Multiplication"] [] 
        ps ps (*) where
          ps=genericTake n primes

p4b n = plotFunc3d [Title "Prime Multiset Pairing"] [] 
        xs ys (curry mset_pair) where
        ps=genericTake n primes
        xs=ps
        ys=ps

p4c n = plotFunc3d [Title "mprod on primes"] [] 
        xs ys (mprod) where
        ps=genericTake n primes
        xs=ps
        ys=ps

p4d n = plotFunc3d [Title "pmprod on primes"] [] 
        xs ys (pmprod) where
        ps=genericTake n primes
        xs=ps
        ys=ps
 
p4f n = plotFunc3d [Title "mprod' x y/ x * y"] [] 
        ps ps (\x y->(mprod' x y) % (x*y)) where
          ps=genericTake n primes
                         
q4c n = plotFunc3d [Title "Prime Pairing"] [] 
        ps ps (curry bitpair) where
          ps=genericTake n primes
                
q5 n = plotLists 
  [Title "Prime Multiplication vs. Prime Pairing curves"] 
  [prods,pairs] where 
    us= map bitunpair [0..2^n-1]
    (xs,ys) = unzip us  
    ps=primes
    xs'=map (from_pos_in ps) xs
    ys'=map (from_pos_in ps) ys
    prods = zipWith (*) xs' ys'
    us'=zip xs' ys'
    pairs= map (ppair mset_pair) us'

plot_gauss_op f m = plotFunc3d title [] zs zs (curry f) where
  title=[Title "Gauss Integer operations through Pairing Functions"]
  zs=[-2^m..2^m-1]

gsum m = plot_gauss_op gauss_sum m
gdif m = plot_gauss_op gauss_dif m

gprod m = plot_gauss_op gauss_prod m

\end{code}

\end{document}